\documentclass{ar-astro2e}
\usepackage{epsfig}
\usepackage{graphics}
\usepackage{amsmath}
\usepackage{ARAstroBib}
\newcommand{\LCDM}{$\Lambda$CDM}

\newcommand{\rvir}{\mbox{$r_{\rm vir}$}}
\newcommand{\vvir}{\mbox{$V_{\rm vir}$}}

\newcommand{\mstar}{\mbox{$m_{\rm star}$}}

\newcommand{\HI}{\ion{H}{i}}

\newcommand{\msolar}{\;{\rm M}_{\odot}}

\newcommand{\gad}{{\sc Gadget-2}}
\newcommand{\arepo}{{\sc Arepo}}

\newcommand{\ion}[2]{\hbox{#1\,{\sc #2}}}

\newcommand{\fgas}{f_{\rm gas}}

\newcommand{\la}{\,\rlap{\raise 0.5ex\hbox{$<$}}{\lower 1.0ex\hbox{$\sim$}}\,}
\newcommand{\ga}{\,\rlap{\raise 0.5ex\hbox{$>$}}{\lower 1.0ex\hbox{$\sim$}}\,}

\newcommand{\Htwo}{\mbox{${\rm H}_{2}$}}

\usepackage{color}

\begin{document}
\input psfig.sty

\title{Physical Models of Galaxy Formation in a Cosmological
  Framework}

\markboth{Somerville \& Dav\'{e}}{Models of Galaxy Formation}

\author{Rachel S. Somerville
\affiliation{Department of Physics and Astronomy, Rutgers University \\
somerville@physics.rutgers.edu}
Romeel Dav\'{e}
\affiliation{Department of Physics, University of the Western Cape, Cape Town\\
South African Astronomical Observatories, Cape Town\\
African Institute for Mathematical Sciences, Cape Town\\
romeeld@gmail.com}
}


\begin{abstract} 
Modeling galaxy formation in a cosmological context presents one of
the greatest challenges in astrophysics today, due to the vast range
of scales and numerous physical processes involved.  Here we review
the current status of models that employ two leading techniques to
simulate the physics of galaxy formation: semi-analytic models and
numerical hydrodynamic simulations. We focus on a set of observational
targets that describe the evolution of the global and structural
properties of galaxies from roughly Cosmic High Noon ($z\sim 2$--3) to
the present.  Although minor discrepancies remain, overall, models
show remarkable convergence between different methods and make
predictions that are in qualitative agreement with
observations. Modelers seem to have converged on a core set of
physical processes that are critical for shaping galaxy properties.
This core set includes cosmological accretion, strong stellar-driven
winds that are more efficient at low masses, black hole feedback that
preferentially suppresses star formation at high masses, and
structural and morphological evolution through merging and
environmental processes.  However, all cosmological models currently
adopt phenomenological implementations of many of these core
processes, which must be tuned to observations. Many details of how
these diverse processes interact within a hierarchical structure
formation setting remain poorly understood. Emerging multi-scale
simulations are helping to bridge the gap between stellar and
cosmological scales, placing models on a firmer, more physically
grounded footing.  Concurrently, upcoming telescope facilities will
provide new challenges and constraints for models, particularly by
directly constraining inflows and outflows through observations of gas
in and around galaxies.
\end{abstract}

\maketitle

\section{INTRODUCTION}
\label{sec:intro}

The past decade has seen remarkable progress in measuring the
properties of galaxies across the electromagnetic spectrum and over
the majority of cosmic history. Wide-field surveys have collected
samples of millions of nearby galaxies, spanning roughly six orders of
magnitude in galaxy mass and a rich range of galaxy types and
environments, from isolated galaxies in voids to rich clusters.
Medium-deep surveys have collected samples of tens of thousands of
galaxies out to $z\sim 6$, and ultra-deep surveys have identified
samples of hundreds to thousands of galaxy candidates at $z\sim 6$--8,
with a few candidates identified (mainly behind lensing clusters) at
$z\sim 9$--10 \citep[for an overview of recent surveys
see][]{Madau:2014}. The pan-chromatic wavelength coverage enabled by
a suite of ground and space based telescopes has allowed detailed
Spectral Energy Distributions (SED) to be constructed for large
samples of galaxies, which make it possible to estimate photometric
redshifts for galaxies that are too faint to readily obtain
spectroscopy, and to estimate intrinsic parameters such as stellar
masses and star formation rates (SFR).

In addition, high spatial resolution imaging, primarily from the
Hubble Space Telescope (HST), and spectroscopy including data from an
increasing number of surveys using Integral Field Spectrographs
have enabled us to study galaxies' internal structure and
kinematics. In particular, the Wide Field Camera 3 on HST has made it
possible to study galaxy structure and morphology in the rest-frame
optical back to ``Cosmic Noon'' --- the peak of cosmic star formation
(SF) and black hole (BH) accretion activity at $z\sim 2$--3
\citep{Madau:2014}.  We are truly living in a golden age of facilities
and databases for studying how galaxies formed and evolved.

Concurrently over the past decade, advances in numerical methodologies
and computing speed have allowed extraordinary progress in our ability
to simulate the formation of structure within the paradigm of the Cold
Dark Matter (CDM) model
\citep[e.g.][]{Springel_mill:2005,Klypin:2011}.  A variety of
techniques have been developed for computing detailed predictions for
the expected observable properties of galaxies based on \emph{ab
  initio} (albeit approximate) treatments of the physical processes
expected to be important in shaping galaxy formation and evolution.
The project of genuinely \emph{ab initio} computational simulation of
galaxy formation is beyond current capabilities, owing to the vast
range of spatial scales involved, from the sub-pc scales of individual
stars and supernovae, and accretion disks of supermassive BH, to the
super-Mpc scales of the ``cosmic web'', and to the wide array of
poorly understood physical processes.  However, by zeroing in on
different scales through many different approaches, models are
providing fundamental insights into the physical processes that are
responsible for molding galaxy properties.  Cosmological galaxy
formation models have now matured into an essential tool for
understanding galaxy evolution, and hence it is timely to review this
topic.

We focus specifically on state-of-the-art physically-motivated
cosmological models of galaxy formation, and ask a series of
questions: 1) How well are these models able to predict or reproduce
the observed distribution functions of global galaxy properties such
as stellar mass, and the evolution of these functions? 2) How well do
the models reproduce global scaling relations such as correlations
between stellar mass, cold gas fraction, SFR and metallicity? 3) What
do these models predict for the demographics of different types of
galaxies (e.g. star forming vs. quiescent, or disk-dominated
vs. spheroid-dominated)? 4) Are models able to reproduce observed
structural scaling relations such as those relating mass with radial
size, density, and internal velocity, and the evolution of these
relations for different types of galaxies? 5) With regard to all of
these questions, what insights have we gained into the process of
galaxy formation from the successes and failures of our current
models?

The plan for the rest of this article is as follows. In
\S\ref{sec:intro:obs}, we give a broad overview of the observational
results that we will target in our review. In \S\ref{sec:intro:cosmo},
we review the cosmological background, mainly pointing the reader to
other sources. In \S\ref{sec:intro:physics} we give a brief overview
of the physical processes that are included in most models of galaxy
formation, and in \S\ref{sec:intro:tools} we introduce different tools
for modeling galaxy properties. In \S\ref{sec:tools} we give a more
detailed description of the methods used in the models that we will
discuss in the remainder of the article, which include numerical
hydrodynamic simulations and semi-analytic models.  In
\S\ref{sec:subgrid}, we discuss the ``sub-grid'' modeling connected
with physical processes that are not directly resolved in cosmological
simulations, including the formation of stars and supermassive black
holes (SMBH), and the impact of ``feedback'' from these objects on
forming galaxies. In \S\ref{sec:results}, we discuss the predictions
of current models and how they measure up to observations for global
properties of galaxies (\S\ref{sec:global}) and galaxy internal
structure and kinematics (\S\ref{sec:structure}).  We conclude with a
summary and outlook in \S\ref{sec:summary}. A glossary of acronyms is
provided at the end of the paper.

\subsection{Observational Targets}
\label{sec:intro:obs}

In this review, we focus on the global and structural properties of
the stellar components of galaxies from roughly Cosmic High Noon
($z\sim 2$--3) to the present.  We acknowledge that there are many
important observations that provide crucial constraints on models that
lie beyond this scope.  The summary here is quite brief; we will refer
the reader to other recent reviews and papers for a more comprehensive
overview.

\subsubsection{Global Properties: Distribution Functions}
Multi-wavelength imaging surveys complemented with photometric or
spectroscopic redshifts yield estimates of familiar global galaxy
properties such as the luminosity and color at various rest-frame
wavelengths from the UV to far-IR. In recent years, it has become
popular to estimate stellar masses by fitting galaxy SEDs with simple
parametric models of galaxy star formation histories combined with
stellar population models \citep{Walcher:2011,Conroy:2013}.  Star
formation rates are also estimated using SED modeling, or roughly
equivalently using extinction-corrected rest-UV measures, but more
reliably by adding mid- to far-IR photometry and/or nebular emission
lines such as H$\alpha$.  We refer to the comoving number density of
galaxies as a function of a global property such as luminosity or
stellar mass as a \emph{distribution function}. It has long been known
that galaxy distribution functions typically have a characteristic
shape often described by a \emph{Schechter
  function}~\citep{Schechter:1976}, which is parameterized by a
normalization, a turn-over, and an asymptotic slope to low masses.
Examples of luminosity functions, stellar mass functions, and the cold
gas (atomic hydrogen) mass function of nearby galaxies from recent
large surveys are presented in the review by \citet{Blanton:2009}.

To higher redshifts, galaxy rest-frame optical-NIR luminosity
functions and stellar mass functions (SMF) have been measured from
medium-deep surveys out to $z\sim 4$.  At higher redshifts, SMF
estimates exist but rely on stellar mass estimates from rest-UV
fluxes, which are likely less robust.  These measurements have yielded
a number of important insights into galaxy assembly: 1) galaxies
appear to be continuously building up their mass over cosmic time, in
accord with the hierarchical formation picture, and inconsistent with
monolithic collapse \citep{Madau:2014}.  2) The number density of
massive galaxies ($\mstar > M_{\rm char}$, where $\mstar$ is the
stellar mass and $M_{\rm char}$ is the characteristic mass in the
Schechter function) increases rapidly from $z\sim 4$--2, but then
stays nearly constant or increases slowly from $z\sim 2$--0,
indicating that massive galaxies formed and assembled their stars
relatively early
\citep{Marchesini:2009,Muzzin:2013,Moustakas:2013}. 3) The comoving
number density of low mass galaxies ($\mstar < M_{\rm char}$)
increases more rapidly than that of more massive galaxies at $z \la
1$--2, indicating that low mass galaxies formed their stars later and
over a longer timescale.  This result is sometimes called ``mass
assembly downsizing'' \citep{Cimatti:2006}.

It has long been known that the color-luminosity distribution of
galaxies is strongly \emph{bimodal} \citep[e.g.][]{Baldry:2004}, with
most galaxies falling onto a relatively narrow (in optical colors)
``red sequence'' or a broader ``blue cloud''. Spectroscopic indicators
of stellar population age as well as UV and IR photometry have
confirmed that in the local Universe, the ``red sequence'' is largely
comprised of ``quiescent'' galaxies with predominantly old stellar
populations, while the ``blue cloud'' represents ``star forming''
galaxies with younger stellar populations and significant ongoing star
formation
\citep{Kauffmann:2003a,Brinchmann:2004,Salim:2007,Schiminovich:2007}.
Because of the strongly bimodal nature of the population, it has
become common to draw a line either in the color-luminosity or
color-mass plane, or in the specific SFR (sSFR $\equiv SFR/\mstar$)
versus $\mstar$ plane, and to speak of ``red'' and ``blue'' galaxies
or ``star forming'' and ``quiescent'' galaxies.

Recent deep surveys have shown that these two populations (star
forming and quiescent) can be clearly identified at least up to $z\sim
2$, and perhaps up to higher redshifts $z\sim 3$--4
\citep{Brammer:2011,Muzzin:2013}.  Intriguingly, it appears that the
comoving number and mass density of \emph{quiescent} galaxies has been
increasing over time since $z\sim 2$, while the number and mass
density of \emph{star forming} galaxies has stayed roughly constant or
decreased during this same interval
\citep{Bell:2004,Bell:2007,Faber:2007,Brammer:2011,Muzzin:2013}. Given
that it is the star forming population that is expected to be growing
more massive due to the birth of new stars, this result has profound
and unexpected implications --- it implies that more and more
star-forming galaxies must be having their star formation extinguished
or ``quenched'' as cosmic time progresses.

\subsubsection{Global Properties: Scaling Relations} Galaxies show
many correlations between their global properties.  We refer to such a
correlation as a ``scaling relation'' when the conditional value of
galaxy property $y$ for a fixed value of another property $x$ has a
relatively small scatter. Stellar mass is often used as the $x$
variable in galaxy scaling relations.  Some well-known examples of
global scaling relations with $\mstar$ are the SFR for star forming
galaxies, sometimes known as the ``star forming main
sequence''~\citep[SFMS; ][]{Noeske:2007,Wuyts:2011}, the mean fraction
of cold gas ($\fgas \equiv m_{\rm gas}/\mstar$) in the interstellar
medium (ISM) \citep{Baldry:2008,Peeples:2011}, and the metallicity of
stars or ISM gas \citep[mass-metallicity relation,
  MZR;][]{Gallazzi:2005,Tremonti:2004,Zahid:2013}.  Furthermore, some
of the tightest known scaling relations in astronomy are those between
galaxy properties and the mass of the SMBH they harbor \citep[see][for
  a comprehensive review]{Kormendy:2013}.

Deep multi-wavelength surveys have provided constraints on the
evolution of these scaling relations.  The normalization of the SFMS
has declined by a factor of $\sim 20$ since $z\sim 2$ \citep[][and
  references therein]{Speagle:2014}, and a fairly tight sequence
appears to be in place up to $z\sim 6$
\citep{Steinhardt:2014,Salmon:2014}.  The MZR seems to have evolved in
the sense that galaxies of a given mass had lower gas-phase
metallicities at high redshift
\citep{Savaglio:2005,Erb:2006,Wuyts:2014,Zahid:2013,Steidel:2014}.
There is evidence from measurements of CO (an indirect tracer of
molecular hydrogen, \Htwo) in fairly massive high redshift galaxies
that the gas fraction of galaxies has decreased significantly over
cosmic time since $z\sim 2$
\citep{Tacconi:2010,Tacconi:2013,Bothwell:2013b,Saintonge:2013,Genzel:2014}.
Indirect estimates of cold gas fractions from inverted star formation
densities, assuming a fixed relationship between star formation
density and cold gas density, also indicate a rapid decrease in cold
gas fraction from $z\sim 2$ to the present \citep[][Popping et al. in
  prep]{Erb:2006,Popping:2012}.

Some scaling relations show clear second-parameter dependences, in the
sense that the {\it scatter} about a given relation is correlated with
some other galaxy property.  For instance, the MZR may show a
second-parameter dependence on star formation, in the sense that more
rapidly star forming galaxies at a given mass have lower
metallicities~\citep{Mannucci:2010,Lara-Lopez:2010}.  The cold gas
content shows a similar correlation, with high \HI-mass galaxies having
lower metallicities~\citep{Bothwell:2013a,Lara-Lopez:2013}.

\subsubsection{Demographics: Correlations with Galaxy Type}

Since the original discovery of fuzzy ``nebulae'' it has been known
that galaxies come in different morphological types
\citep{Hubble:1926}. There are many different methods for quantifying
and classifying galaxy morphology, and this subject is reviewed in
\citet{Conselice:2014}; see also the more nearby-Universe focussed
discussion in \citet{Buta:2013}.  Although galaxy morphology
encompasses many complex facets of galaxy structure including the
presence of bulges, thin and thick disks, bars, spiral arms, etc., for
the purposes of this article we focus on a single simplified metric:
the fraction of a galaxy's light or mass contributed by a flattened,
rotationally supported \emph{disk}, and that contained in an oblate or
triaxial, pressure supported \emph{bulge} or \emph{spheroid} (often
denoted by the bulge-to-disk ratio $B/D$ or bulge-to-total ratio
$B/T$). The bulge-to-disk ratio is broadly correlated with classical
Hubble type \citep{Simien:1986}. We will further simplify much of our
discussion by referring to just two classes of galaxies,
``disk-dominated'' and ``spheroid dominated''\footnote{In this paper
  we use the term ``spheroid'' to mean galaxies or galaxy components
  that structurally and kinematically resemble classical giant
  ellipticals, not to be confused with ``dwarf spheroidals'' or
  ``spheroidal galaxies''. See \S\ref{sec:strucscale} for further
  explanation}. Unfortunately, there is no standard value for the
critical value of $B/T$ used to divide these populations, with values
from $0.3<(B/T)_{\rm crit} <0.7$ used in the literature. As it is
difficult to robustly decompose the light of observed galaxies into a
spheroid and disk component, other metrics such as the concentration
(the ratio of the radius containing 90\% of the light to the radius
containing 50\% of the light) or the ``Sersic index'' \citep[$n_s$;
  another measure of the `slope' of the light profile;
  e.g.][]{Blanton:2009} are frequently used as rough proxies for
morphology. We deliberately avoid using the terms ``early type'' and
``late type'' as they are sometimes used to refer to galaxy classes
divided by morphology and sometimes to those divided according to
their stellar populations (star forming vs. quiescent).

Regardless of how galaxies are classified, there are robust
demographic trends for disk-dominated vs. spheroid-dominated
galaxies. There is a very strong trend between morphology and color or
star formation activity, such that disk-dominated galaxies are
predominantly blue and star forming, while spheroid-dominated galaxies
are largely red and quiescent, with nearly uniformly old stellar
populations
\citep[e.g.][]{Roberts:1994,Kauffmann:2003a,Blanton:2009}. This trend
appears to hold up to $z \sim 2$, with the caveat that red optical
color becomes a less robust tracer of old stellar populations, as many
star-forming galaxies at high redshift are reddened by dust. The
characteristic Schechter function mass $M_{\rm char}$ is larger for
quiescent or spheroid-dominated galaxies, and the slope is much
shallower \citep[e.g.][]{Bernardi:2010}. Put another way, the fraction
of spheroid-dominated galaxies increases strongly with stellar mass
and luminosity. Furthermore, as emphasized by \citet{Binggeli:1988},
different types of galaxies can have luminosity functions that deviate
considerably from the Schechter form.

A number of studies have shown that the probability for a galaxy to be
quiescent depends on both its stellar mass and large-scale environment
or halo mass \citep{Balogh:2004,Hogg:2004,Peng:2010,Woo:2013}.  Recent
works have shown that the correlation between quiescence and other
internal properties such as spheroid fraction, velocity dispersion,
and central density is even stronger than that with stellar mass
\citep{Bell:2012,Cheung:2012,Lang:2014,Bluck:2014}.

\subsubsection{Structural Scaling Relations}
\label{sec:strucscale}

Both disks and spheroids exhibit correlations between their stellar
mass or luminosity, their radial size, and their internal velocity
\citep{Faber:1976,Kormendy:1977,Tully:1977,Shen:2003,Courteau:2007,Bernardi:2010}. For
disk-dominated galaxies, the radial size is usually characterized by
the \emph{scale radius} $r_s$ (the scale radius in the exponential
function characterizing the radial light profile; e.g.
\citet{MvdBWbook}, Eqn. 2.29 p. 50) and the characteristic velocity is
the rotation velocity at the maximum of the rotation curve $V_{\rm
  rot}$, which usually occurs at around $2r_s$. For spheroid-dominated
galaxies, the radial size is characterized by the half light radius or
\emph{effective radius} $r_e$ (the radius that contains half of the
total luminosity), and the internal velocity is characterized by the
(line of sight) velocity dispersion $\sigma$. Several of these
relationships have names, such as the Tully-Fisher relation for disks
\citep[$L$-$V_{\rm rot}$;][]{Tully:1977}, and the Faber-Jackson
\citep[$L$-$\sigma$;][]{Faber:1976}, and Kormendy
\citep[$L$-$r_e$;][]{Kormendy:1977} relations for spheroids.  A
combination of these three quantities forms a \emph{Fundamental
  Plane}; i.e., galaxies populate a relatively thin plane in
$L$-$r$-$V$ space, or rescaled versions of these variables
\citep{Djorgovski:1987,Faber:1987,Bender:1992,Burstein:1997}. The
familiar named bivariate relations are projections of this plane.

The slope, scatter, and evolution of these structural scaling
relationships for spheroids and disks carry important clues about the
formation history and relationship between these objects. For example,
1) the size-mass relationship is considerably steeper for spheroids
than for disks at all redshifts
\citep{Shen:2003,Bernardi:2010,vanderwel:2014}; 2) since $z\sim 2$, the
size-mass relation for spheroids has evolved much more rapidly than
that for disks \citep{Trujillo:2006,vanderwel:2014}; 3) the size
distribution at fixed mass is narrower for spheroids than for disks
\citep{vanderwel:2014} 4) the evolution of the Tully-Fisher \emph{and}
Faber-Jackson relation has been relatively mild 
\citep{Kassin:2007,Miller:2011,Miller:2012,Cappellari:2009,Cenarro:2009}. We
note that many high redshift studies present the scaling relations for
galaxies divided according to whether they are star forming or
quiescent, rather than spheroid or disk dominated, but this seems to
make little difference to the qualitative results
\citep{vanderwel:2014}.

Another illustration of the importance of structural-kinematic scaling
relations is demonstrated by the distinction between ``classical''
bulges and ``pseudo''-bulges
\citep{Kormendy:2004,Kormendy:2013b}. Classical bulges have centrally
concentrated light profiles with Sersic indices $n_s \sim 2$--3 (where
a de Vaucouleur profile has $n_s = 4$), and lie on an extension of the
Fundamental Plane for giant ellipticals. Pseudobulges have more
extended light profiles that are more similar to disks ($n_s \sim 1$)
and lie on a different Fundamental Plane from classical bulges and
giant ellipticals \citep{Kormendy:2009}. Furthermore, classical bulges
and pseudobulges have different correlations with SMBH mass
\citep{Kormendy:2013}. Similarly, the dwarf galaxies that are
confusingly termed ``dwarf spheroidals'' and ``dwarf ellipticals''
obey very different Fundamental Plane relations than do classical
bulges and ellipticals of all luminosities
\citep{Kormendy:2009,Kormendy:2012}. In fact, dwarf spheroidals and
dwarf ellipticals are indistinguishable from dwarf irregulars in their
structural parameter correlations. These diverse scaling relations
hint at different formation mechanisms for these objects, as reviewed
in \citet{Kormendy:2013b}.

\subsection{Cosmological Background}
\label{sec:intro:cosmo}

Our modern theory of cosmology is based on the ansatz that the
Universe is homogeneous and isotropic on large scales (the
\emph{cosmological principle}), and Einstein's theory of General
Relativity (GR) that says that the structure of space-time is
determined by the mass and energy content of the Universe. Together
these allow us to derive equations that describe the evolution of the
\emph{scale factor} (or characteristic size and density) of the
Universe in terms of the parameters specifying the mass and energy
density. Observations have shown that the Universe started from a much
denser, hotter, and nearly homogeneous state and has been expanding
for approximately the past thirteen and a half billion years
\citep[e.g.][hereafter MvdBW]{MvdBWbook}.

In this standard picture, quantum fluctuations in the very early
Universe were processed during a period of very rapid expansion called
\emph{inflation} to create the small inhomogeneities that are detected
via temperature fluctuations in the Cosmic Microwave Background. These
tiny fluctuations, viewed at the time when free electrons combined
with nuclei to form neutral atoms at a redshift $z \simeq 1100$, have
now been studied in exquisite detail with a large number of
experiments, including the {\it Wilkinson Microwave Anisotropy Probe}
and {\it Planck} satellites. When combined with other observations
such as the distance-redshift relation from Type~Ia supernovae,
abundances of galaxy clusters, constraints on the present-day
expansion rate (Hubble parameter $H_0$) from nearby Cepheid stars, and
galaxy clustering (e.g. Baryon Acoustic Oscillations), these
measurements yield stringent constraints on the fundamental
cosmological parameters \citep{Hinshaw:2013,Planck:2013}.

These combined observations point to a Universe that is geometrically
flat and dominated by Dark Matter and Dark Energy, which together
account for more than 95\% of the energy density of the Universe.  The
physical nature of both of these mysterious substances is unknown,
although there are numerous candidates. In the most popular variant of
the standard model, which we will refer to as \LCDM, the dark matter
is ``cold'' and collisionless and makes up $\sim 25$\% of the cosmic
mass-energy density, and the dark energy is in the form of a
``cosmological constant'' $\Lambda$ (as expected in the most general
form of Einstein's equations of General Relativity), comprising $\sim
70$\%. The remaining 4\% is in {\it baryons} (which in this context
include leptons), i.e. normal atoms that make up stars, gas, and heavy
elements (``metals").  Although these cosmological parameters are
still uncertain by up to perhaps ten percent, for the purposes of
understanding how galaxies form and evolve, this level of uncertainty
is largely irrelevant.

With the initial conditions specified, if we neglect ``baryonic''
physics, it is relatively straightforward to compute how the density
field of the dominant dark matter component evolves as the Universe
expands. If we imagine the matter density field as a mountain range,
the landscape in the CDM picture is extremely craggy, with many small
scale peaks superimposed on top of the medium and large scale peaks
and valleys. As the Universe expands, the background density
decreases. When a peak exceeds a critical over-density relative to the
background, the region within that peak stops expanding and becomes
gravitationally self-bound. Numerical $N$-body techniques have been
used to extensively study and characterize the growth of structure in
dissipationless (dark matter only) \LCDM\ simulations, as we discuss
in \S\ref{sec:nbody}. The gravitationally bound structures that form
in these simulations are commonly referred to as \emph{dark matter
  halos}, and the abundance, internal structure, shape, clustering,
and angular momentum of these halos over cosmic time has been
thoroughly quantified (see MvdBW Ch. 6 and 7 and references
therein). Based on these dark matter (DM) only simulations, the
standard \LCDM\ paradigm has been judged to be extremely successful at
explaining and reproducing observations on scales larger than a few
kpc \citep[e.g.][]{Primack:2005}, thereby providing a robust framework
upon which to build models of galaxy formation and evolution.

\subsection{Overview of Physical Processes}
\label{sec:intro:physics}

In this section we briefly overview the main physical processes that
are commonly included in current models of galaxy formation. We
discuss these processes and their implementation in more detail in
\S\ref{sec:tools} and \S\ref{sec:subgrid}.

\noindent $\bullet$ {\bf Gravity} -- Gravity plays a crucial role in
building the ``skeleton'' for galaxy formation. The shape and
amplitude of the primordial power spectrum of density fluctuations
depends on the cosmological parameters and the properties of dark
matter. This spectrum, processed by gravity, determines the number of
dark matter halos of a given mass that have collapsed at any given
time, and how quickly these halos grow over cosmic time via merging
and accretion. It also determines how dark matter halos cluster in
space. In the standard paradigm, every galaxy is born within one of
these dark halos. When halos merge, each containing their own
``central'' galaxy, gravity and dynamical friction gradually cause the
orbits to decay, until the galaxies merge. Mergers can have important
effects on galaxies, including triggering bursts of star formation and
accretion onto central supermassive black holes, and transforming
galaxy structure and morphology.
 
\noindent $\bullet$ {\bf Hydrodynamics and Thermal evolution} -- When
an over-dense region composed of gas and dark matter collapses, strong
shocks form, increasing the entropy of the gas. The subsequent
evolution of the gas is then determined by how efficiently the gas can
cool and radiate away its thermal energy. The primary cooling
processes relevant for galaxy formation over most of cosmic history
are two-body radiative processes. Gas that is hotter than $T \ga 10^7$
K is fully collisionally ionized and cools predominantly via
bremsstrahlung (free-free emission). In the temperature range $10^4 <
T < 10^7$ K, collisionally ionized atoms can decay to their ground
state, and electrons can recombine with ions. Below temperatures of
$10^4$ K, cooling occurs through collisional excitation/de-excitation
of heavy elements (metal line cooling) and molecular cooling.

Following collapse and shock-heating, if radiative cooling is
inefficient, a pressure-supported quasi-hydrostatic gaseous halo may
form. This gas will then gradually cool in what is often referred to
as a \emph{cooling flow}. This is also sometimes referred to as ``hot
mode'' accretion. Once the gas cools and loses pressure support, it
will collapse until it is supported by its angular momentum. If the
cooling time of the gas is short compared to the dynamical time, the
gas may accrete directly onto the proto-galaxy without ever forming a
hot quasi-hydrostatic halo
\citep{White:1991,Birnboim:2003}. Cosmological hydrodynamic
simulations have shown that this sort of ``cold mode'' accretion tends
to occur when gas flows in along relatively cold, dense filaments
\citep{Keres:2005}.

\noindent $\bullet$ {\bf Star formation} -- Once gas has collapsed
into the central regions of the halo, it may become
self-gravitating, i.e. dominated by its own gravity rather than that
of the dark matter. As gas cools more rapidly the higher its density,
if cooling processes dominate over heating, then a run-away process
can ensue whereby Giant Molecular Cloud (GMC) complexes form, and
eventually some dense cloud cores within these complexes collapse and
reach the extreme densities necessary to ignite nuclear
fusion. However, many details of this process remain poorly
understood. Moreover, most cosmological simulations are not able to
resolve even the scales on which GMC form, much less individual cores.
Therefore all existing cosmological simulations implement empirical
sub-grid recipes to model star formation.

\noindent $\bullet$ {\bf Black Hole Formation and Growth} -- The first
``seed'' BH may have formed in the early universe either as the
remnants of Population III (metal free) stars, via direct collapse of
very low angular momentum gas, or via stellar dynamical processes
\citep{Volonteri:2010}. These seed BH may grow by accreting gas that
either has negligible angular momentum, or by forming an accretion
disk that drains the gas of angular momentum via viscosity
\citep{Netzer:2013}. These processes are, again, poorly understood and
virtually impossible to model explicitly in cosmological simulations,
so are modeled via sub-grid recipes.

\noindent $\bullet$ {\bf Star Formation Feedback} -- Observations show
that less than 10\% of the global baryon budget today is in the form
of stars. However, in CDM models without some sort of ``feedback'' (or
suppression of cooling and star formation), we would expect most of
the gas to have cooled and formed stars by the present day. Even the
pioneers of the earliest models of galaxy formation within a CDM
framework recognized this ``overcooling problem", and suggested that
energy generated by supernova explosions could heat gas and perhaps
blow it out of galaxies, making star formation inefficient
\citep{White:1978,White:1991,Dekel:1986}. It is now recognized that
there are many processes associated with massive stars and supernovae
(e.g. photo-heating, photo-ionization, winds) that could contribute to
making star formation inefficient and to driving large-scale winds
that reduce the baryon fractions in galaxies \citep[see][for an
  overview]{Hopkins:2012a}. Once again, most cosmological simulations
cannot resolve these physical processes in detail, so nearly all
current models implement sub-grid recipes to attempt to capture their
effect on galaxy scales.

\noindent $\bullet$ {\bf AGN Feedback} -- There is strong
observational evidence that most or perhaps all spheroid-dominated
galaxies (which comprise the majority of all massive galaxies) contain
a supermassive black hole \citep[see][for a recent
  review]{Kormendy:2013}. A simple calculation indicates that the
amount of energy that must have been released in growing these black
holes must exceed the binding energy of the host galaxy, suggesting
that it \emph{could} have a very significant effect on galaxy
formation \citep{Silk:1998}, however, it is still uncertain how
efficiently this energy can couple to the gas in and around
galaxies. Observational signatures of feedback associated with Active
Galactic Nuclei (AGN) include high-velocity winds, which may be
ejecting the cold ISM from galaxies, and hot bubbles apparently
generated by giant radio jets, which may be heating the hot halo gas
\citep[see][for recent reviews]{Fabian:2012,Heckman:2014}. AGN
feedback is also treated with sub-grid recipes in current cosmological
simulations.

\noindent $\bullet$ {\bf Stellar populations and chemical evolution}
-- In order to make direct comparisons between models and
observations, many modelers convolve their predicted star formation
histories with \emph{simple stellar population} models, which provide
the UV-Near IR SED for stellar populations of a single age and
metallicity \citep{Conroy:2013}, folding in an assumed stellar Initial
Mass Function (IMF)\footnote{Note that all cosmological simulations to
  date, as far as we are aware, have assumed that the IMF is
  universal. However, there is mounting evidence that this assumption
  may not be valid \citep[see the recent review by][]{Bastian:2010},
  which could have important implications for galaxy formation.}. Many
models now include the important contribution of gas recycling from
stellar mass loss self-consistently within simulations
\citep{Leitner:2011}. In addition, as stars evolve and go supernova,
they produce and distribute heavy elements throughout the gas that
surrounds galaxies, evidently polluting the intergalactic medium (IGM)
out to fairly large distances from galaxies. Chemical evolution is a
critical part of galaxy formation for several reasons: (i) cooling
rates at intermediate temperatures are highly enhanced in
metal-enriched gas; (ii) the luminosity and color of stellar
populations of a given age are sensitive to metallicity; and (iii)
heavy elements produce dust, which dims and reddens galaxies in the UV
and optical and re-radiates the absorbed energy in the mid-to-far
IR. Most cosmological models of galaxy formation now include a
treatment of chemical evolution.

\noindent $\bullet$ {\bf Radiative Transfer} -- The radiation emitted
by stars and AGN can have an important impact on galaxy formation.
Radiation can directly heat gas, and can also modify cooling rates
(especially for metal-enriched gas) by changing the ionization state
of the gas.  Moreover, the transmission of radiation of different
wavelengths through and scattering by dust can greatly impact the
measured total luminosity, color, and observationally determined
morphological and structural properties of galaxies, especially in the
rest-frame UV and optical, which are often all that is available at
high redshift. Most current cosmological simulations that are run to
low redshift ($z\la 6$) do \emph{not} include radiative transfer
self-consistently due to the added computational expense. However,
with sufficiently high resolution, radiative transfer through a dusty
ISM can be computed in post-processing to estimate the observed
pan-chromatic properties of galaxies \citep[e.g.][]{Jonsson:2010} and
their line emission~\citep[e.g.][]{Narayanan:2008}.

\subsection{Overview of Basic Tools}
\label{sec:intro:tools}

\begin{figure}
\resizebox{\textwidth}{!}{\includegraphics{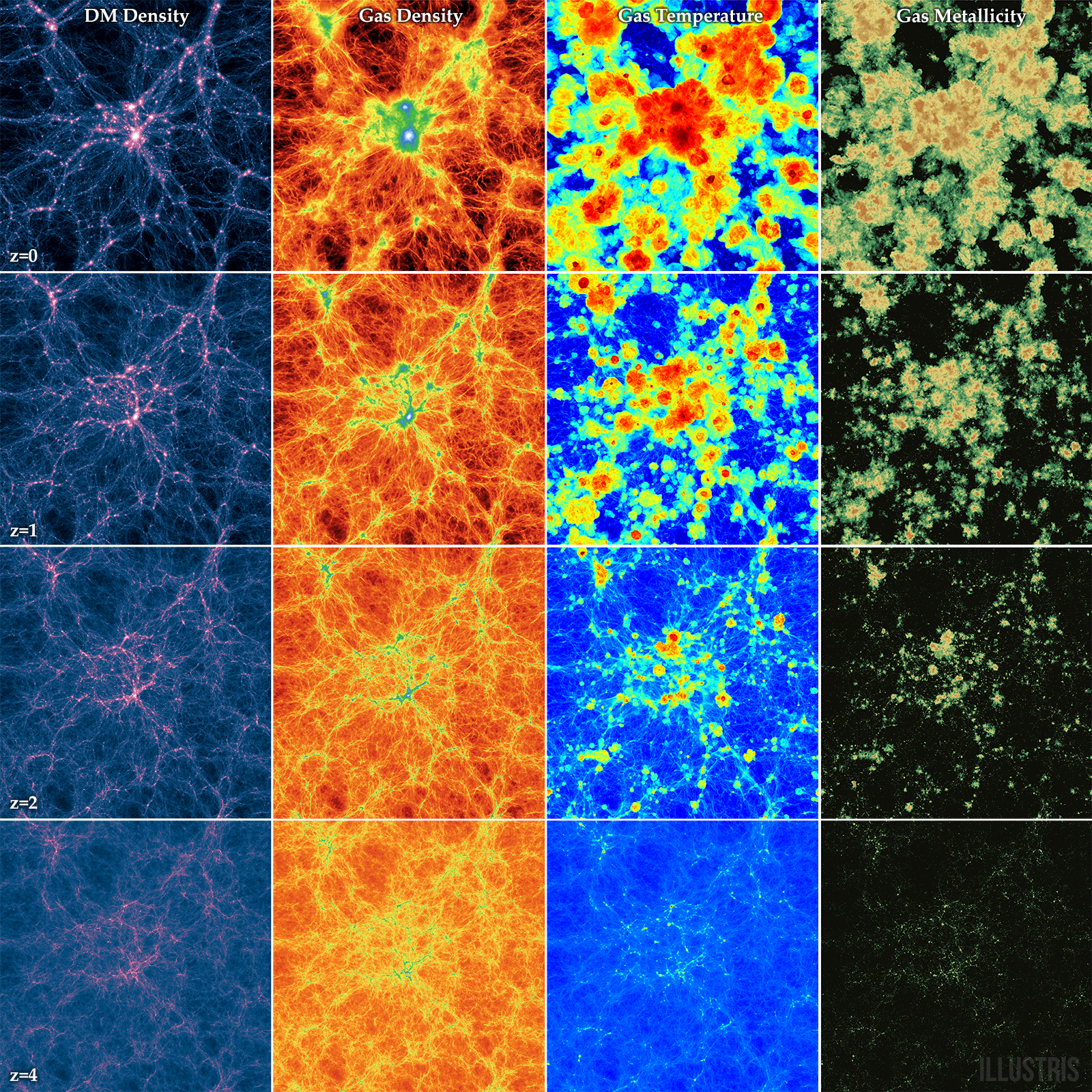}}
\caption{\footnotesize Visualization of representative quantities
  computed by numerical hydrodynamic simulations, from the Illustris
  project. From left to right, the dark-matter density, gas density,
  gas temperature, and gas metallicity are shown at different cosmic
  times (from top to bottom: $z=0$, $z=1$, $z=2$, $z=4$). The slice
  shown has a projected thickness of 21.3 cMpc and shows the whole
  Illustris simulation box which is 106.5 cMpc on a side.  Reproduced
  from \protect\citet{Vogelsberger:2014a}. }
\label{fig:hydroexamp1}
\end{figure}

\begin{figure}
\resizebox{\textwidth}{!}{\includegraphics{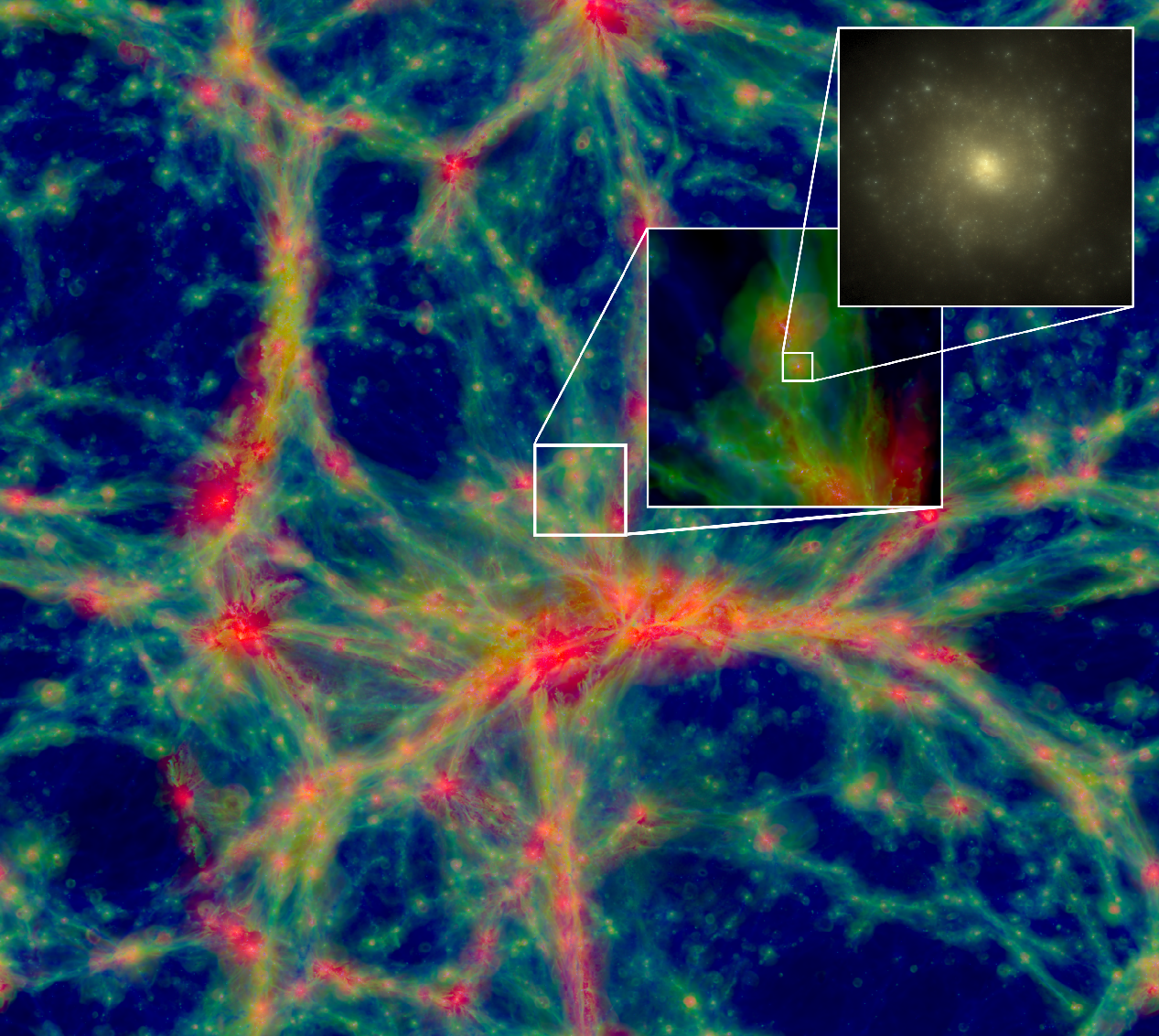}}
\caption{\footnotesize A 100 $\times$ 100 $\times$ 20 cMpc slice
  through the EAGLE simulation, illustrating the dynamic range that is
  attainable with state-of-the-art numerical hydrodynamic
  simulations. The intensity represents the gas density while the
  color indicates the gas temperatures (blue through green through red
  from cooler to hotter). The inset shows a region 10 cMpc and 60 ckpc
  on a side.  The zoom in to an individual galaxy with stellar mass $3
  \times 10^{10} \msolar$ shows the optical band stellar
  light. Reproduced from \protect\citet{Schaye:2014}. }
\label{fig:hydroexamp2}
\end{figure}

\begin{figure}
\centerline{\includegraphics[width=\textwidth]{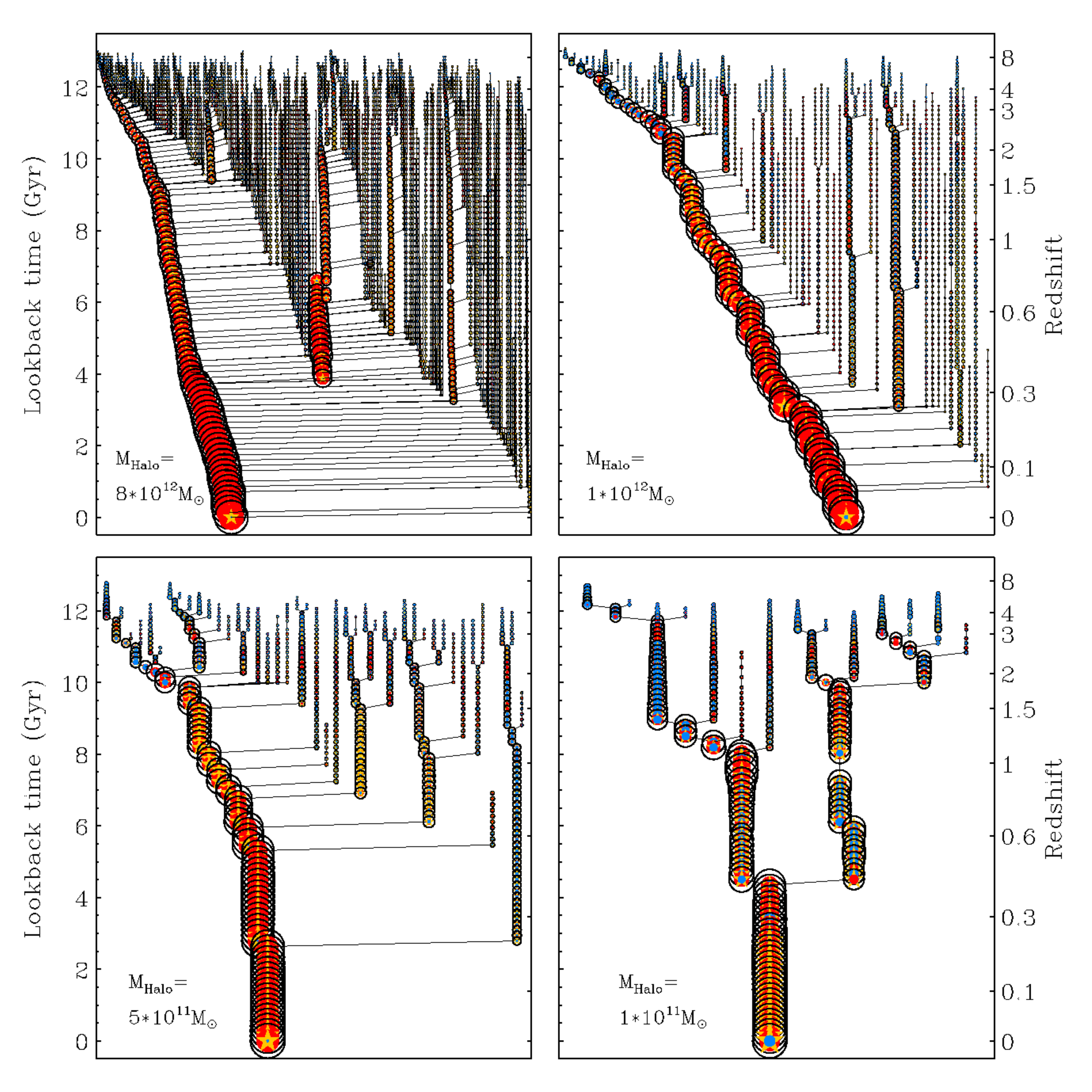}}
\caption{\footnotesize Visualization of representative predictions
  from a semi-analytic model. Symbol sizes represent the mass of the
  host dark matter halo; the x-axis is arbitrary. Symbols connected by
  lines represent halo mergers. Colors represent the mass of different
  galaxy components (red: hot gas; blue: cold gas; yellow:
  stars). Several different final host halo masses are shown as
  indicated on the figure panels. Halos with the same virial mass can
  have a diversity of merger histories (not shown). Reproduced from
  \protect\cite{Hirschmann:2012a}. }
\label{fig:samexamp}
\end{figure}

Theorists have developed a wide range of different tools for modeling
galaxy formation and evolution. Here we briefly summarize the most
commonly used tools and highlight some significant differences between
them. We provide a more detailed description of the methods used in
the modeling tools that are the subject of this review in
\S\ref{sec:tools}.

There is a popular class of what are generally called ``models'',
including Halo Occupation Distribution (HOD) models
\citep[e.g.,][]{Berlind:2002,Zheng:2005}, Conditional Luminosity
Function models \citep{vandenbosch:2007}, and sub-halo abundance
matching (SHAM) models and related techniques
\citep[e.g.,][]{Conroy:2006,Tasitsiomi:2004,Moster:2010,Behroozi:2010}. These
techniques derive mappings between observable properties of galaxies
and predicted properties of dark matter halos, and in general contain
no actual modeling of \emph{physical processes}. Although this family
of techniques is extremely useful for gaining insights into the
required connection between observable galaxies and dark matter halos,
we will not discuss these types of ``models'' in detail in this review.

The most explicit way to model galaxy formation is using
\emph{numerical hydrodynamic techniques}, in which the equations of
gravity, hydrodynamics, and thermodynamics are concurrently solved for
particles and/or grid cells representing dark matter, gas, and
stars. The advantage of these techniques is that, within the
limitations of the adopted numerical resolution, one obtains
predictions of the density of each of these three components (as well
as that of heavy elements) over cosmic time. One also obtains
predictions for the velocities of the stars and dark matter, and the
temperature of the gas.  Thus the structure and kinematics of galaxies
as well as their global properties and spatial distribution can be
studied in great detail (see Fig.~\ref{fig:hydroexamp1} and
\ref{fig:hydroexamp2} for examples). The main limitation of these
techniques is that computational exigencies restrict the dynamic range
that can be explicitly simulated. This, combined with our still
imperfect understanding of the physics that governs ``small-scale''
processes such as star formation, black hole growth, and feedback
processes, means that (as already discussed), many important processes
must be treated using uncertain and somewhat arbitrary sub-grid
recipes. Moreover, computational limitations have historically made it
difficult to experiment extensively with different sub-grid recipes or
to explore the multi-dimensional space of the variables that
parameterize these recipes.

The other technique that has been widely used to model galaxy
formation in a cosmological context is known as ``semi-analytic
modeling'' (SAM). This method does not explicitly solve fundamental
equations for particles or grid cells, but rather adopts a set of
simplified flow equations for bulk components \citep[see][for
  reviews]{Baugh:2006,Benson:2010}. For example, a typical SAM tracks
how much gas accretes into halos, how much hot gas cools and turns
into stars, how feedback processes remove cold gas from the galaxy or
heat the halo gas, how mergers transform disks into spheroids,
etc. Fig.~\ref{fig:samexamp} shows graphical representations of some
of the quantities that can be tracked in a SAM for several example
halo ``merger trees''.

The computational requirements of these models are enormously reduced
compared with fully numerical simulations. This makes it possible to
make predictions for very large volumes, or to simulate galaxies over
a larger range of halo mass, and also to extensively explore different
sub-grid recipes treating the most uncertain aspects of galaxy
formation.  Recently, several groups have coupled SAMs with a Bayesian
inference approach, and used Markov Chain Monte Carlo techniques to
sample the posterior probability distribution of the multi-dimensional
space of the model parameters \citep{Henriques:2009,Lu:2011}. This is
a powerful approach for exploring parameter degeneracies and obtaining
more rigorous statistical assessments of the ``goodness of fit'' of
specific models or model families with observational data (see
\citet{Bower:2010} for an alternative approach using Bayesian Emulator
methods). As well, the less explicit nature of SAMs has allowed
modelers to bypass some of the numerical issues which for many years
caused difficulties in reproducing basic properties of galaxies in
numerical simulations.

The field has now reached an interesting point where numerical
simulations have started to be able to reproduce fundamental
observations at a similar level as semi-analytic
models. Interestingly, much of this success has been achieved by
adopting a similar approach to the one that has long been used by
SAMs, namely, 1) parameterizing the physical processes that can't be
simulated explicitly, and tuning these parameters to match a subset of
observations, 2) experimenting with different sub-grid recipes to
achieve the best match to a set of observations. Even the recipes
themselves are in many cases very similar to the ones that are
commonly implemented in SAMs. Encouragingly, the two techniques have
arrived at the same qualitative conclusions about galaxy formation and
evolution for all of the topics that we will discuss in this
article. For this reason, we structure this article largely in terms
of the physical processes and general insights into how they shape
galaxy formation, giving examples from both SAMs and numerical
simulations.

\section{TOOLS AND METHODS}
\label{sec:tools}

\subsection{Gravity}
\subsubsection{Numerical $N$-body methods}
\label{sec:nbody}
Gravity solvers, or $N$-body codes, provide the backbone for galaxy
formation models, be they SAMs or hydrodynamic simulations.
Fundamentally, these codes must determine the force on each mass element
from all others by solving Poisson's equation. Numerically
solving Poisson's equation to evolve large-scale structure has a
long and storied history that has been extensively reviewed elsewhere
\citep[e.g.][]{Bertschinger:1998,Bagla:2005,Dehnen:2011}, so we
greatly limit our discussion here.

The basic approach is to subdivide a representative portion of the
universe into many particles, compute the forces on these particles
from all others, and evolve the system forward in discrete time-steps.
In cosmological $N$-body simulations, the equations are solved within
a comoving frame, and the volume is typically evolved with periodic
boundaries, under the assumption that there are a space-filling set of
identical volumes that approximately represent the larger-scale matter
distribution. The expansion rate of the comoving frame is computed
using the Friedmann equation (obtained from the Einstein equations
within GR, see e.g. MvdBW Ch. 3.2), but the equations actually solved
are the familiar Newtonian versions since GR corrections are generally
negligible.

$N$-body methods are either particle-based, mesh-based, or a hybrid.
In galaxy formation, the most popular particle-based approach is
the tree code~\citep{Barnes:1986}, in which the force from distant
groups of particles are approximated via their multipole moments.
The particle-mesh (PM) method, in contrast, computes the potential
on a grid via a Fourier transform of the density field, and moves
particles along potential gradients~\citep{Hockney:1988}.  Both
scale with particle number N as $\cal{O}(N\log N)$, though PM is
considerably faster.  Moreover, PM codes intrinsically account for
all periodic replicas of the volume, while tree codes must use
add-on techniques such as Ewald summation~\citep{Hernquist:1991}.

The advantage of tree codes is that the forces on particles can be
accurately represented down to the chosen {\it force softening length}
$\epsilon$, while PM codes are limited in resolution to their cell
size.  The ratio of the box length to $\epsilon$ defines the {\it
  dynamic range} of the calculation.  A hybrid Tree-PM approach is
thus a popular method to increase dynamic range, in which the
small-range forces are more accurately calculated using a tree, while
large-range (and periodic) forces are computed via a faster PM
method~\citep[e.g. \gad;][]{Springel_gadget:2005}.  The largest
$N$-body simulations today evolve $\sim 10^{12}$ particles, with a
dynamic range exceeding a million.  With the advent of new computing
technologies such as Graphics Processing Units, there is the potential
for even larger computations if such highly-threaded, cache-limited
hardware can be effectively utilized; so far this has proved
challenging, but progress is being made.

\subsubsection{Dark Matter Halos and Sub-halos}

A basic ansatz of our current picture of galaxy formation is that
galaxies form within dark matter halos. Identifying these objects in
$N$-body simulations is the first step in constructing the \emph{merger
trees} (see below) that form the gravitational backbone for SAMs.
On-the-fly halo finding is carried out within some hydro codes as
well, in order to use halo properties for sub-grid recipes.

The halo mass (or ``virial mass'') is usually defined as the mass
within a sphere that encloses an average density $\Delta_{\rm vir}$
relative to the background density of the Universe. Similarly, the
virial radius is defined as the radius within which the overdensity is
equal to this critical value. The actual value of $\Delta_{\rm vir}$
is unfortunately not standardized, and is based on a simple model of
the collapse of a uniform spherical overdensity. In an Einstein-de
Sitter universe, after collapse and virialization, such a uniform
spherical perturbation will have an average density $\simeq 178$ times
that of the background (or critical) density (MvdBW Ch. 5.1). Many
works use a fixed value of $\Delta_{\rm vir}=200$, which is just a
rounding up of $178$; some apply it relative to the critical density
and some relative to the background density. Some works use a redshift
and cosmology-dependent value of $\Delta_{\rm vir}$, as given by the
fitting function from \citet{Bryan:1998}. These different conventions
introduce redshift-dependent differences of as much as a factor of two
in halo virial masses, radii, and internal velocities, to which
readers must be alert when comparing results from the literature.

One of the generic features of the \LCDM\ paradigm is that halos have
a great deal of ``sub-structure''. This sub-structure arises from
objects that collapse and become bound at an earlier time, then get
subsumed into a larger virialized structure.  A ``sub-halo'' is a halo
that was once a distinct halo but is now contained within another
virialized halo.

Methods used to identify halos in $N$-body simulations include
``friends-of-friends'' (FOF), Spherical Overdensity (SO), and 6D
phase-space based methods.  See \citet{Knebe:2011} for a comprehensive
description and comparison of the results of different halo finders.
Different halo finders tend to agree fairly well (within $\sim 10$\%)
for basic halo properties such as mass and peak circular velocity of
distinct halos; the cumulative $z=0$ halo mass function differs by
$\pm 10$\% across the 16 halo finders tested in
\citet{Knebe:2011}. However, \citet{Klypin:2011} point out that much
larger differences between FOF and SO-based finders can arise at high
redshift. Identifying substructure is more halo-finder dependent; here
6D phase-space based halo finders such as ROCKSTAR
\citep{Behroozi:2013a} were found to perform significantly better.

\subsubsection{Merger Trees}

In semi-analytic models, the formation of structure through
gravitational instability in an expanding Universe is represented via
merger trees. A merger tree records the masses of dark matter halos
and the times at which these progenitor halos merge together to form a
larger halo (see Fig.~\ref{fig:samexamp}). Merger trees may either be
extracted from N-body simulations or constructed using semi-analytic
methods.

The first proposed methods for constructing merger trees
semi-analytically \citep{Kauffmann:1993,Cole:1994,SK:1999} used
statistical methods based on the Extended Press-Schechter model
\citep{Lacey:1993}.  More recent methods apply empirical corrections
to achieve better agreement with numerical simulations
\citep[e.g.][]{Parkinson:2008}. These methods provide an important
complement to merger trees extracted from $N$-body simulations, as
they are extremely flexible, and can be used to efficiently explore
different cosmologies and power spectra and large dynamic ranges in
halo mass. Moreover, $N$-body based merger trees have their own
limitations, as discussed below.

Extracting merger trees from an $N$-body simulation appears
straightforward on the face of it --- one identifies dark matter halos
at a series of redshifts or output times, and then identifies which
halos at earlier times are ``progenitors'' of a given halo identified
at some later time. In practice, however, there are complications. 1)
Results may be sensitive to the method used for identifying halos, as
discussed above. 2) The definition of progenitor is not unique, since
the particles from a halo at some time $t_1$ may end up in different
halos at a later time $t_2$. 3) A halo may be identified as a sub-halo
in one timestep, then move outside of the virial radius of the host
again at some later time. 4) Sub-halos are tidally stripped as they
orbit within their host halos, and eventually become difficult to identify ---
most halo finders can no longer robustly identify sub-halos when they
drop below 30-40 particles \citep{Knebe:2011}.

For semi-analytic merger trees and to track sub-structure once the
sub-halo can no longer be identified in the $N$-body simulation, most
SAMs include a procedure to estimate the time for a satellite's orbit
to decay due to dynamical friction. A variation of the Chandrasekhar
formula (MvdBW, \S12.3.1) is generally used for this purpose. Many
SAMs use refined versions of this formula based on numerical
simulations \citep[e.g.][]{Boylan_kolchin:2008}. Some sub-halos may be
tidally destroyed before they reach the center, and their stars added
to a diffuse stellar halo. Satellites that survive until they reach
the center are assumed to merge with the central galaxy. SAMs then
implement a set of recipes for the effect of the merger, which
generally include an enhanced ``burst'' phase of star formation as
well as some sort of morphological transformation (e.g. moving stars
from the disk to the spheroid component).

\subsection{Hydrodynamics: Numerical Techniques}
To directly model the visible component of the Universe requires
modeling gas physics, i.e. solving the equations of
hydrodynamics and evolving them concurrently with the chosen gravity
solver.  Doing so enormously increases the complexity of the code,
resulting in longer calculations with greater intrinsic uncertainties.
Most hydro codes are based on solving the Euler equations (e.g. MvdBW
p. 366), representing mass, momentum, and energy conservation,
typically closed by assuming a non-relativistic ideal gas equation of
state.  The Euler equations are a form of the Navier-Stokes equations
assuming no viscosity or conduction.  In most cases, it is necessary
to add an artificial viscosity term in order to properly handle
convergent flows and shocks.  Some experimentation has also been done
with adding other physics whereby it is necessary to solve the
Navier-Stokes equations directly; see \citet{Springel:2010} for more
discussion.

\subsubsection{Lagrangian Methods}

In galaxy formation, historically the most popular Lagrangian method
is Smoothed Particle Hydrodynamics \citep[SPH; see reviews
by][]{Monaghan:1992,Springel:2010}. Briefly, in SPH, the particles
themselves carry the information about the fluid, which is obtained
via a kernel-weighted sum over neighboring particles closer than a
{\it smoothing length} ($h$):
\begin{equation}
X_i = \Sigma_j m_j (X_j/\rho_j) W(|{\bf r}_i-{\bf r}_j|,h_i,h_j).
\end{equation}
Here, $X_i$ is the quantity to be estimated, $m$ and $\rho$ are
the particles' mass and density, and $W$ is the kernel, which is
some spherical function of the distance between particles in units
of the smoothing length.  $X_i$ can also be a gradient of a quantity,
in which case the gradient propagates through to the kernel which
becomes $\nabla W$.  The efficiency and simplicity of evaluating the 
Euler equations based on these local kernel-smoothed quantities gives
SPH many of its key advantages, including natural spatial adaptivity 
and trivial implementation in three dimensions.

``Classic" SPH~\citep[e.g.][]{Hernquist:1989,Monaghan:1992} evaluates
the density first as a kernel-smoothed average over nearby masses,
then the thermal energy to update the pressure, then the hydrodynamic
acceleration.  A variant of this method is used in the code GASOLINE
\citep{Wadsley:2004}.  A key drawback is that this method does not
explicitly conserve energy and entropy in adiabatic flows in the case
of variable smoothing lengths.  Entropy-conserving
(EC-)SPH~\citep{Springel_gadget:2005} mitigated this flaw by
explicitly including variational terms in $h$ as derived from the
Lagrangian, and was formulated using entropy as the evolved variable.
EC-SPH is employed in the widely-used code
\gad~\citep{Springel_gadget:2005}.  Subsequently, it was noted that a
side-effect of EC-SPH is to create an artificial pressure between cold
and hot phases, resulting in a surface tension that causes for example
cold clumps moving through a hot medium to be significantly more
resistant to disruption than they are in grid-based
codes~\citep{Agertz:2007}.  Ironically, classic SPH performs somewhat
better in this regard (at the cost of increased particle
interpenetration), but all of these versions fail to realistically
capture Kelvin-Helmholtz instabilities.

Several promising approaches have been developed recently to mitigate
these issues.  \citet{Read:2012} proposed SPHS, in which they showed
that the error in the momentum equation in classic SPH can be reduced
by using a different kernel shape with a larger number of neighbors,
along with a modification of artificial viscosity to include a
higher-order dissipation switch that anticipates shocks.  SPHS can
yield much better results for a range of surface instability tests,
but the required increase in the number of SPH neighbors (442 vs.
$\sim 40$) slows the calculation and lowers the resolution.

A different approach was pursued by \citet{Saitoh:2013}, following on
\citet{Ritchie:2001}.  They argued that difficulties in classic or
EC-SPH arose from the requirement that the density distribution be
differentiable, which is violated at contact discontinuities.  They
proposed a new formulation that was ``density-independent" (DI-SPH),
which used kernel sums to separately obtain the energy density and
internal energy, from which the density is inferred.  DI-SPH was shown
to remove the artificial surface tension and enable improved treatment
of surface instabilities (among other tests).  \citet{Hopkins:2013}
re-formulated DI-SPH in terms of entropy to incorporate the improved
conservation properties of EC-SPH.  This pressure-entropy (PE-)SPH
provides much improved handling of surface instabilities versus
classic SPH, with fewer numbers of neighbors than for SPHS. A modified
treatment of artificial viscosity has also been widely implemented,
and helps improve the performance of SPH in this regard
\citep[e.g.][]{Hu:2014}. Thanks to such improvements, current
formulations of SPH can now track surface instabilities and associated
phenomena to an accuracy that, not long ago, were widely regarded to
be challenging for SPH.

\subsubsection{Eulerian Methods}

A time-honored approach to solving hydrodynamics is to discretize
the fluid onto grid cells, and then compute the advection of
properties across the cell boundaries.  This is the basis of Eulerian
approaches, which formulate the solution to the Euler equations in
the fixed frame, rather than in the fluid frame as in the Lagrangian
approach.

Most current cosmological Eulerian hydro codes employ a high-order
Godunov scheme.  Here, the Riemann problem is solved across cell
faces, which yields a pressure at each cell face, thereby giving
the force on the fluid across the cell.  The fluid, with all its
associated properties, is then advected across the cell face.  If
the cell is assumed to have uniform properties within it, this is
called a (first-order) Godunov solver.  Modern codes employ parabolic
interpolation, known as the Piecewise Parabolic Method (PPM).  Note
that while higher order interpolation provides a more accurate
solution, it requires using information from neighboring cells which
effectively lowers the spatial resolution.

Given the dynamic range involved in modeling galaxy formation, a key
development was the implementation of Adaptive Mesh Refinement (AMR).
Here, cells satisfying some local criteria (typically based on mass)
are split into sub-cells, enabling improved resolution in those
regions.  This effectively achieves some of Lagrangian methods'
primary advantage of being naturally adaptive in space and time.
Current AMR hydro codes for galaxy formation include
Enzo~\citep{Bryan:2014}, RAMSES~\citep{Teyssier:2010},
FLASH\footnote{\tt http://flash.uchicago.edu/site/flashcode/}, and
Hydro-Adaptive Refinement Tree \citep[H-ART;][]{Kravtsov:1997}.

\subsubsection{Arbitrary Lagrangian-Eulerian Methods}

Optimally, one would like to unite the advantages of PPM in handling
shocks and contact discontinuities with SPH's natural adaptivivity.
One approach is to use a deformable mesh, in which the mesh follows
the fluid.  Such arbitrary Lagrangian-Eulerian codes have
historically not played a large role in astrophysics~\citep[see
  e.g.][]{Pen:1998}, but this has recently changed with the
introduction of \arepo~\citep{Springel_arepo:2010}.

\arepo\ uses a Voronoi tesselation to subdivide space around particles.
A Voronoi tesselation is a space-filling set of polyhedral cells
where the space within a given cell is closer to one particle than
any other.  The Riemann problem is then solved across the cell faces
in order to obtain the force on the particle.  The mesh is re-generated
as the particles move.  In this way, \arepo\ is able to naturally
follow the fluid like a Lagrangian code, while retaining the
advantages of Godunov solvers such as excellent handling of contact
discontinuities, surface instabilities, and shocks, and the lack
of artificial viscosity.

\subsubsection{Advantages and Disadvantages}

Traditionally, Eulerian methods have enjoyed a superiority in
handling strong shocks and surface instabilities, while Lagrangian
methods like SPH are more adaptive and provide increased dynamic
range for a given CPU expense.  However, in recent times the gaps
are closing in both directions.  \arepo\ has some important advantages
over both Lagrangian and Eulerian methods, particularly
EC-SPH~\citep{Vogelsberger:2012}.

An advantage of a particle-based approach such as SPH is that the
movement of mass is directly tracked.  This makes it more
straightforward to follow the mass as it assembles into galaxies, and
to track where material ejected from galaxies ends up.  It is also
straightforward to implement kinetic winds, which as we will discuss
below has had substantial success as a sub-grid prescription for
galactic outflows.  Nonetheless, in mesh codes it is possible to use
tracer particles for these purposes.  For instance, \arepo\ has
implemented kinetic winds by spawning particles that are decoupled
from the hydro mesh, which then later rejoin.

AMR offers the key advantange that the mesh can be refined to
arbitrarily high resolution, while particle-based methods are limited
in resolution by the particle mass.  This allows individual systems to
be examined in great detail, albeit at great computational cost.  For
example, Enzo merger simulations by \citet{Kim:2009} and H-ART
cosmological zoom simulations by \citet{Ceverino:2014} both achieved a
dynamic range of $\ga 10^6$, while the most ambitious current SPH
simulations can only achieve $\sim 10^5$.

More broadly, since all modern codes generally yield similar answers
in basic tests relevant to galaxy formation where the answer is
approximately known, at this stage it is difficult to identify one
code or methodology that is clearly superior to the others.  For
most properties, differences in sub-grid prescriptions yield much
larger variations than differences in hydrodynamical techniques.

\subsection{Thermal evolution}\label{sec:thermal}

\subsubsection{Cooling and Heating in Numerical Simulations}

The key difference between baryons and dark matter in galaxy formation
is that baryons can dissipate their potential energy via radiative
processes.  Radiative cooling and photo-ionization heating are thus
implemented in essentially all codes, while radiation transport is a
growing subfield with specific applications to the epoch of
reionization (EoR) and line emission.

Most simulations today also include cooling from metal line emission,
which dominates particularly at $10^5\la T\la 10^7$ K for typical
warm-hot gas metallicities.  Early works employed cooling rates
assuming collisional ionization equilibrium~\citep{Sutherland:1993},
but more recent work by \citet{Wiersma:2009} better account for the
photo-ionization of metals by the metagalactic radiation field.

Simulations focusing on the post-EoR universe typically account for
photo-ionization heating by assuming all the gas is optically thin and
in ionization equilbrium with a spatially-uniform metagalactic
radiation field~\citep[e.g.][]{Faucher:2009,Haardt:2012}.  During the
EoR, these assumptions break down, and continuum radiative transfer is
necessary in order to properly model the feedback from
photo-ionization heating on galaxy growth.  Two approaches are used:
applying radiative transfer in post-processing to existing density
distributions~\citep{Iliev:2006}, which is useful for evolving large
volumes to study the final stages of EoR; and full radiative
hydrodynamic codes that evolve the ionizing field together with the
baryons, including modeling star formation to self-consistently
predict the properties of the
sources~\citep{Iliev:2009,Pawlik:2011,Finlator:2011,Wise:2011}.  Given
that this review focuses on the post-reionization Universe, we will
not discuss this further here.

\subsubsection{Cooling and Cosmological Accretion in SAMs}
\label{sec:sam_cooling}

Most semi-analytic models implement some variant of the self-similar
cooling flow model originally proposed by \citet{White:1991} to track
the thermal evolution of gas. As the gas enters the halo, it is
assumed to be shock-heated to the virial temperature $T_{\rm vir} =
35.9 [V_{\rm vir}/({\rm km/s})]^2$ K, where $V_{\rm vir}$ is the halo
virial velocity. One may then calculate the cooling time, which is the
time it would take for the gas to radiate away all of its energy:
\begin{equation}
t_{\rm cool} = \frac{\frac{3}{2} \mu m_p kT}{\rho_g(r)
  \Lambda(T, Z_h)} \, .
\label{eqn:tcool}
\end{equation}
Here, $\mu m_p$ is the mean molecular mass, $T$ is the temperature of
the gas, $\rho_g(r)$ is the radial density profile of the gas,
$\Lambda(T, Z_h)$ is the temperature and metallicity dependent cooling
function \citep[e.g.][]{Sutherland:1993}, and $Z_h$ is the metallicity
of the hot halo gas.

The hot gas is assumed to be distributed with a smooth spherically
symmetric density profile. Most models assume that the density profile
is described by a singular isothermal sphere ($\rho_g(r) \propto
r^{-2}$), although some use different density profiles, such as a
Navarro-Frenk-White (NFW) profile \citep{Navarro:1997} or a cored NFW
profile \citep{Cole:2000}.

One can then solve for the ``cooling radius'', within which gas has
had time to dissipate all of its thermal energy by cooling. To do
this, one must adopt a timescale over which cooling is assumed to have
taken place. Common choices for this timescale are the time since the
halo has experienced a ``major'' (at least 2:1) merger
\citep[e.g.][]{SP:1999}, or the halo dynamical time $t_{\rm dyn} =
\rvir/\vvir$ \citep[e.g.][]{Springel:2001}. It may happen that the
model predicts $r_{\rm cool} > \rvir$, indicating that the cooling
time is shorter than the dynamical time, corresponding to the ``cold
flow'' regime described in \S\ref{sec:intro:physics}. In this case,
most modelers generally assume that gas can flow into the halo on a
dynamical time.
Although this model is very simple, several studies have shown that
the predicted cooling and accretion rates are in surprisingly good
agreement with those from numerical hydrodynamic simulations
\citep{Benson:2001,Yoshida:2002,Hirschmann:2012a,Monaco:2014}.

\subsection{Chemical evolution}
\label{sec:chemev}

Tracking the enrichment of gas with heavy elements is important for
cooling calculations, and for predictions of galactic chemical
evolution.  Most numerical hydro simulations now include a model for
chemical enrichment.  Early models tracked only Type~II supernova (SN)
enrichment, which is closely related to the oxygen abundance.  To
track other key elements such as carbon and iron, it is necessary to
model asymptotic giant branch (AGB) stars whose ejecta dominate the
present-day carbon budget, and Type~Ia SN that produce the bulk of the
iron in stellar-dominated systems.  Such delayed feedback sources are
now included in most codes, which track a suite of individual
elements~\citep{Oppenheimer:2008,Wiersma:2009b}.  The dominant
uncertainty typically comes from the metal yield models from SN and
stellar evolution, particularly at low metallicities and high masses.
Hence at present, absolute abundance predictions should be considered
accurate to only a factor of two, but relative trends of metallicity
versus other galaxy properties such as stellar mass are likely more
robust.

Most SAMs use a simple \emph{instantaneous recycling} approximation in
which a yield $y$ of heavy elements is produced by stars in each
timestep: $dM_Z = y \, dm_{\rm star}$, where $dM_Z$ is the mass of
metals produced and $dm_{\rm star}$ is the mass of stars formed.  In
general these metals are deposited into the cold ISM, although some
models deposit some of the metals directly in the hot halo gas. Metals
may then be ejected from the cold gas reservoir by winds, and are
either deposited in the IGM or in the hot gas halo.  Most SAMs treat
the yield $y$ as a free parameter rather than taking it from SN yield
calculations, and neglect enrichment by Type Ia SNae and AGB stars (so
again, the predicted metallicities most closely trace $\alpha$
elements such as oxygen). However a few SAMs in recent years have
incorporated more detailed treatments of chemical enrichment, tracking
multiple individual elements, and the finite timescales for enrichment
and gas recycling from AGB stars, Type Ia, and Type II SNae
\citep{Arrigoni:2010,Nagashima:2005,Yates:2013}.

\subsection{Initial conditions and zoom simulations}

The generation of standard cosmological initial conditions involves
(1) generating a linear matter power spectrum via a transfer
function~\citep[e.g.][]{Eisenstein:1999}; (2) Gaussian-random
sampling the power spectrum for modes within the simulation volume;
and (3) evolving the modes forward in the linear regime via the
Zel'dovich approximation; see \citet{Bertschinger:1998} for more
details.  This generates particle positions and velocities sampling
the matter field within a specified volume for a specified cosmology,
at some specified high redshift that is optimally just before
structure within the volume first goes nonlinear~\citep[see e.g.
MUSIC; ][]{Hahn:2011}.

An increasingly popular and useful technique is {\it zoom}
simulations.  In zooms, a sub-volume within a cosmologically
representative region is evolved at much higher resolution, together
with surrounding regions of coarser resolution that provide the tidal
field from large-scale structure.  After an initial coarse-grained
run, a halo or region of interest is selected, and its particles are
tracked back to the original initial conditions to define the
\emph{zoom region}.  Particles within the zoom region are sampled to
finer resolution, including the requisite small-scale power, and the
entire volume is run again, typically with hydrodynamics turned on
only in the zoom region.  In this way, zooms provide an increased
dynamic range at a manageable computational cost, albeit only for a
single galaxy or halo and its environs.

Simulations of idealized isolated galaxies, or mergers thereof,
provide a valuable testbed to explore detailed physical processes,
particularly in the ISM.  Initial conditions are typically created in
a stable disk configuration \citep{Hernquist:1993a}, and then
dynamical perturbations grow either from tides induced by a merger
or internal stochasticity.  Such models can achieve extremely high
resolution (by cosmological standards) and can serve to isolate
physics of particular interest, hence they remain useful even if they
do not fully represent the cosmological baryon cycle.

\section{SUB-GRID PHYSICS}\label{sec:subgrid}

\subsection{Star Formation and the ISM}
\label{sec:sfism}
A huge body of observations from UV through near-IR light traces the
emission from stars. In order to make contact with these observations,
models must attempt to compute how gas in galaxies turns into
stars. The ISM is a complex place, with multiple gas phases
co-existing at very different densities and temperatures
\citep{McKee:1977}. Cosmological simulations of more than a single
galaxy are still orders of magnitude away from capturing the spatial
scales, temperatures, and densities where stars actually
form. Moreover, physical processes that are not typically included or
captured well in cosmological simulations, such as magnetic fields and
turbulence, are thought to play important roles on the scales of dense
molecular cloud cores and protostars \citep{McKee:2007}. However,
advances in our theoretical understanding of star formation as well as
better observational characterization of key scaling relations
\citep[see][ for a review]{Kennicutt:2012} have enabled the
development of empirical recipes that smooth over much of the
small-scale complexity.

Stars are observed to form in the dense, cold, molecular phase of the
ISM, and current observations support a (nearly) universal star
formation efficiency in molecular gas, where about 1\% of the gas
is converted into stars per free fall time
\citep{Bigiel:2008,Bigiel:2011,Leroy:2013,Krumholz:2012b}. Thus the
ability to track where \emph{molecular} gas forms should lead to a
more physical approach to modeling star formation. The ISM is observed
to become H$_2$-dominated at $\sim 1$--100 atoms cm$^{-3}$. Because
gravitational instability is thought to be one of the driving forces
in the formation of GMC \citep{Dobbs:2013}, simply requiring a density
threshold for star formation of a few atoms cm$^{-3}$ may be a good
first approximation. However, this also requires high enough
resolution ($\la 100$ pc) to attain these densities, which is
currently achievable only in zoom simulations.

In more detail, \Htwo\ formation is catalyzed by dust, and destroyed
by Lyman-Werner radiation, so one would expect that \Htwo\ production
is thus roughly proportional to metallicity, while destruction
depends on the ability to self-shield against interstellar radiation.
Some zoom simulations now include a simplified phenomenological
treatment of chemical networks and \Htwo\ dust- and self-shielding
\citep{Gnedin:2009,Christensen:2012}.  Fitting functions that attempt
to capture the essence of \Htwo\ formation and dissociation and the
resulting dependence of \Htwo\ fraction $f_{H2}$ on gas density,
metallicity, and local UV background have been presented based on
these and on idealized (non-cosmological) disk simulations and
analytic models \citep{Krumholz:2009,McKee:2010,Gnedin:2011,Gnedin:2014}.

An alternative approach for partioning gas into \HI\ and \Htwo\ is to
use the empirical relationship between $f_{\rm H2}$ and the disk
mid-plane pressure, pointed out by \citet[][BR]{Blitz:2004}. They
found that the molecular fraction $R_{\rm mol} \equiv \Sigma_{\rm
  H2}/\Sigma_{\rm HI}$ was correlated with the disk hydrostatic
mid-plane pressure $P$: $R_{\rm mol} =
\left(\frac{P}{P_0}\right)^{\alpha_{\rm BR}}$, where $P_0$ and
$\alpha_{\rm BR}$ are free parameters that are obtained from a fit to
the observational data. The hydrostatic pressure as a function of
radius in the disk can be estimated based on the cold gas surface
density, the stellar surface density, and the ratio of the vertical
velocity dispersions of the gas and stars \citep{Elmegreen:1989}. This
approach can be used to estimate $f_{\rm H2}$ either self-consistently
(see below) or in post-processing in numerical simulations or
SAMs~\citep[][]{Duffy:2012,Obreschkow:2009}.

\subsubsection{Numerical Implementation}
The basic recipe for star formation in many cosmological
simulations has not changed markedly from the pioneering work of
\citet{Katz:1992}.  Gas that is dense and converging is assigned a SFR
based on a \citet{Schmidt:1959} law, namely
\begin{equation}
\dot\rho_* = \frac{\epsilon_* \rho_{\rm gas}}{t_{\rm ff}} \propto \rho_{\rm gas}^{1.5}
\label{eqn:sflaw}
\end{equation}
where the last proportionality arises because the local free-fall time
$t_{\rm ff}\propto \rho^{-0.5}$.  The free parameter $\epsilon_*$ is
typically calibrated to match the amplitude of the observed
\citet{Kennicutt:1998} relation in simulations of idealized, isolated
disks.  Long-term SF histories are generally insensitive to
$\epsilon_*$ within reasonable choices \citep{Katz:1996,Schaye:2010},
because as discussed later, globally, SF is driven primarily by gas
accretion, and over cosmological timescales is not limited by the rate
of conversion of gas into stars in the ISM.
A somewhat different approach was proposed by \citet{Schaye:2008}:
they analytically recast the Kennicutt-Schmidt relation as a function
of pressure rather than density, assuming a self-gravitating disk. 

Stars are generally only allowed to form when the density exceeds some
critical value, the choice of which is another free parameter.
\citet{Springel:2003a} incorporated a density threshold based on where
the Jeans mass became lower than the particle mass, at which point a
sub-grid ISM model is required; this value turned out to be $\approx
0.1$~cm$^{-3}$ for typical mass resolutions adopted in cosmological
volumes at the time.

This simple SF prescription applied to individual disk galaxies was
found to quickly collapse gas down to the (artificial) Jeans scale in
the simulations, which produced highly clumpy disks that looked
nothing like local grand-design spirals.  The solution, introduced in
cosmological runs by \citet{Springel:2003a} was to artificially
overpressurize the ISM, by implementing a sub-grid ISM model based on
\citet{McKee:1977} that tracked the balance between SN energy input
and cooling within a multi-phase ISM.  The temperature of the ISM gas
(defined as gas above the SF threshold density) is then raised up to
as high as $10^6$ K.  \citet{Robertson:2004} extended this model to an
arbitrary ISM effective equation of state, and showed that with
appropriate overpressurization, this approach can reproduce smooth,
stable, gas-rich spirals as observed today.  
Ironically, as we discuss further in \S\ref{sec:structure:disks}, it
turns out that simulations with no or minimal ISM pressure
\citep{Ceverino:2010,Ceverino:2014} do well at reproducing the clumpy
disks that are now known to be common at high redshift \citep[$z \sim
  2$; ][]{Genzel:2011}, though simulations with pressurization can
also reproduce these~\citep{Genel:2012a}.

It is clear that real stars do not form at densities of $\sim 0.1$
atoms~cm$^{-3}$.  Moreover, since the Kennicutt relation is only
observed to hold when the ISM is averaged over scales of $0.5-1$~kpc,
once simulations resolve smaller scales, it becomes dubious to use a
SF prescription that is calibrated to match this relation.  Thus much
recent effort has gone into incorporating more realistic treatments of
the ISM into cosmological simulations. High-resolution zoom
simulations that simply adopt a higher star formation threshold ($\sim
5$ atoms cm$^{-3}$) and efficient stellar-driven winds (see
\S\ref{sec:structure:disks} for further discussion) show marked
improvement in their ability to produce realistic disks
\citep[e.g.][]{Governato:2007,Guedes:2011}. Other simulators
\citep[e.g.][]{Kuhlen:2012,Agertz:2014} have incorporated sub-grid
recipes to compute the density of molecular hydrogen $\rho_{H2}$ and
then use that in an equation similar to Eqn.~\ref{eqn:sflaw} in place
of $\rho_{\rm gas}$ --- no arbitrary density threshold need then be
applied.

An exciting development is that cosmological zoom simulations are
starting to be able to resolve the Jeans mass/length of gas,
corresponding to the scale of molecular cloud complexes, allowing more
direct modeling of the multi-phase ISM, \citep[e.g. the FIRE
  simulations,][]{Hopkins:2014}.  Concurrently, ISM simulations
including detailed treatments of non-equilibrium chemistry and
turbulence are pushing outwards in scale to start ``bridging the gap''
with the cosmological
runs~\citep[e.g.][]{Walch:2011,MacLow:2012}. Continuing interactions
between the galaxy formation and ISM/star formation communities will
soon allow us to place our sub-grid recipes on a more secure physical
foundation.

\subsubsection{Implementation in Semi-Analytic Models}

The usual approach to modeling star formation in SAMs is very similar
to the approach used in numerical hydro simulations, described
above. Gas that has ``cooled'' according to the cooling model
described in \S\ref{sec:sam_cooling} loses its pressure support and
collapses further, until it is supported by its angular momentum,
forming a disk.  The initial angular momentum of the halo gas can then
be used to estimate the radial size of the disk, as described in
\S\ref{sec:results:structure}. Some SAMs track the radial structure of
the disk in cylindrically symmetric annuli
\citep{Kauffmann:1996a,Avila-Reese:1998,Dutton:2009,Fu:2010}, while
most models assume the disk radial surface density distribution to be
an exponential, as is generally the case in observed disk galaxies.

Different SAMs use different but roughly physically equivalent
variants of Eqn.~\ref{eqn:sflaw}. Early SAMs typically used an
expression of the form
\[ \dot{m}_*= \epsilon_* \frac{m_{\rm cold}}{\tau_*} \]
where $\dot{m}_*$ is the total star formation rate in the galaxy,
$m_{\rm cold}$ is the total cold gas mass in the galaxy, $\tau_*$ is a
characteristic timescale for star formation, and $\epsilon_*$ is a
parameter representing the global star formation efficiency. The SF
timescale is often assumed to scale with the dynamical time of the
dark matter halo, $\tau_* \propto \tau_{\rm dyn} \propto r_H/V_H$,
where $r_H$ is the characteristic halo radius and $V_H$ is the
characteristic halo circular velocity. However, it was quickly
realized that a SF law of this form could not reproduce the observed
trend of increasing cold gas fractions with decreasing stellar mass in
the low redshift universe. Thus, modelers either introduced a SF
threshold, such that only the fraction of the cold gas above this
threshold was eligible to participate in star formation, or made
$\tau_*$ an explicit function of halo properties, e.g. of $V_H$
\citep{Cole:2000}, such that the star formation timescale is made
longer in lower mass galaxies.

Models that track disk structure in more detail are able to use
empirical laws that are closer to what is actually observed. For
example, \citet{Somerville:2008b} adopted a ``Kennicutt''-like
expression, where the star formation rate surface density of the disk
is calculated according to $\Sigma_{\rm SFR} = A_{\rm SF}\Sigma_{\rm
  gas}^{N_{\rm SF}}$ for $\Sigma_{\rm gas} > \Sigma_{\rm crit}$ (and
zero otherwise).  The parameters $A_{\rm SF}$ and $N_{\rm SF}$ are
taken directly from observations \citep[e.g.][]{Kennicutt:1998}, and
$\Sigma_{\rm crit}$ is treated as a free parameter.  A similar
approach, but with a radius and circular velocity dependent
$\Sigma_{\rm crit}$ based on the Toomre condition for gravitational
instability, is adopted in the MPA SAMs
\citep[e.g.][]{Kauffmann:1999,Croton:2006,Guo:2011}.

Several groups have recently developed SAMs that attempt to track
atomic and molecular gas separately
\citep{Fu:2010,Lagos:2011a,Popping:2014a,Somerville:2014}. Various
recipes for \Htwo-formation, either employing the metallicity-based
fitting functions of \citet{Krumholz:2009} and \citet{Gnedin:2011}, or
alternatively the empirical pressure-based recipe from
\citet{Blitz:2004}, have been implemented in these SAMs. Again, the
computed density of \Htwo\ may then be used in an empirically
calibrated SF law with no need to assume a density threshold ---
essentially removing all free parameters from the SF recipe (within
the observational uncertainties on the slope and normalization of the
relationship between $\Sigma_{\rm SFR}$ and $\Sigma_{H2}$).  Overall,
it appears that the main predictions of SAMs, especially for stellar
properties of galaxies, are quite insensitive to the details of the
gas partitioning recipe
\citep{Fu:2010,Lagos:2011a,Popping:2014a,Somerville:2014,Berry:2014}.

It is well known that galaxy interactions and mergers can trigger
starbursts with enhanced star formation efficiency (SFE), and most
SAMs implement a ``burst mode'' of star formation in galaxies that
have experienced a recent merger. Studies based on hydrodynamic
simulations of binary galaxy mergers have shown that the enhancement
in the SFE above that in an isolated galaxy is a fairly strong
function of the mass ratio of the merger. Many SAMs implement the
fitting function introduced by \citet{Cox:2008}, who parameterized the
burst efficiency as $e_{\rm burst} = e_{\rm burst, 0}\, \mu^\gamma$,
where $\mu$ is the merger mass ratio, and $e_{\rm burst}$ is defined
as the fraction of the total gas reservoir that is consumed in the
burst.

Subsequent studies have shown that $e_{\rm burst}$ and the burst
timescale also depend on the implementation of stellar feedback and
the treatment of the ISM
\citep{Cox:2008,Robertson:2006b}. \citet{Hopkins:2009a} found that the
burst efficiency depended strongly on the cold gas fraction in the
progenitors, with lower burst efficiencies in mergers with higher
progenitor gas fractions. However, \citet{Moster:2011} did not find a
strong correlation with the progenitor cold gas fraction when they
including a hot halo in the merger progenitors. Although there have
been numerous studies of star formation enhancement in mergers using
numerical hydrodynamic simulations of binary mergers
\citep[e.g.][]{Mihos:1996a,Springel:2000,Cox:2006b,Cox:2008}, these
simulations are not in a cosmological context, and therefore must
assume idealized initial conditions. Furthermore, most have not
included cosmological accretion or cooling from a hot gas halo. To our
knowledge, there has not been a systematic exploration of the
enhancement of star formation activity in mergers using
cosmologically-situated hydrodynamic simulations.

SAMs predict that burst-mode star formation makes a relatively minor
contribution to the overall global star-formation rate density at any
epoch \citep[e.g. ][]{Baugh:2005,Somerville:2008b}, in agreement with
observations \citep{Rodighiero:2011,Schreiber:2014} and cosmological
hydro simulations~\citep{Murali:2002,Keres:2005}. However,
merger-triggered bursts may be important for producing certain
populations such as ultra-luminous infrared galaxies (ULIRGS) and
high-redshift sub-mm detected galaxies
\citep{Niemi:2012,Hayward:2013}, in agreement with observations that
suggest a strong connection between major mergers and
starbursts~\citep[e.g.][]{Sanders:1996,Kormendy:2009}.

\subsection{Black Hole Growth}
\label{sec:bhgrowth}

The first ``seed'' black holes may have been left behind after the
explosion of massive stars formed out of primordial gas in the early
universe. These ``Pop III'' seed BH are expected to have masses of
$\sim 100\, M_\odot$; however, such seeds cannot grow into the $10^9\,
M_\odot$ black holes required to power observed quasars at $z\sim 6-7$
if their growth is Eddington-limited. Recently, several mechanisms for
creating more massive seed BH ($10^4$--$10^6 \msolar$) have been
proposed \citep[see][for a review]{Volonteri:2010}. However, in
cosmological simulations, the usual approach is to place seed BH by
hand in halos above a critical mass ($M_H \ga 10^{10}$--$10^{11}
\msolar$). In some cases, seeds of a fixed mass are used, in others,
the seed mass is chosen to place the BH on the local $M_{\rm
  BH}-\sigma$ relation. The results that we will discuss here are
generally insensitive to the details of the seeding procedure.

One must then calculate how rapidly these seed BH will accrete gas and
grow in mass. The currently predominant model relies on the idea that
black hole growth is limited by Bondi accretion of mass within the
sphere of influence \citep{Bondi:1952}, given by
\begin{equation}\label{eq:bondi}
\dot{M}_{\rm Bondi} = \alpha \, \frac{4\pi \, G^{2} \,
M_{\rm BH}^{2} \, \rho}{(c_{\rm s}^{2}+v^{2})^{3/2}},
\end{equation}
where $M_{\rm BH}$ is the mass of the BH, $c_s$ is the sound speed of
the gas, $v$ is the bulk velocity of the BH relative to the gas,
$\rho$ is the density of the gas, and $\alpha$ is a boost parameter
included because models typically lack the spatial resolution to
resolve the Bondi radius \citep{Booth:2009,Johansson:2009b}.  Early
models took $\alpha$ to be constant (typically $\sim 100$), but 
some simulators make $\alpha$ a function of density
\citep[e.g.][]{Booth:2009} and some recent simulations resolve the
Bondi radius so can adopt $\alpha=1$.
Typically, the accretion rate is capped at the Eddington rate. As
galaxies merge, their BHs are assumed to merge when they come within
some distance of each other, typically a softening length (thereby
ignoring GR timescales for BH inspiral).

The Bondi accretion model predicts fairly low accretion rates when
galaxies are undisturbed, but when strong torques drive gas towards
the nucleus as in a major merger, accretion rates can be boosted to
levels sufficient to power quasars
\citep{Springel_agnfb:2005,Dimatteo:2005}. This is consistent with the
observation that local ULIRGs, which are mostly major mergers, also
show strong AGN activity~\citep{Sanders:1996}. In one paradigm, low
accretion rates ($\la 0.01 \dot{M}_{\rm Edd}$, where $\dot{M}_{\rm
  Edd}$ is the Eddington rate) are associated with \emph{radiatively
  inefficient} accretion, as in an Advection Dominated Accretion Flow
\citep{Narayan:1994,Blandford:1999}. In this case, most of the energy
is advected into the BH and little emerges as radiation. BH powered at
higher accretion rates are radiatively efficient and give rise to the
population of observed X-ray, UV, and optically luminous quasars and
AGN.

The assumption of Bondi accretion requires accompanying strong
feedback to obtain BHs that follow the $M_{\rm BH}-\sigma$ relation,
as this simple argument demonstrates~\citep{Angles:2013}. Consider two
BHs of mass $M_a$ and $M_b$.  If they grow according to the general
prescription $\dot{M}_{\rm BH}=D(t)M_{\rm BH}^p$, then
\begin{equation}\label{eq:conv}
\frac{d}{dt} \left ( \frac{M_{\rm a}}{M_{\rm b}} \right ) = D(t) \, \frac{M_{\rm a}^{p}}{M_{\rm b}} \left [ 1 - \left ( \frac{M_{\rm a}}{M_{\rm b}} \right )^{1-p} \right ].
\end{equation}
It is easy to show that the two masses will diverge if $p>1$, and they
will converge if $p<1$.  For Bondi accretion $p=2$; hence for BHs to
converge onto an $M_{\rm BH}-\sigma$ relation, some strongly
self-regulating feedback process must counteract Bondi accretion and
make $p$ effectively less than unity.  We will discuss possible
feedback processes in \S\ref{sec:agnfb}, but in general
such tuned self-regulation is not so straightforward to arrange,
for the usual reason that outward energetic processes tend to escape
through paths of least resistence whereas inflows typically arrive
through the dense, harder-to-disrupt gas.

It is worth emphasizing that the widely used Bondi model implicitly
assumes that the accreting gas has negligible angular momentum, which
is unlikely to be a good assumption in general. Recently, the problem
of dissipating angular momentum to enable BH accretion has received
more attention in the cosmological milieu.  Hopkins \& Quataert
(\citeyear{Hopkins:2010c,Hopkins:2011d}) studied angular momentum
transport in disks with non-axisymmetric perturbations both
analytically and in simulations, showing that such secular processes
can significantly fuel BH growth, as also suggested by
\citet{Bournaud:2011} and \citet{Gabor:2013}.  Implementing this
analytic work into zooms and cosmological runs, \citet{Angles:2013}
and \citet{Angles:2014a} showed that this torque-limited accretion
behaves qualitatively differently than Bondi accretion, since in the
\citet{Hopkins:2011d} model, the exponent of BH growth is
$p=\frac{1}{6}$.  Hence while this model also must incorporate
feedback, such feedback does not have to strongly couple to the inflow
in order to achieve self-regulation.

Black hole accretion in SAMs is of necessity more schematic. In one of
the first semi-analytic models that incorporated BH growth in the
framework of a cosmological model of galaxy formation,
\citet{Kauffmann:2000} assumed that all BH growth is triggered by
major mergers. Following such an event, they assumed that some
fraction of the cold gas was accreted by the BH, with this fraction
being a function of the halo circular velocity. A similar recipe
is incorporated into the later generations of MPA-SAMs
\citep[e.g.][]{Croton:2006,Delucia:2007,Guo:2011,Henriques:2013}. Other
SAMs additionally trigger accretion following minor mergers and disk
instabilities
\citep{Somerville:2008b,Hirschmann:2012b,Bower:2006}. Some models
allow an additional growth channel through a ``Bondi-like'' accretion
from the hot halo \citep{Somerville:2008b,Fanidakis:2011}. In the
Santa Cruz SAMs \citep{Somerville:2008b}, black hole growth is
parameterized based on the results of hydrodynamic binary merger
simulations \citep{Cox:2006b,Cox:2008,Robertson:2006b} as
characterized by \citet{Hopkins:2005a}. In this model, rapid black
hole accretion is triggered following a major or minor merger. The BH
accretes at the Eddington rate until the BH reaches a critical mass,
where the energy being radiatied is sufficient to halt further
accretion. The accretion rate then declines in a power-law ``blow
out'' phase until the BH switches off \citep{Hopkins:2005b}. 

All SAMs and numerical cosmological hydrodynamic simulations that
explicitly include BH growth use the local $M_{\rm BH}-\sigma$ or
$M_{\rm BH}-M_{\rm spheroid}$ relation to calibrate the free parameters
in the BH accretion recipes. A wide variety of BH growth recipes
appear to be able to successfully reproduce this relationship.

\subsection{Feedback Processes}
\label{sec:feedback}

Feedback can be divided into two general classes, preventive and
ejective.  Preventive feedback retards star formation by stopping gas
from accreting into the ISM, while ejective feedback describes
processes that remove the gas from the ISM after it has been accreted.
Current wisdom suggests that preventive feedback dominates when the
majority of halo gas is near the halo's virial temperature (as in very
small dwarfs or massive galaxies), while ejective feedback dominates
when most of the halo's gas is well below its virial temperature (as
in typical star-forming galaxies).  However, individual physical
processes can potentially act in both ejective and preventive ways.

\subsubsection{Squelching: Photoionization Suppression}
\label{sec:squelch}

Photons above 13.6~eV that ionize hydrogen typically add an
$\sim$eV-scale amount of latent heat, corresponding to a temperature
increase of $\sim 10^4$ K.  Hence the post-reionization optically-thin
IGM has a temperature around this value, which means that gas in halos
whose virial temperatures are comparable to $10^4$ K will be unable to
cool their gas.  This temperature corresponds to a halo mass of $\sim
10^8\ M_\odot$, implying that photoionization will strongly reduce the
baryon content and hence suppress galaxy formation in halos below this
mass.  This suppression has sometimes been called
\emph{squelching} \citep{Somerville:2002}.

Squelching can have a residual impact on halos much larger than
$10^8\, M_\odot$, since they are hierarchically assembled in part from
squelched halos.  \citet{Gnedin:2000} showed that the characteristic
mass below which halos contain substantially less than their fair
share of baryons is well represented by a filtering scale that smooths
the baryonic perturbations.  Hence one can define a {\it filtering
  mass}, which describes the halo mass that on average contains half
the cosmic fraction of baryons.

The filtering mass depends on the intricate interplay between
photoionization, cooling, and hierarchical growth, which is
challenging to model.  Early work suggested a roughly constant
circular velocity below which baryon accretion is suppressed, of
around $30-50$~km/s~\citep[e.g.][]{Thoul:1996,Quinn:1996}.  If
extrapolated to today, this would imply halos up to several times
$10^{10}\,M_\odot$ would be significantly suppressed in baryon
content~\citep{Gnedin:2000}.  More recent simulations by
\citet{Okamoto:2008} found a smaller filtering mass scale, $M_F \sim
4\times 10^{9}\, M_\odot$ today, but these simulations still assumed
ionization equilibrium, did not include metal line cooling, and
adopted a uniform meta-galactic ionizing background.  Observations of
late-type dwarfs with circular velocities $\la 42$~km/s suggest that
their baryon content is much smaller than expected from scaling
relations based on larger galaxies~\citep{Kormendy:2014}, thus
providing direct constraints on the filtering mass.

An additional complication can arise when galactic outflows are
included along with squelching, as the two can combine to produce an
``amplification of suppression" that is stronger than the product of
the individual effects~\citep{Pawlik:2009,Finlator:2012}.  The
magnitude of the effect depends on the outflow model implemented, but
can be up to a 60\% amplification during the EoR.  Unfortunately, the
high expense of these calculations that include radiative transfer
while resolving very small halos prohibits their evolution down to
$z=0$; hence it is not clear how significant this effect is at later
epochs.

In semi-analytic models, photoionization squelching is generally
implemented by assuming that reionization occurs instantaneously
throughout the Universe, at a fixed input redshift. At all later
times, the gas that is allowed to accrete into halos is reduced by a
factor $f_{\rm coll}(M_H, z)$. This function is parameterized based on
the results of numerical hydrodynamic simulations, and is expressed as
a function of the filtering mass
\citep{Gnedin:2000,Kravtsov:2004b,Okamoto:2008}.

\subsubsection{Star Formation Feedback}
\label{sec:sffb}

Stars, massive ones in particular, deposit copious amounts of energy
and momentum into the ISM during their life and in death. Stellar
feedback is invoked to explain two kinds of inefficiencies in
galaxies: 1) The efficiency of the conversion of gas into stars
\emph{within} GMC's is puzzlingly low, only about 1\% per free fall
time \citep{Krumholz:2012b}; 2) the stellar and baryon fraction within
galactic-sized halos is much less than the universal value, ranging
from a few to twenty percent \citep{Moster:2010,Behroozi:2010}. The
first inefficiency has been ascribed to turbulence generated by stars
and SNe within GMCs \citep[e.g.][and references
  therein]{Krumholz:2012b}. For purposes of cosmological simulations,
since observations suggest that this efficiency is nearly universal,
this can largely be folded into the normalization of the star
formation recipe.

The second inefficiency is generally ascribed to large-scale galactic
outflows powered by massive stars and SNae. Signatures of such
outflows, with mass loss rates likely of the same order as the star
formation rate or larger, are ubiquitously observed in star forming
galaxies \citep{Veilleux:2005}. Modeling galactic outflows has
therefore become a central challenge for recent simulations.

Early work attempted to model stellar feedback via the deposition of
thermal energy from SNae in the surrounding gas
\citep[e.g.][]{Katz:1996}.  It was quickly realized that this had
almost no effect, because the short cooling times meant that the
energy was radiated away very quickly, adding negligible ISM pressure,
let alone driving an outflow.  Since then, most cosmological models
have adopted some sub-grid prescription to enable effective ejective
feedback that typically involves either implementing ad hoc
``tricks'', such as turning off cooling for some time or super-heating
the gas, that attempt to mimic the ISM processes that allow
stellar-driven winds to develop in real galaxies, or else implementing
outflows via kinetic energy injection.

A variant on the ISM heating model called ``blast wave" feedback was
developed by \citet{Stinson:2006} and has been extensively used in
{\sc Gasoline} and RAMSES~\citep{Bournaud:2010}.  Here, after the gas
is heated, radiative cooling is shut off for the lifetime of the
SN-driven blastwave as predicted by a spherical Sedov solution.  This
enables the gas to ``feel" the higher pressure and develop a coherent
large-scale outflow.  While successful in many regards, this model
still predicted too much early star formation, so \citet{Stinson:2013}
added ``early stellar feedback" intended to mimic the energy input
from young stellar winds and radiation.

Another variant of a purely thermal stellar feedback model was
proposed by \citet{DallaVecchia:2012} and implemented in the EAGLE
simulations \citep{Schaye:2014}. Instead of turning off cooling, the
trick for preventing the energy from immediately cooling away involves
making the energy deposition stochastic. The mean amount of energy
injected per mass of stars formed is set by stellar population models
and supernova energetics.  The temperature jump of particles receiving
a boost is specified ($\Delta T=10^{7.5}$~K, typically), and a
parameter $f_{\rm th}$ determines the probability that a given SPH
particle in the vicinity of a star-forming particle will get heated.
Hence the gas is heated to much higher temperatures than would be the
case if the same amount of energy were continuously added to all of
the SPH neighbors, increasing the cooling time and mitigating energy
losses.  The overall efficiency of the feedback can be adjusted by
varying $f_{\rm th}$. \citet{Schaye:2014} made $f_{\rm th}$ a function
of the local gas metallicity and density, as they found that this most
successfully reproduced the observed SMF and galaxy sizes.

A popular approach introduced by \citet{Navarro:1993} and implemented
into \gad\ by \citet{Springel:2003a} is to simulate outflows by giving
gas ``kicks", rather than trying to overpressurize ISM gas by adding
thermal energy.  Such kinetic outflows are less directly tied to the
physics generating outflows, but enable greater control over outflow
parameters in order to both mimic observed outflows more closely and
assess the impact of varying the outflow parameters.  In such models,
hydrodynamics is sometimes shut off (``decoupled") for some period of
time to mimic the collective power of supernovae blowing a chimney
through the ISM; it is unclear whether this provides a more physical
description of outflow propagation through the ISM, but it generally
does result in better resolution
convergence~\citep{DallaVecchia:2008}.  These models are parameterized
by a mass loading factor $\eta \equiv \dot{M}_{\rm out}/\dot{M}_*$ and
a wind velocity $v_{\rm wind}$, which together determine how many
particles to kick and how hard to kick them. \citet{Springel:2003a}
assumed a constant mass loading factor and constant wind velocity, and
showed that this yielded a cosmic star formation history in much
better agreement than a model without outflows.
\citet{Oppenheimer:2006} showed that adopting scalings motivated by
analytic ``momentum-driven'' wind models \citep{Murray:2005} produced
better agreement with many galaxy and IGM properties including the
galaxy mass-metallicity relation, the enrichment history of the IGM,
and the galaxy stellar mass function \citep[see
  also][]{Oppenheimer:2008,Finlator:2008,Dave:2011,Dave:2013}.  For
momentum-driven winds, the mass loading factor scales as $\eta \propto
\sigma^{-1}$, and the wind velocity scales as $v_{\rm wind} \propto
\sigma$, where $\sigma$ is the velocity dispersion of the galaxy.  The
Illustris simulations \citep{Vogelsberger:2014a} also employ kinetic
winds by creating and launching decoupled wind particles, and
rejoining them back into the gas mesh after recoupling.  They adopt
scalings expected for ``energy-driven" winds, namely $\eta \propto
\sigma^{-2}$.

Most semi-analytic models parameterize star formation feedback in a
similar manner, based on the approach introduced in \citet{White:1991}
and \citet{Kauffmann:1993}. In each timestep, the SAM computes the
rate at which cold gas is ejected from the disk by a wind:
\[ \dot{m}_{ej} = \epsilon_{\rm w} \left (\frac{V_0}{V_c}\right)^{\alpha_w} \dot{m}_*\]
where $\dot{m}_*$ is the star formation rate in the galaxy, $V_c$ is
the circular velocity of the galaxy, $V_0$ is an arbitrary
normalization parameter, and $\epsilon_{\rm w}$ and $\alpha_w$ are
treated as tunable free parameters. For $\alpha_w=1$ or $\alpha_w=2$,
this is equivalent to the ``momentum driven'' or ``energy driven''
wind scalings discussed above. One must then decide what happens to
the ejected gas, and here different modelers diverge more widely. Some
fraction of the ejected gas may escape the dark matter halo, and may
be tracked in an ``ejected'' reservoir from which it is allowed to
accrete into the halo again over a longer timescale. Otherwise, the
ejected gas is added to the halo hot gas reservoir. SAMs generally
implement some sort of model, of varying complexity, to arrange that
the fraction of ejected gas that escapes the halo is larger at lower
halo $V_H$, and assymptotes to unity above $V_H \simeq $ 120-150 km/s
(or a halo mass of a few $\times 10^{12} \msolar$).

\subsubsection{AGN Feedback}
\label{sec:agnfb}

Observational phenomena associated with accreting black holes include
electromagnetic radiation, relativistic jets, and less-collimated
non-relativistic outflows \citep{Krolik:1999}. There are several
different physical mechanisms whereby the large amounts of energy and
momentum produced by AGN can couple with the gas in and around
galaxies, possibly regulating the growth of the black hole itself, and
potentially suppressing cooling and star formation on galactic
scales. At the most basic level, AGN can heat gas up (thermal
feedback), drive winds that eject gas (kinetic feedback), and ionize
or photo-dissociate gas (radiative feedback). The main heating
mechanisms are Compton, photo-ionization, and photo-electric
heating. Radiation may also drive winds via pressure on spectral
lines, free electrons, or dust. These winds may originate in the torus
or accretion structure near the black hole, the broad line region
(BLR), larger nuclear scales ($\sim$ kpc), or all of the above. Winds
arising on ``small'' (BLR/accretion disk) scales may drive
galaxy-scale winds by shocking and sweeping up ISM gas --- or they may
simply vent out of the galaxy without ejecting much mass. In addition,
highly relativistic giant radio jets may heat the intra-cluster medium
through bubbles, weak shocks, and sound waves
\citep{Mcnamara:2007,Fabian:2012}.

Focussing first on the processes associated with the radiatively
efficient (``radiative mode'', sometimes called ``quasar mode'' or
``bright mode'') of BH growth, one of the major dynamical questions is
whether AGN-driven winds are primarily ``energy driven'' or ``momentum
driven''. As in the case of stellar driven winds, the question is how
quickly and efficiently is the thermal energy generated when the wind
shocks the surrounding gas radiated away. Momentum cannot, of course,
be radiated away, and so if most of the thermal energy is quickly
dissipated, we term the wind ``momentum driven''. If radiative losses
are negligible, we term it ``energy driven''. Clearly real winds may
often be somewhere in between. The significance of this distinction is
that the momentum flux of swept-up material in an energy-conserving
outflow is ``boosted'' due to work done by the hot shocked gas (an
effect familiar from the Sedov-Taylor phase in supernova remnants).

It has been argued that in the dense cold gas that must surround
rapidly accreting black holes, cooling times are short and winds must
be predominantly momentum-driven
\citep{King:2005,Ostriker:2010,Debuhr:2011a}. However, observations of
AGN-driven outflows suggest ``boost'' factors of $\dot{p}/\dot{p}_{\rm
  rad} \sim 2$--30 \citep[e.g.][]{Sturm:2011,Moe:2009}, with an
average probably around 10, where $\dot{p}_{\rm rad} = L_{\rm AGN}/c$
is the radiative momentum flux output by the AGN. \citet{Faucher:2012}
argued recently based on analytic calculations that AGN-driven
outflows are likely to be largely energy-conserving in many situations
relevant to observed systems, particularly for ``fast'' ($v_w \sim
10,000$-30,000 km/s) winds.

One of the earliest three dimensional simulations of AGN feedback in
galaxies was presented in \citet{Springel_agnfb:2005} and
\citet{Dimatteo:2005}. Here, the BH accretion rate was
modelled using the Bondi approach outlined above, and the resulting
bolometric luminosity was assumed to be proportional to the BH
accretion rate. A fixed fraction of the bolometric luminosity was
deposited into the neighboring gas particles as thermal energy. These
simulations did not use cosmological initial conditions, but
considered binary mergers of idealized disk galaxies without hot gas
halos. This work showed that deposition of about 5\% of the bolometric
luminosity was able to drive strong outflows that eventually halted
further accretion onto the BH and also removed nearly all cold
gas from the galaxy, resulting in quenching of star formation
\citep{Springel_redellip:2005}. Furthermore, the models produced
self-regulated BH growth, leading to a tight $M_{\rm
  BH}-\sigma$ relationship in agreement with the observed one. A
similar approach has been used in a large number of subsequent
studies. Although these studies, taken at face value, suggest that
purely energy driven winds can regulate BH growth and drive
large-scale outflows, it is likely that radiative losses were
artificially suppressed due to the highly pressurized ISM model
adopted in these simulations.

Moreover, these simulations neglected the expected momentum
deposition. Several recent works have implemented momentum-driven
winds in hydrodynamic simulations via radiation pressure on dust
\citep{Debuhr:2010,Debuhr:2011a} and via BLR winds
\citep{Choi:2012,Choi:2014a} and found that these winds can play a
significant role in modulating the growth of the black hole and the
galaxy. The star formation remains quenched over a much longer
timescale in the simulations that include momentum-feedback, because
the density of hot gas near the center of the halos is significantly
reduced \citep{Choi:2014b}.

The other important class of feedback processes is connected with
highly collimated jets of relativistic particles (``jet mode'' or
``radio mode''; see the recent reviews by \citealt{Fabian:2012} and
\citealt{Heckman:2014}). The kinetic energy in these jets can exceed
the total bolometric luminosity of the AGN by several orders of
magnitude. While jets may be observed at many wavelengths, there is a
class of sources detected at radio wavelengths that do not exhibit the
classical signatures of ``radiatively efficient'' AGN --- no UV,
X-ray, or IR excess, and no highly ionized emission lines. Optically,
these objects resemble normal massive early type galaxies. They are
associated with radiatively inefficient accretion, and with extremely
low accretion rates onto the central BH. The radio jets are observed
to correspond, in many cases, with ``bubbles'' visible in X-ray images,
regions filled with hot plasma presumably heated by shocks from the
jet's interaction with the ICM. Studies of the bubble energetics have
shown that there is easily enough energy deposited in the ICM to
offset cooling; in fact, in groups and low-mass clusters the energy
probably exceeds the requirements for balancing cooling by up to an
order of magnitude. Radio galaxies are common in massive early type
galaxies in groups and clusters, and bubbles and/or radio sources are
seen in 95\% of ``cool core'' clusters (clusters with short central
cooling times).

Once again, the energetics are such that one expects this ``jet mode''
feedback to have a significant impact on galaxy formation, but many
details of the physics remain unclear. The main puzzle is how such
highly columnated bi-polar jets can nearly isotropically heat a large
volume of intragroup or cluster gas \citep{Vernaleo:2006}. The bubbles
provide an important clue --- these bubbles rise buoyantly in the hot
atmosphere, reaching fairly large radii. Heating may occur via
turbulent mixing of bubbles with the ICM \citep{Scannapieco:2008},
viscous dissipation of weak shocks \citep{Ruszkowski:2004}, or cosmic
ray heating \citep{Sharma:2009}. Although some recent simulations that
attempt to explicitly model jet heating in 3D have claimed greater
success at averting the cooling flow problem
\citep{Gaspari:2011,Li:2014}, all of these simulations neglect a
cosmological formation history, with merging and accretion, as well as
star formation and stellar feedback. A detailed physical understanding
of how the jets couple to the surrounding hot gas and how effective
they are in regulating cooling flows over long timescales remains
lacking \citep[see also][]{Babul:2013,Cielo:2014}.

\citet{Sijacki:2007} were the first to attempt to include both the
``radiative'' and ``jet'' modes of AGN feedback in numerical
cosmological simulations, albeit in a simplified way, necessitated by
the relatively coarse numerical resolution. 
Above a critical black hole accretion rate ($\sim 0.01$ times the
Eddington rate), the AGN was assumed to be radiatively efficient and a
fraction of the AGN bolometric luminosity was deposited in the gas as
thermal energy. Below the critical accretion rate, the AGN is assumed
to be radiatively inefficient and to produce jets which inflate
bubbles --- however they do not directly simulate the jet. Instead
they insert bubbles by hand, with energy and radius scaled to the
black hole mass as motivated by analytic models for radio cocoon
expansion.

Similar approaches have now been implemented in a few sets of
cosmological simulations.  The Illustris simulations, using the
\arepo\ moving mesh code, also use the Bondi accretion model and a
similar feedback scheme to that of \citet{Sijacki:2007}. In addition,
the Illustris simulations include a simplified treatment of
photo-ionization and photo-heating due to the AGN radiation field.  A
somewhat different approach is taken in the EAGLE \citep{Schaye:2014}
and OWLS \citep{Schaye:2010} simulations --- they adopt a variant of
the ``stochastic thermal feedback'' model used for star formation
feedback, described in \S\ref{sec:sffb}, in which an average energy
injection rate is given by $E_{\rm BH} \propto \dot{m}_{\rm acc} c^2$,
where $\dot{m}_{\rm acc}$ is the accretion rate onto the BH and $c$ is
the speed of light. The injected energy is stored by each BH until it
can stochastically heat some minimum number of particles with a
temperature increase $\Delta T$. The value of $\Delta T$ may depend on
resolution, and is higher than for the stellar feedback model $\Delta
T \sim 10^{8.5}$--$10^9$ K \citep{Schaye:2014}.  Other simulations do
not explicitly follow black hole growth and associated feedback, but
include heuristic ``quenching'' mechanisms based on surrogate galaxy
or halo properties \citep{Gabor:2011,Gabor:2012}.

A large number of groups have also implemented ``radiative mode''
AGN-driven winds and ``jet mode'' AGN heating in semi-analytic
models. 
Although the details differ from model to model, there are a number of
common elements that are widely adopted: 1) A distinction is made
between black hole fueling via cold gas (which is typically assumed to
be driven into the nucleus by mergers and/or disk instabilities; see
\S\ref{sec:bhgrowth}), and hot gas which is generally assumed to
accrete via a cooling flow. 2) BH accretion fueled by the merger/disk
instability driven mode is associated with radiatively efficient
accretion at a significant fraction of the Eddington rate; accretion
fueled by hot gas is assumed to lead to very sub-Eddington,
radiatively inefficient accretion associated with the ``jet mode''. 3)
The ``jet mode'' is assumed to be activated only in the presence of a
quasi-hydrostatic hot halo, i.e. when the halo is predominantly
accreting via the ``hot mode'' discussed earlier. 4) The ``jet mode''
is able to extract a certain fraction of the BH mass in the form of
energy, which is used to offset cooling, or is assumed to be able to
establish heating-cooling balance when the BH mass exceeds a critical
value.

For example, in the \citet{Croton:2006} model, the ``jet mode''
accretion rate is modeled as:
\[ \dot{m}_{\rm BH, R} = \kappa_{\rm AGN} \left(\frac{m_{\rm BH}}{10^8 \msolar}\right) \left(\frac{f_{\rm hot}}{0.1}\right) \left(\frac{V_{\rm vir}}{200\, {\rm km/s}}\right)^3 \]
where $f_{\rm hot}$ is the fraction of the total halo mass in the form
of hot gas, $m_{\rm BH}$ is the mass of the black hole, and $V_{\rm
  vir}$ is the virial velocity of the halo. 
The cooling rate computed as described in \S\ref{sec:sam_cooling} is
offset by a heating term, such that the effective cooling rate is:
\[ \dot{m}_{\rm cool, eff} = \dot{m}_{\rm cool} - \frac{L_{\rm AGN}}{\frac{1}{2} V_{\rm vir}^2} \]
where $L_{\rm AGN} = \epsilon_{\rm rad} \dot{m}_{\rm BH} c^2$ with
$\epsilon_{\rm rad}=0.1$ the conversion of accreted rest mass into
energy. Other SAMs use similar scalings, some with more attempted
explicit connection with the invoked physical processes and/or with
observations \citep[e.g.][]{Somerville:2008b,Monaco:2007}, but these
appear to produce similar results at $z=0$, and even for the redshift
evolution of massive galaxies \citep{Fontanot:2009}.

In addition to ``jet mode'' feedback, some SAMs implement AGN-driven
winds. \citet{Somerville:2008b} adopted momentum driven wind scalings
associated with ``radiative mode'' AGN activity:
\[
\frac{dM_{\rm out}}{dt} = \epsilon_{\rm wind}\, \epsilon_{\rm rad} \frac{c}{V_{\rm
    esc}} \dot{m}_{\rm acc} \,
\]
where $\epsilon_{\rm wind}$ represents the effective coupling
efficiency, $V_{\rm esc}$ is the escape velocity of the galaxy, and
$\dot{m}_{\rm acc}$ is the BH accretion rate in the radiatively efficient mode. See
also \citet{Fontanot:2006} for an alternate implementation of ``radiative
mode'' wind feedback in SAMs.

Examples of SAMs that do not follow BH growth explicitly, but instead
implement more heuristic halo-based quenching include
\citet{Cattaneo:2006} and \citet{Lu:2011}. In these models, cooling is
simply switched off when the halo mass exceeds a critical value, which
may depend on redshift.

\section{RESULTS FROM CURRENT MODELS: INSIGHTS AND PUZZLES}
\label{sec:results}
We return for a moment to Fig.~\ref{fig:hydroexamp1}
and~\ref{fig:hydroexamp2} to illustrate some general insights into the
process of galaxy formation and evolution in the \LCDM\ framework that
have arisen from numerical simulations. Starting with the left column
of Fig.~\ref{fig:hydroexamp1}, we see that structure formation in the
dark matter component proceeds via the formation of sheets or giant
walls, which form filaments where they intersect. Dark matter halos
form at the intersection of filaments, which funnel dark matter and
gas into halos like tributaries flowing into a lake. Comparing the
first and second columns of Fig.~\ref{fig:hydroexamp1}, one can see
that there is a very strong correspondence between the dark matter and
gas density fields on large scales. This illustrates that gas flows on
large scales are dominated by gravity. Moving to the third column of
Fig.~\ref{fig:hydroexamp1}, we can see that the gas surrounding
massive halos is hot, and larger regions become heated as time
progresses. This heating is in part due to shock heating as halos
collapse, but in these simulations is in large part due to star
formation and AGN feedback. Finally, examining the rightmost column of
Fig.~\ref{fig:hydroexamp1}, we see that metals are dispersed to quite
large distances from galaxies, and polluted regions again fill a
larger comoving volume over time. Fig.~\ref{fig:hydroexamp2} shows how
filaments of relatively cold gas can sometimes penetrate some distance
into hot halos -- these supply the ``cold mode'' accretion discussed
earlier (sometimes called ``stream fed'' accretion). The inset in
Fig.~\ref{fig:hydroexamp2} emphasizes how small galaxies are compared
with the structures seen in the ``cosmic web''.

\subsection{Global Properties}
\label{sec:global}
\subsubsection{Stellar Mass Assembly Over Cosmic Time}
\label{sec:stellmass}

\begin{figure}
\resizebox{\textwidth}{!}{\includegraphics{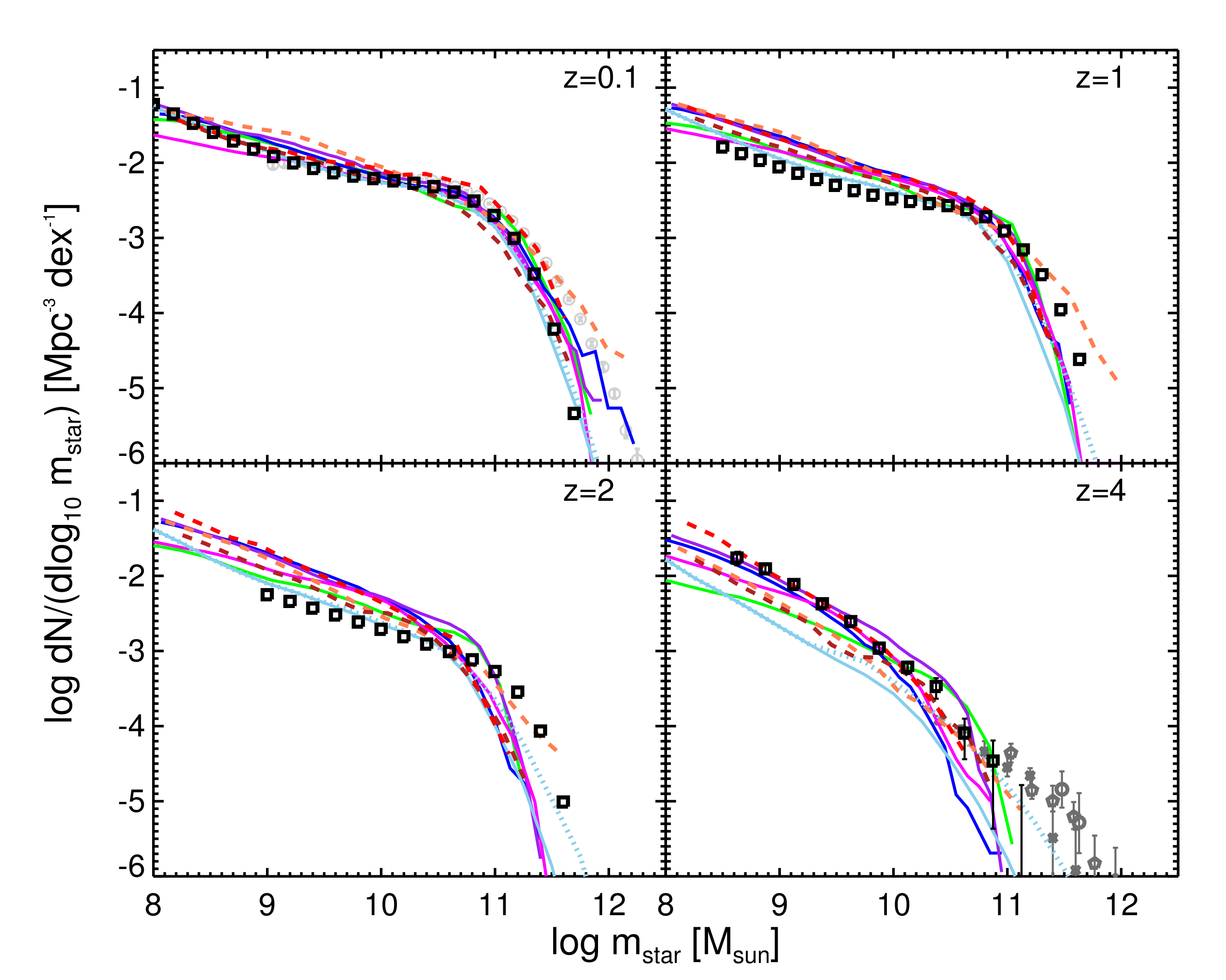}}
\caption{\footnotesize Galaxy stellar mass function at redshifts
  $z\sim 0$--4. In the $z=0.1$, $z=1$, and $z=2$ panels, black square
  symbols show a double-Schechter fit to a compilation of
  observational estimates. Observations included in the fit are:
  $z=0.1$ -- \protect\citet{Baldry:2008},
  \protect\citet{Moustakas:2013}; $z=1$ and $z=2$ panels --
  \protect\citet{Tomczak:2014}, \protect\citet{Muzzin:2013}. The fits
  shown at $z=1$ and $z=2$ are interpolated to these redshifts from
  adjacent redshift bins in the original published results. The formal
  quoted 1$\sigma$ errors on the estimates shown in these three panels
  are comparable to the symbol size, and are not shown for clarity
  (the actual uncertainties are much larger, but are
  difficult to estimate accurately). In the $z=0.1$ panel, the
  estimates of \protect\citet{Bernardi:2013} are also shown (open gray
  circles). In the $z=4$ panel we show estimates from
  \protect\citet[][squares]{Duncan:2014},
  \protect\citet[][crosses]{Caputi:2011}, \protect\citet[][circles,
    for $z=3$--4]{Marchesini:2010}, and \protect\citet[][pentagons,
    $z=3$--4]{Muzzin:2013}. Solid colored lines show predictions from
  semi-analytic models: SAGE (Croton et al. in prep, dark blue), Y. Lu
  SAM \protect\citep[][magenta]{Lu:2013}, GALFORM
  \protect\citep[][green]{Gonzalez-Perez:2014}, the Santa Cruz SAM
  \protect\citep[][purple]{Porter:2014}, and the MPA Millennium SAM
  \protect\citep{Henriques:2013}. The dotted light blue line shows the
  \protect\citet{Henriques:2013} SAM with observational errors
  convolved (see text). Colored dashed lines show predictions from
  numerical hydrodynamic simulations: EAGLE simulations
  \protect\citep[][dark red]{Schaye:2014}, ezw simulations of Dav\'{e}
  and collaborators \protect\citep[][bright red]{Dave:2013} and the
  Illustris simulations \protect\citep[][orange]{Vogelsberger:2014}. }
\label{fig:mfcomp}
\end{figure}

A fundamental observational target for modelers is reproducing the
statistical distributions of global properties for galaxy populations
at different cosmic epochs, such as luminosity functions (LF), stellar
mass functions, and cold gas mass functions. It has been realized for
some time that the observed local LF or SMF is not `naturally'
reproduced by galaxy formation models based within the
\LCDM\ paradigm: CDM models generically predict that the slope of the
mass function of dark matter halos has a slope of $\alpha_H \sim -2$,
while the slope of the observed galaxy SMF locally is much shallower
($\alpha_g \simeq -1.3$). A number of authors suggested that supernova
feedback could flatten out the low-mass slope by suppressing star
formation in low-mass halos \citep{Larson:1974,Dekel:1986,White:1991}.
Furthermore, although \LCDM\ predicts an exponential cut-off or
``knee'' in the halo mass function, with a similar functional form to
that of the observed SMF, the halo mass function turn-over is at much
larger masses. Although the cooling times in these massive, group and
cluster-sized halos are predicted to be somewhat longer than in
low-mass halos \citep{Rees:1977,bfpr:84}, this turns out to be
insufficient to explain the very inefficient star formation required
to reconcile the abundance of massive galaxies with that of dark
matter halos.

After decades of effort, theoretical models of galaxy formation are
now fairly successful at reproducing the SMF of galaxies at $z\sim 0$
by invoking a plausible, if still in most cases schematic, set of
physical processes. Fig.~\ref{fig:mfcomp} shows a compilation of
predictions of recent numerical hydrodynamic simulations and
semi-analytic models for the SMF from $z=4$ to $z\sim 0$. These models
are all taken directly from the original publications and no attempt
has been made to calibrate them to the same set of observations or to
correct for the slight differences in cosmology\footnote{The stellar
  masses in the GALFORM models have been multiplied by a factor of
  1.23 to convert from a Kennicutt to a Chabrier IMF
  \protect\citep{Mitchell:2013}.}.
This success has been obtained by ``tuning'' not only free parameters
but also the recipes associated with the sub-grid physics (star
formation, stellar feedback, AGN feedback). Predictions of the
build-up of stellar mass over cosmic time, with these recipes and
parameters held fixed, present a more stringent test of the models. 

In a broad brush sense the model predictions are generally
encouraging. A very general prediction of \LCDM-based models is that
galaxies built up their stellar mass gradually over time, which is
supported by observations. All models predict efficient early star
formation ($z\ga 4$) in low mass halos, and steep stellar mass and
rest-UV luminosity functions at these early epochs, in agreement with
observations. Models including AGN feedback or heuristic quenching
predict that massive galaxies formed earlier and more rapidly than
lower mass galaxies, again in qualitative agreement with
observations. Most models even demonstrate very good quantitative
agreement, within the errors on stellar mass estimates, between
predicted and observed SMF and LF for massive galaxies ($\mstar>M_{\rm
  char}$). Note that in Fig.~\ref{fig:mfcomp}, most of the theoretical
predictions for the stellar masses have \emph{not} been convolved with
the expected uncertainties that are inherent in the observational
estimates. Including these in a simplified manner brings the model
predictions into better apparent agreement with the observations on
the massive end \citep[e.g.][]{Lu:2013,Henriques:2013}, as shown here
for the MPA SAM as an illustration. For a more detailed study of this
issue see \citet{Mitchell:2013}.

As can be seen as well in Fig.~\ref{fig:mfcomp}, models currently have
greater difficulties reproducing the abundances and assembly histories
of low-mass galaxies at intermediate redshifts. \citet{Fontanot:2009}
demonstrated that three independently developed SAMs overproduce
galaxies with $\mstar \la 10^{10} \msolar$ by a factor of $\sim 2$--3
over the redshift range $4 \la z \la 0.5$. \citet{Weinmann:2012}
showed that a qualitatively similar problem exists for SAMs and for
hydrodynamic simulations. This problem appears to persist even in the
latest state-of-the-art cosmological hydrodynamic simulations, as seen
in Fig.~\ref{fig:mfcomp}, and already pointed out in the case of
Illustris by \citet{Torrey:2014}. As discussed in
\citet{Fontanot:2009}, several different sets of observations suggest
that massive galaxies form early and rapidly, while low-mass galaxies
form later and with a more extended timescale --- the phenomenon that
is often referred to as ``downsizing'' or ``staged'' galaxy formation
\citep{Noeske:2007b}. The overproduction of low-mass galaxies is a
symptom of the failure of current models to reproduce this
mass-dependence in the star formation histories of galaxies.  Low-mass
dark matter halos actually have \emph{earlier} formation times than
high-mass halos --- the opposite of the trend seen in observations
\citep{Conroy:2009}. In current simulations, the star formation
histories closely trace the DM mass accretion histories, thus
similarly failing to reproduce the observed trend.

It seems clear that the sub-grid recipes controlling star formation
and/or stellar feedback need to be modified in order to solve this
problem. \citet{Henriques:2013} found that making the stellar feedback
stronger and modifying the timescale for the re-accretion of ejected
gas led to significant improvement in the MPA-SAM for the predicted
abundances of low-mass galaxies as well as other observed properties
at $z\la 3$. \citet{White:2014} investigated several classes of
empirical solutions to this problem, including modifying the
efficiency of stellar driven galaxy outflows, modifying the timescale
for gas to turn into stars, and modifying the timescale for gas to be
accreted (or re-accreted) into galaxies. They concluded that solutions
that modified the outflow efficiencies and accretion timescales were
the most promising. Moreover, \citet{Torrey:2014} experimented with
changing the coupling strength and velocity of the stellar driven
winds, and found that this can change the normalization of the SMF at
the low-mass end, but cannot change the evolutionary shape, which is
what is required to solve this problem.

A convenient way to assess the success of a cosmological simulation in
reproducing the galaxy SMF or LF is via empirical constraints on the
relationship between stellar mass (or luminosity) and halo mass, as
derived by ``galaxy-halo mapping'' techniques such as SHAM and HOD,
and other methods such as galaxy-galaxy lensing, clustering, satellite
kinematics, and X-ray observations \citep[][and references
  therein]{Moster:2010,Behroozi:2010,Moster:2013,Behroozi:2013b}.
Different methods and groups are generally all in broad agreement that
star formation feedback plays a crucial role in shaping this
relationship for halos with $M_H \la 10^{12} \msolar$, with
photo-ionization squelching perhaps also playing a significant role
below halo masses of about a few $10^{10} \msolar$ (this mass scale
remains uncertain; see \S\ref{sec:squelch}). At larger halo masses,
$M_H \ga 10^{12} \msolar$, there is a general consensus that AGN
feedback probably plays an important role, although other processes
(such as gravitational heating) may contribute as well
\citep{Khochfar:2008,Johansson:2009a,Birnboim:2011}. In order to
reproduce the slope of the stellar mass-halo mass ($m_{\rm
  star}-M_{\rm halo}$) relation at low masses, most models adopt
stellar feedback recipes that either assume or result in mass loading
factors that increase fairly strongly with decreasing halo mass or
circular velocity, similar to the energy- or momentum-driven wind
scalings discussed in \S\ref{sec:sffb}.

There is a broad though not universal consensus that AGN feedback
implemented purely via deposition of thermal energy associated with
the radiatively efficient mode of BH growth \citep[as in
  e.g.][]{Dimatteo:2005} does not by itself suppress cooling and star
formation in massive halos enough (or on long enough timescales) to
satisfy observational constraints.  Although thermal energy deposition
can temporarily slow or halt cooling, after several Gyr, the gas
starts to re-cool and form stars~\citep{Gabor:2012,Choi:2014b}.  An
exception is the stochastic thermal feedback model implemented in
EAGLE, which reproduces the observed stellar fractions very well,
though there is still tension between the predicted temperatures of
the hot gas in group- and cluster-sized halos and X-ray observations
\citep{Schaye:2014}.  Another solution on the high-mass end is nearly
constant injection of energy via ``jet mode'' feedback, although as
discussed in \S\ref{sec:agnfb}, implementations of this process in
cosmological simulations remain schematic. Inclusion of the momentum
deposition associated with the radiatively efficient mode also appears
to be able to suppress cooling for longer \citep{Choi:2014b}.

\subsubsection{Global Scaling Relations: Gas, Star formation and Metals}

\begin{figure}
\resizebox{\textwidth}{!}{\includegraphics{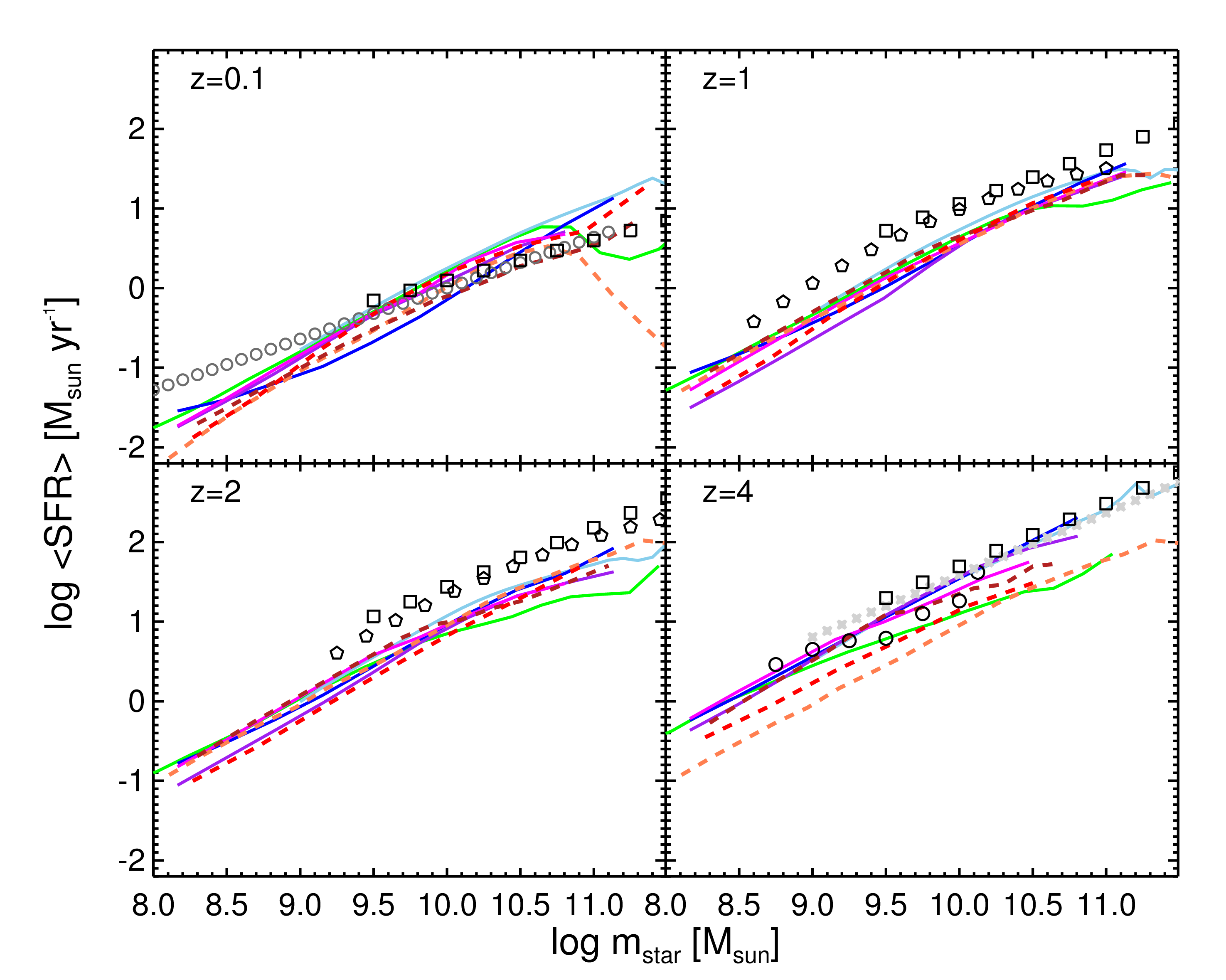}}
\caption{\footnotesize The average star formation rate in bins of
  stellar mass, for redshift bins from $z=0$--4. Grey and black
  symbols show observational estimates: $z=0.1$ --
  \protect\citet[][open circles]{Salim:2007}; $z=1$ and $z=2$ --
  \protect\citet[][pentagons, interpolated in redshift from the
    published results]{Whitaker:2014}; $z=4$--
  \protect\citet[][crosses]{Steinhardt:2014};
  \protect\citet[][circles]{Salmon:2014}; all panels -- fit to data
  compilation from \protect\citet[][squares]{Speagle:2014}. Colored
  lines show predictions from semi-analytic models and numerical
  hydrodynamic simulations; key is the same as in
  Fig.~\protect\ref{fig:mfcomp}. Note that the observational estimates
  shown are for star forming galaxies; different methods have been
  used to isolate the ``star forming sequence'' from ``quiescent''
  galaxies. Some of the modelers have applied a cut to select star
  forming galaxies, but some have not. }
\label{fig:sfrcomp}
\end{figure}

\begin{figure}
\resizebox{\textwidth}{!}{\includegraphics{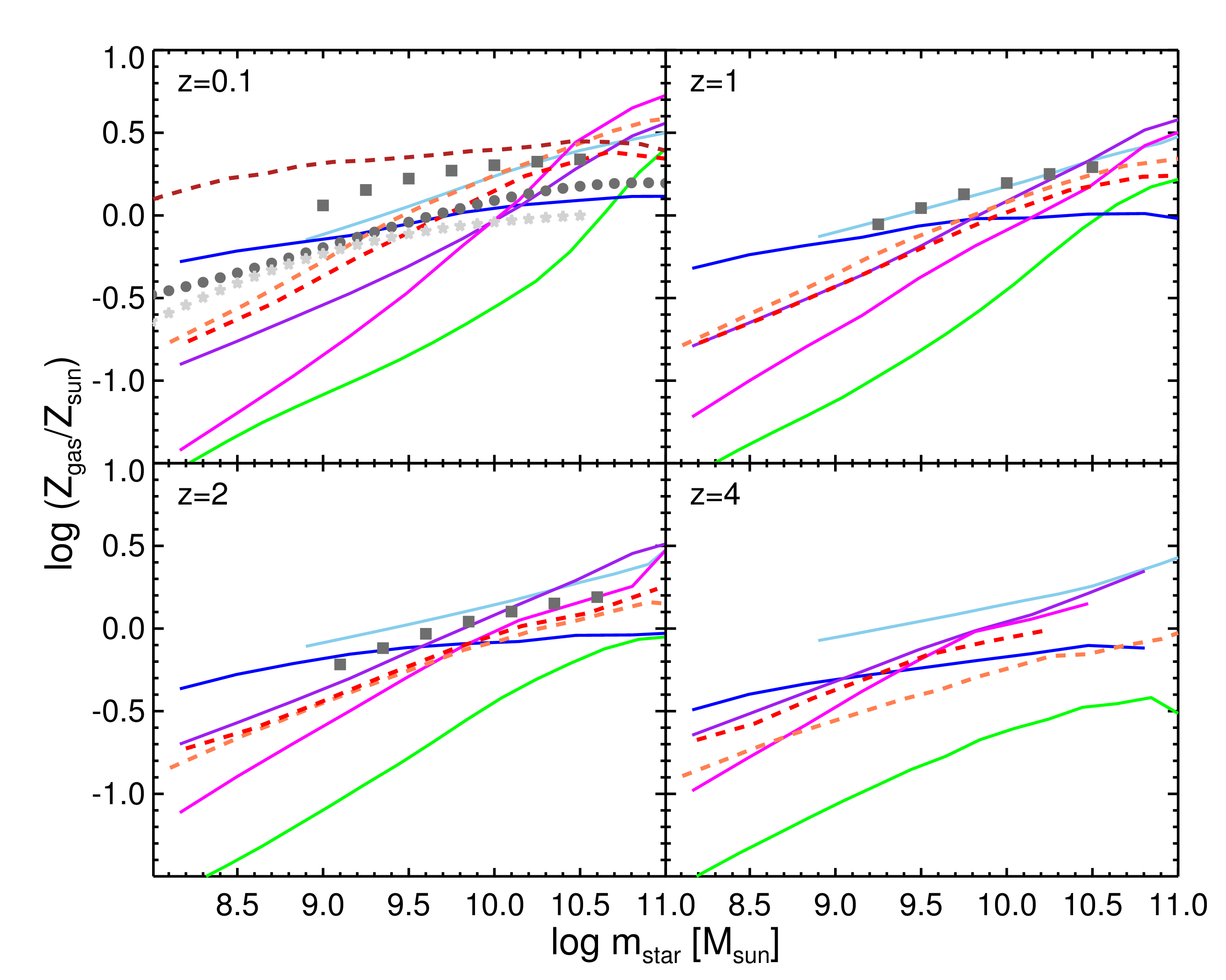}}
\caption{\footnotesize The average metallicity of cold gas in bins of
  stellar mass, for redshift bins from $z=0$--4. Grey and black
  symbols show observational estimates: $z=0.1$ --
  \protect\citet[][filled circles]{Peeples:2014}; \protect\citet[][
    stars]{Andrews:2013}. In all panels, the filled squares show the
  compilation of \protect\citet{Zahid:2013}. Colored lines show
  predictions from semi-analytic models and numerical hydrodynamic
  simulations; key is the same as in Fig.~\protect\ref{fig:mfcomp}. }
\label{fig:zgascomp}
\end{figure}

\begin{figure}
\resizebox{\textwidth}{!}{\includegraphics{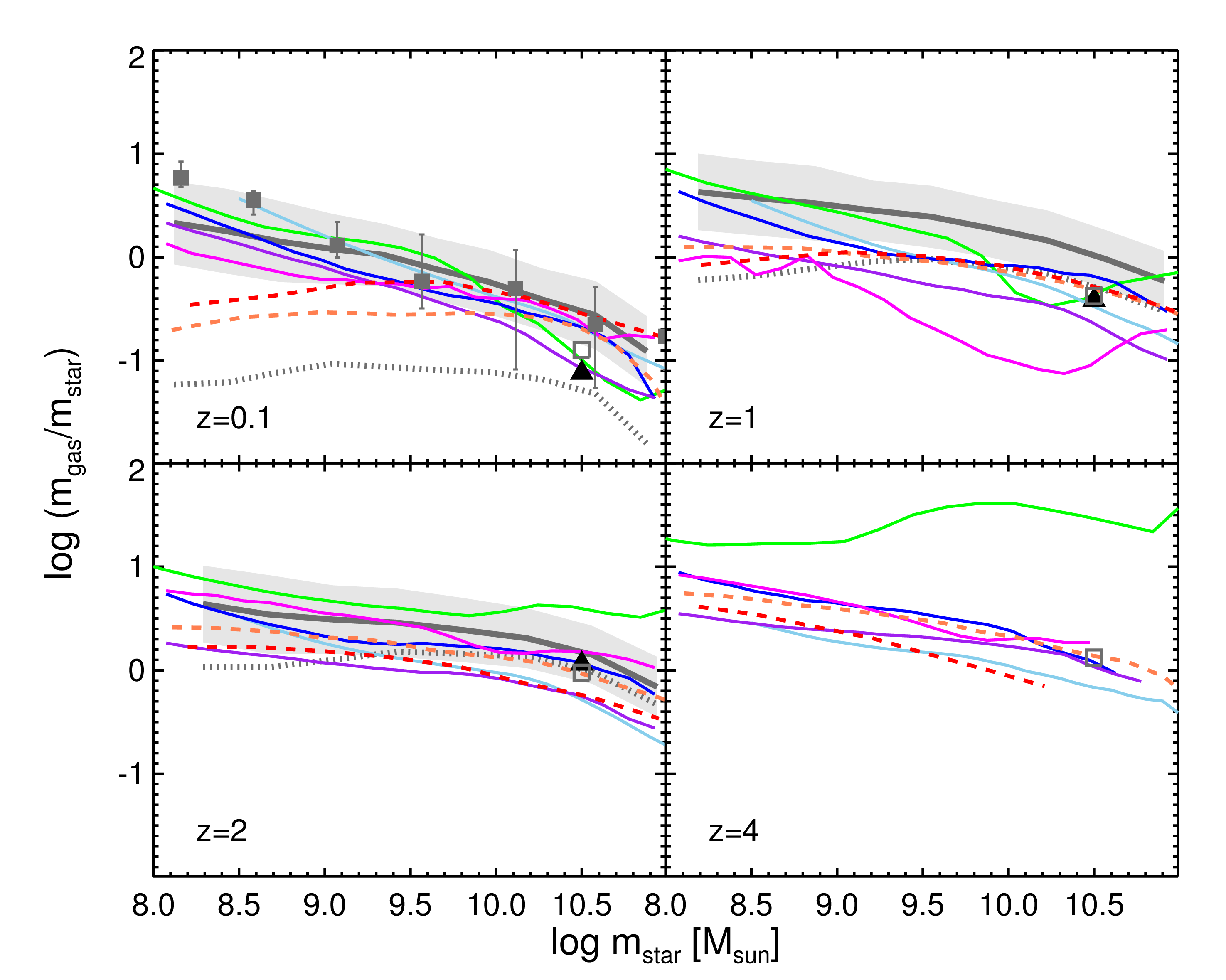}}
\caption{\footnotesize The average cold gas fraction in bins of
  stellar mass, for redshift bins from $z=0$--4. Grey and black lines
  and symbols show observational estimates: $z=0.1$ -- binned results
  from the compilation of \protect\citet[][filled
    squares]{Peeples:2014}.  In all panels, the open squares show the
  predictions of the ``equilibrium model'' presented in
  \protect\citet{Saintonge:2013}, which are in good agreement with
  their data compilation and with the estimates of \citet[][shown as
    solid triangles]{Genzel:2014}, extending up to $z\sim 3$. The
  solid gray lines and light gray shaded areas show the empirical
  total cold gas fraction ($(M_{HI}+M_{H2})/\mstar$) estimates from
  \protect\citet{Popping:2014b}. Dotted gray lines show the molecular
  fraction estimates ($M_{\rm H2}/\mstar$) from
  \protect\citet{Popping:2014b}. Note that the $z=0$ observational
  estimates shown are for \HI+\Htwo, while the $z>0$ estimates are
  based on CO and most closely trace \Htwo. Colored lines show
  predictions for the total cold gas fraction from semi-analytic
  models and numerical hydrodynamic simulations; color key is the same
  as in Fig.~\protect\ref{fig:mfcomp}. }
\label{fig:fgascomp}
\end{figure}

Galaxies are comprised of stars, gas, metals, black holes, and dark
matter.  The scaling relations between these properties as a function
of mass and redshift provide crucial constraints on galaxy growth, and
are in principle among the most direct ways to constrain baryon
cycling processes. 

The basic origin of many scaling relations can be understood in a
simple framework based on mass balance in the ISM (alternately called
an ``equilibrium'', ``bathtub'', or ``gas regulator" model):
\begin{equation}
\dot{M}_{\rm inflow} = \dot{M}_* + \dot{M}_{\rm outflow} + \dot{M}_{\rm gas},
\label{eq:massbal}
\end{equation} 
where the terms are the baryonic mass inflow rate, SFR, mass outflow
rate, and rate of change of the mass in the gas reservoir.  When
averaged over cosmological timescales, $\dot{M}_{\rm gas}$ is expected
to be small compared to the other
terms~\citep{Finlator:2008,Dave:2012,Dekel:2014}, though its evolution
can have important effects~\citep{Lilly:2013}.  Inflow into halos is
driven primarily by gravitational accretion from the
IGM~\citep{Keres:2005,Dekel:2009}. The rate at which dark matter halos
grow, the {\it halo mass accretion rate} ($\dot{M}_{\rm halo}$), is
well-characterized in $\Lambda$CDM, and roughly given by $\dot{M}_{\rm
  halo}\propto M_{\rm
  halo}(1+z)^{2.5}$~\citep{Dekel:2009,Faucher:2011}.  However,
preventive feedback within galaxy halos can retard gas accretion into
the ISM, and outflows can remove fuel for star formation even after it
enters the ISM, so $\dot{M}_{\rm inflow}$ may not trace $\dot{M}_{\rm
  halo}$.

We can rewrite equation~\ref{eq:massbal} as
\begin{equation}
{\rm sSFR} = \frac{\zeta (1+z)^{2.5}}{(\mstar/M_{\rm halo})\times (1+\eta)},
\end{equation}
where $\zeta$ is the fraction of material entering the virial radius
that makes it into the ISM, and $\eta\equiv \dot{M}_{\rm
outflow}/\dot{m}_{\rm star}$ is the outflow mass loading factor.
The dependence of sSFR on $\mstar$ and $z$ therefore reflects the
evolving combination of accretion and feedback.

Figure~\ref{fig:sfrcomp} shows a comparison of SFR vs. $\mstar$ for
the SAMs and simulations shown in Figure~\ref{fig:mfcomp}, along with
a compilation of recent observational determinations as described in
the figure caption. All models generally reproduce the near-unity
slope, at all redshifts.  Most models match the amplitude at $z\sim
0$, although the turnover at high masses due to quenching can vary
significantly (and can be sensitive to the definition of ``star
forming'' galaxies), and models tend to predict a steeper trend at low
masses.  By $z\sim 1-2$, it is clear that most models fall below the
observations, a long-standing discrepancy first highlighted in
\citet{Daddi:2007}. The redshift dependence of the sSFR is generically
difficult to match in models because it differs strongly in the
intermediate redshift regime ($4 \la z \la 0.5$) from the dependence
predicted by $\dot{M}_{\rm halo}$~\citep{Dave:2008,Sparre:2014}.  By
$z=4$, some models are able to match the data, though others continue
to fall substantially short. The normalization of the predicted SFR
vs. $\mstar$ relation depends on resolution and the calibration of the
sub-grid parameters --- e.g. \citet{Schaye:2014} show (their Fig. 11)
that a higher resolution simulation in the EAGLE suite, re-calibrated
to the SMF, predicts a higher SFR at $\mstar \la 10^{10}$, in better
agreement with the observations. However, the redshift dependence of
the sSFR is roughly unchanged \citep{Furlong:2014}.

These trends can be generally understood in the mass balance
framework.  From abundance matching, the $m_{\rm star}-M_{\rm halo}$
relation is constrained to evolve mildly with
redshift \citep{Behroozi:2013b,Moster:2013}.  If $\zeta$ and $\eta$
also evolve slowly, then sSFR should evolve as $\sim (1+z)^{2.5}$.
This is indeed roughly the evolution observed out to $z\sim
2$~\citep[e.g.][]{Lilly:2013,Speagle:2014,Whitaker:2014}.  However, to
higher redshift the evolution slows, suggesting that either the
$m_{\rm star}-M_{\rm halo}$ relation or $\eta$ is higher, or that
$\zeta$ is lower.  The assumption of $\dot{M}_{\rm gas}\approx 0$ may
be faulty at very early epochs if the inflowing gas cannot be
processed in the ISM fast enough, which observations suggest may be
the case at $z\ga 4$~\citep{Papovich:2011}.  It had been suggested
that the efficiency of converting ISM gas into stars is reduced owing
to the lower metallicity at early epochs which results in less
efficient formation of molecular gas~\citep{Krumholz:2012}, but
hydrodynamic simulations and SAMs incorporating \Htwo\ formation
modeling suggest that this effect is not large enough to solve the
problem at the observed mass
scales~\citep{Christensen:2012,Somerville:2014}. Moreover, the EAGLE
simulations also include a metallicity-dependent density threshold for
star formation as proposed in \citet{Schaye:2004}, representing the
same physical effect, but still suffer the same problem. Hence
although it is encouraging that most models are now able to predict
the sSFR evolution to within a factor of 2--3 and predict roughly the
right qualitative trend, such discrepancies, if real, could be
pointing to a need to revise our basic understanding of the physical
processes regulating star formation at these epochs.

The mass dependence of the sSFR also poses interesting challenges to
models.  In detail, the halo mass accretion rate has a super-linear
dependence on $M_{\rm halo}$, which would naively translate into a
positive slope for sSFR$(m_{\rm star})$.  Observations indicate a
sub-unity slope, becoming shallower with time (see
Figure~\ref{fig:sfrcomp}).  Part of this may be due to the fact that
for $M_{\rm halo}\ga 10^{11}M_\odot$, an increasing amount of halo
inflow is gravitationally shocked into hydrostatic
equilibrium~\citep{Birnboim:2003,Keres:2005,Gabor:2012}.  Simulations
show this is sufficient to explain the mildly negative slope in
moderate-sized halos~\citep{Faucher:2011,Dave:2011}, but is
insufficient to explain the rapid increase in quenched galaxies at
high masses, which requires additional feedback, likely associated
with AGN.

Moreover, models with outflows tuned to reproduce the observed SMF
(hence $m_{\rm star}-M_{\rm halo}$) predict a flat or positive slope
for sSFR$(m_{\rm star})$ at $M_{\rm halo}\la 10^{11}M_\odot$, while
observations show a negative slope.  
The stochasticity of star formation in dwarf galaxies
\citep{Tolstoy:2009} may result in a duty cycle whereby observed
samples preferentially select dwarfs that are in a high sSFR state,
but observations show that essentially all isolated dwarfs in the
nearby universe are star-forming~\citep{Geha:2012}.  This is an aspect
of the ``dwarf galaxy conundrum'' highlighted in \citet{Weinmann:2012}
and \citet{White:2014} and discussed above, which remains a puzzle: in
current models that are normalized to fit the present-day SMF, dwarf
galaxies not only form their stars too early (resulting in the
low-mass excess at intermediate redshift seen in
Fig.~\ref{fig:mfcomp}), they also have sSFR that are well below the
observed values \citep[see also][]{Torrey:2014}.  At higher
redshifts, the mass dependence is in good agreement with existing
observations, but deeper near-IR data is needed to test if a similar
discrepancy occurs in mass-selected samples of dwarfs at $z\ga 1$.

Another key scaling relation is the stellar mass-gas phase metallicity
relation (MZR), which can also be understood from
equation~\ref{eq:massbal}.  Given a metal yield $y$ per unit star
formation, the metallicity will be the yield times the SFR, divided by
the amount of accreting gas, i.e.
\begin{equation}
Z = \frac{y\dot{M}_*}{\dot{M}_{\rm inflow}} \approx \frac{y}{1+\eta},
\label{eq:mzr}
\end{equation}
where this approximation is valid in the limit of no recycled (i.e.
previously enriched) accretion into the ISM~\citep{Finlator:2008}.
Wind recycling is generally more rapid in more massive
galaxies~\citep{Oppenheimer:2010}, which will tend to make the MZR
steeper.  Also, outflows that are significantly more enriched than the
ISM can result in a lower metallicity.  Hence the MZR primarily
reflects galactic outflows, modulated by wind recycling and the
``metal loading factor" \citep{Peeples:2011}.

Figure~\ref{fig:zgascomp} shows the predicted MZR in our suite of SAMs
and simulations, compared with observations.  We emphasize that, due to
uncertainties in the theoretical yields of at least a factor of $\sim
2$, and differences of $\sim 30$ percent in the adopted value of solar
metallicity in different simulations, the absolute normalizations of
the predicted MZR should not be given as much weight as the trends
with mass and redshift. We also show a recent compilation of
observational estimates. Gas-phase abundance measures are sensitive to
calibration~\citep{Kewley:2008}, but it is usually the case that {\it
  relative} abundances are more consistent among various
indices. Hence the slope of the observed MZR is more robustly known
than the amplitude, though the amplitude should still be accurate to
within a factor of $2-3$. We show MZR determinations at $z\sim 0.1$,
$z\sim 1$, and $z\sim 2$ converted to the same calibration, from
\citet{Zahid:2013}. We also show the local MZR from
\citet{Peeples:2014}, which uses the average of all of the
calibrations presented in \citet{Kewley:2008}, and the local ``direct
method'' MZR from \citet{Andrews:2013}.

At $z=0$, most models produce roughly the correct metallicity for
galaxies with stellar masses of a few $\times 10^{10}\, M_\odot$, but
predicted MZRs are typically steeper than the observed relations to
low masses and have a less pronounced turnover to high masses (EAGLE,
which produces a very shallow MZR, is a notable exception).  To higher
redshifts, models generally predict slow evolution, about a factor of
two at a given stellar mass from $z=2\rightarrow 0$, which is roughly
consistent with available observations.

To explore the origin of the slope discrepancy, note that
equation~\ref{eq:mzr} shows that (in the absence of recycling and
metal-enriched outflows), when $\eta\gg 1$ as is generally the case at
low masses in these models, the observed MZR $Z \propto
m_*^{0.3}$~\citep{Tremonti:2004} implies $\eta\propto m_*^{-0.3}$.
When such a scaling (which is similar to the momentum driven wind
scaling) is implemented into hydrodynamic simulations, this produces
good agreement with the observed MZR~\citep{Finlator:2008,Dave:2011b},
but the predicted SMF is somewhat too steep at the faint
end~\citep{Dave:2011}.  Ameliorating this by incorporating a steeper
mass dependence of $\eta$ results in an MZR that is too
steep~\citep{Dave:2013}.  Accounting for wind recycling does not help
this problem-- \citet{Oppenheimer:2010} highlighted the importance of
wind recycling in shaping the SMF at intermediate masses, but in
general wind recycling is more important at higher masses, which
further steepens the MZR.  In general, current simulations have
difficulty simultaneously matching the low-mass ends of the SMF and
the MZR, suggesting that enrichment in low-mass galaxies is not fully
understood. This problem was also discussed in the context of the
Illustris simulations by \citet{Torrey:2014}, who speculated that this
tension may suggest that preventative, rather than ejective, feedback
is dominant in low-mass galaxies.

Cold gas scaling relations provide information on the fuel for star
formation.  CO measurements are currently the best tracer of molecular
gas content, although there remain significant uncertainties in the
conversion factor from CO to $H_2$~\citep[$X_{CO}$;][]{Bolatto:2013},
particularly to higher redshifts.  Observations show that low-mass
galaxies are more gas-rich, with $f_{\rm gas}\propto
m_*^{-0.57}$~\citep{Peeples:2011}. Direct estimates of the
\Htwo\ fraction of galaxies to high redshift from CO and dust-based
methods (corrected for selection effects) indicate a rise in
$m_{H2}/(m_{H2}+\mstar)$ back in cosmic time to $z\sim 2$, then a
plateau or possibly a slight decline
\citep{Geach:2011,Tacconi:2013,Saintonge:2013,Scoville:2014,Genzel:2014}.
Empirical estimates of \Htwo\ and total gas fraction based on extended
SHAM modeling \citep{Popping:2014b} indicate a similar behavior.

Figure~\ref{fig:fgascomp} shows a comparison of the cold gas fractions
($\equiv m_{\rm cold}/\mstar$) in models, where we defined cold gas in
the numerical simulations as that having a hydrogen number density
$n_H>0.13$~cm$^{-3}$.  At $z=0$, models generally reproduce the
steeply rising gas fractions to low masses, though some have gas
fractions significantly below the observed ones at the lowest masses.
Gas fractions tend to be fairly sensitive to the prescription used to
turn that gas into stars, which varies significantly between
models. This generic trend of gas fraction with galaxy mass in the
models arises from two physical effects that make the global SFE lower in
low-mass galaxies: stronger stellar feedback \citep{Brooks:2007}, and
less efficient formation of
\Htwo\ \citep{Christensen:2012,Popping:2014a}. Models generally
predict rising gas fractions at a given mass to earlier epochs, in
broad agreement with observational and SHAM-based estimates out to
$z\sim 2$, though gas fractions from the \emph{a priori} models tend
to be lower than the empirical SHAM predictions at $z\sim 2$--1;
perhaps this is another manifestation of the ``dwarf galaxy
conundrum'' discussed above. Models that track \Htwo\ formation
generally predict that galaxies become increasingly \Htwo-dominated at
higher masses and at high redshift
\citep{Popping:2014a,Lagos:2011b,Fu:2010}.

Atomic hydrogen (\HI) can be detected in emission in nearby galaxies,
and in distant galaxies via absorption. Since \HI\ represents a
transient phase of accretion from the ioinized IGM to the molecular
ISM, it is necessary to include both self-shielding and molecular gas
formation physics in order to model it, neither of which are
straightforward at typical cosmological, or even zoom, resolutions.
Nonetheless, SAMs and simulations can broadly reproduce \HI\ mass
functions and scaling
relations~\citep{Popping:2009,Obreschkow:2009,Lagos:2011b,Duffy:2012,Dave:2013,Popping:2014a}.
\HI\ may be a particularly good tracer of environmental processes
including satellite quenching \citep{Lagos:2014,Rafieferantsoa:2014},
because it is usually arises in the more loosely-bound outskirts.  In
addition, SAMs and numerical simulations are being used to study the
nature of \HI\ seen in absorption (Lyman-limit and Damped
Lyman-$\alpha$ systems), and its connection with galaxies identified
in emission
\citep{Berry:2014,Rahmati:2013,Rahmati:2014,Bird:2014}. These studies
provide important complementary constraints on disk formation and
feedback processes.

So far, we have only considered mean scaling relations, which can be
understood in terms of the average accretion rate into the ISM. In the
accretion-driven scenario, galaxies fluctuate around the scaling
relations, and the timescale to return to the mean is comparable to
that required to double the mass of the galaxy.  Hence the scatter of
the scaling relations reflects the frequency and efficacy of
``perturbing" events. In particular, mergers can drive significant
departures from the mean scalings. For example, galaxies that lie
significantly above the SF main sequence are observed to have
concentrated, spheroid-like (high Sersic) light profiles
\citep{Wuyts:2011}, as expected if they are driven by major
mergers. 
Reproducing the scatter in the observed scaling relations over cosmic
time is a stringent challenge that models are only beginning to tackle
\citep[e.g.][]{Sparre:2014}.

For the mass-metallicity relation, the scatter is seen to be
well-correlated with SFR, in the sense that galaxies at a given mass
with low metallicity have high
SFR~\citep{Mannucci:2010,Lara-Lopez:2010} and high
\HI\ content~\citep{Lara-Lopez:2013,Bothwell:2013a}.  This is a
natural outcome of the accretion rate fluctuation scenario, since a
galaxy that undergoes an uptick in accretion will increase its SFR and
gas content, owing to a larger gas supply, and lower its metallicity
since the accreted gas (or infalling galaxy) will tend to have lower
metallicity~\citep{Dave:2011b}.  This so-called ``fundamental
metallicity relation" has two aspects, namely this second-parameter
trend, and the claim by \citet{Mannucci:2010} that it is invariant
with redshift from $z\sim 0-2.5$.  However, calibration uncertainties
in metallicity measures owing to evolving ISM
conditions~\citep{Kewley:2013} make the redshift independence
difficult to robustly confirm, and even the existence of this
second-parameter trend with SFR is not as clear at higher
redshifts~\citep{Steidel:2014,Sanders:2014}.

\subsubsection{Demographics of Star-Forming and Quiescent Galaxies}

The existence of quiescent galaxies, that almost entirely ceased
forming stars many billions of years ago, is an additional indication
of the need for some sort of ``quenching'' mechanism --- processes
that prevent gas from cooling and/or forming stars. \citet{Peng:2010}
coined the terms ``mass'' and ``environmental'' quenching. In view of
the strong correlations between quiescence and other galaxy internal
properties (see \S\ref{sec:intro:obs}), we prefer the terms
``internal'' and ``environmental'' quenching. Some of the discussion
here will mirror that in \S\ref{sec:stellmass}, however, the
requirements for producing the correct internal and environmental
statistical correlations for quiescent galaxies are more stringent
than simply reproducing the stellar mass function -- models that
reproduce the latter are not guaranteed to reproduce the former.

The massive galaxies that are predominantly early type and quiescent
in the observed universe are expected to reside within massive dark
matter halos ($\ga 10^{12} \msolar$). These halos are expected
theoretically, and known observationally through X-ray observations,
to be filled with hot gas at virial temperatures of a few $\times
10^6$--$10^8$~K that is gravitationally shock-heated on infall, and
enriched to about a third of solar.  Assuming hydrostatic equilibrium,
this gas should be cooling fairly rapidly, at rates of hundreds to
thousands of solar masses per year. The absence of the signatures of
gas cooling below about one-third of the virial temperature in
clusters, along with the absence of large amounts of cold gas or young
stars, constitutes the classical ``cooling flow'' problem \citep[][and
  references therein]{Mcnamara:2007}. This problem has its counterpart
in theoretical models, in that it has proven difficult to find
plausible physical mechanisms that can suppress cooling and keep
galaxies in massive halos as quiescent as they are observed to be.

Simulations without any sort of ``quenching'' mechanism (such as AGN
feedback) produce inverted color-magnitude relations (more massive and
luminous galaxies are more likely to be blue and star forming) without
any hint of bimodality \citep{Somerville:2008b,Gabor:2011}. The first
generation of SAMs that included AGN feedback were able to
\emph{qualitatively} reproduce the observed bimodality of the color
and sSFR distribution and the fraction of quiescent galaxies as a
function of stellar mass
\citep{Croton:2006,Bower:2006,Somerville:2008b,Kimm:2009}; certainly,
including AGN feedback greatly improved the results relative to the
old models. In these models, the mechanism that was primarily or
entirely responsible for quenching was the ``jet mode'' type of
feedback described in \S\ref{sec:agnfb}, in which star formation dies
out because the hot gas halo is continually heated so the supply of
new cold gas is cut off. More heuristic models, in which cooling is
simply shut off when the dark matter halo exceeded a certain critical
mass, performed nearly as well as models that explicitly implemented
``jet mode'' AGN feedback \citep{Cattaneo:2006,Kimm:2009}. Some recent
SAMs reproduce the observed tighter correlation of the quiescent
fraction with $B/T$ than with stellar mass, while others do not,
suggesting that this could provide constraints on quenching mechanisms
\citep{Lang:2014}. 

\citet{Springel_redellip:2005} showed that including thermal AGN
feedback in hydrodynamic simulations of isolated binary mergers was
able to drive powerful winds that evacuated most of the cold gas from
the galaxy, leading to strong quenching of star formation. These
results motivated semi-empirical models positing that quenching
associated with mergers, rapid black hole growth, and ``radiative
mode'' AGN feedback could explain the growth of the quiescent early
type population \citep{Hopkins:2008a,Hopkins:2008b}. However,
subsequent work with semi-analytic models and cosmologically-based
hydro simulations suggested that thermal feedback associated with the
``radiative mode'' of BH accretion, when implemented using algorithms
similar to those of \citet{Springel_agnfb:2005}, is not able to
produce long-lived quiescent galaxies, since it fails to prevent
subsequent accretion which reactivates star formation within a Gyr or
two \citep[see e.g.][]{Gabor:2011,Choi:2014b}.

Following the approach presented in \citet{Sijacki:2007}, the
Illustris simulations explicitly included both local thermal heating
associated with black hole accretion above a critical rate
(representing ``radiative mode''), and more distributed heating
associated with low BH accretion rates (representing ``jet
mode''). They produced a bimodal color magnitude diagram, with a red
galaxy fraction as a function of stellar mass and environment in good
agreement with observations by \citet{Peng:2010} and
others~\citep{Vogelsberger:2014a}.  The observed red sequence colors
have proven difficult to reproduce quantitatively in all types of
models~\citep{Guo:2011,Gabor:2012,Vogelsberger:2014a}, but significant
uncertainties remain in this regime in the stellar population models
that are used to predict such colors from models.  In contrast, the
``stochastic thermal'' AGN feedback model as implemented in EAGLE does
not explicitly have two distinct modes, and can still reproduce
quenched galaxy observations at a similar level \citep{Schaye:2014}.
Cosmological zoom-in simulations including fast momentum-driven AGN
winds also appear to be able to quench and maintain quiescence over
long timescales without any explicit ``jet mode'' type feedback
\citep[][Choi et al. in prep]{Choi:2014b}.  Conversely,
\citet{Gabor:2014} suggested that the presence of a hot halo kept hot
by AGN feedback is sufficient to quench a galaxy, without the need for
additional radiative mode feedback, showing that this reproduces both
internal and environmental quenching as observed.  Hence there remains
much debate over the relative importance of these two AGN feedback
modes, whether one or both are required, and even whether they are
distinct.

Reproducing the observed patterns of ``environmental quenching'' has
provided another challenge to models. \citet{Peng:2012} showed that
when SDSS galaxies were identified as ``satellites'' or ``centrals''
using a group catalog, the fraction of quiescent centrals depended
only on stellar mass, while the fraction of quiescent satellites
depended on both mass and environment.  Certainly there are many
candidate processes that could preferentially quench satellites, such
as harrassment, tidal stripping, or ram pressure stripping.  In many
SAMs, galaxies are not allowed to accrete any new gas from the hot
halo or the IGM once they become satellites (sometimes called
``strangulation''). This is known to produce far too high a fraction
of quiescent satellites
\citep{Weinmann:2006b,Font:2008,Kimm:2009}. Instead, satellite
quenching seems to take a surprisingly long time, perhaps many
Gyr~\citep{Wetzel:2012}.  Hydrodynamic simulations indeed show that
infalling satellites remain star-forming for at least a
Gyr~\citep{Simha:2009}, as it takes time for the hot gas and dark
matter from the halo in which the satellite galaxy was born to be
stripped away. Including this delayed stripping of the hot gas halo,
without including any other environmental effects (e.g. tidal or ram
pressure stripping of the cold gas in satellites) improves satellite
statistics in SAMs \citep{Font:2008,Weinmann:2010}, though some
tension with observations remains \citep{Hirschmann:2014a}.  A
particularly curious observational result is ``galaxy conformity", in
which halos with red central galaxies preferentially have red
satellites~\citep{Weinmann:2006a}.  This effect extends even beyond
the virial radius to surrounding centrals, and it is not reproduced at
the observed level in SAMs~\citep{Kauffmann:2013}.  It can be
reproduced in ``age abundance matching'' models, an extension of
abundance matching that uses halo formation times to assign SFR or
colors \citep{Hearin:2014}, but the physics that drives conformity
remains unclear.  Satellite and environmental quenching has not yet
been extensively investigated in self-consistent cosmological
simulations, but there is clearly much to be learned by doing so and
this is an area where much progress can be made in the near future.

\subsection{Internal Structure and Kinematics}
\label{sec:structure}
\subsubsection{Formation of Galactic Disks}
\label{sec:structure:disks}
\begin{figure}
\centerline{\includegraphics[width=0.9\textwidth]{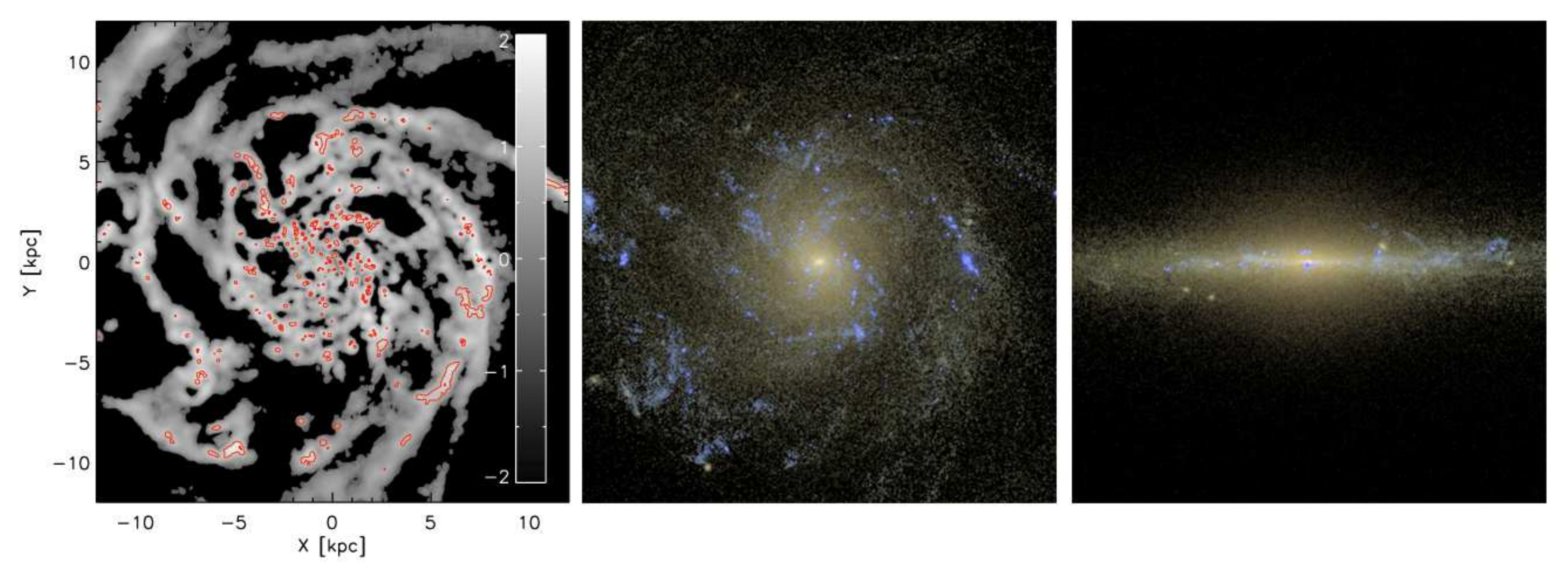}}
\centerline{\includegraphics[width=0.45\textwidth]{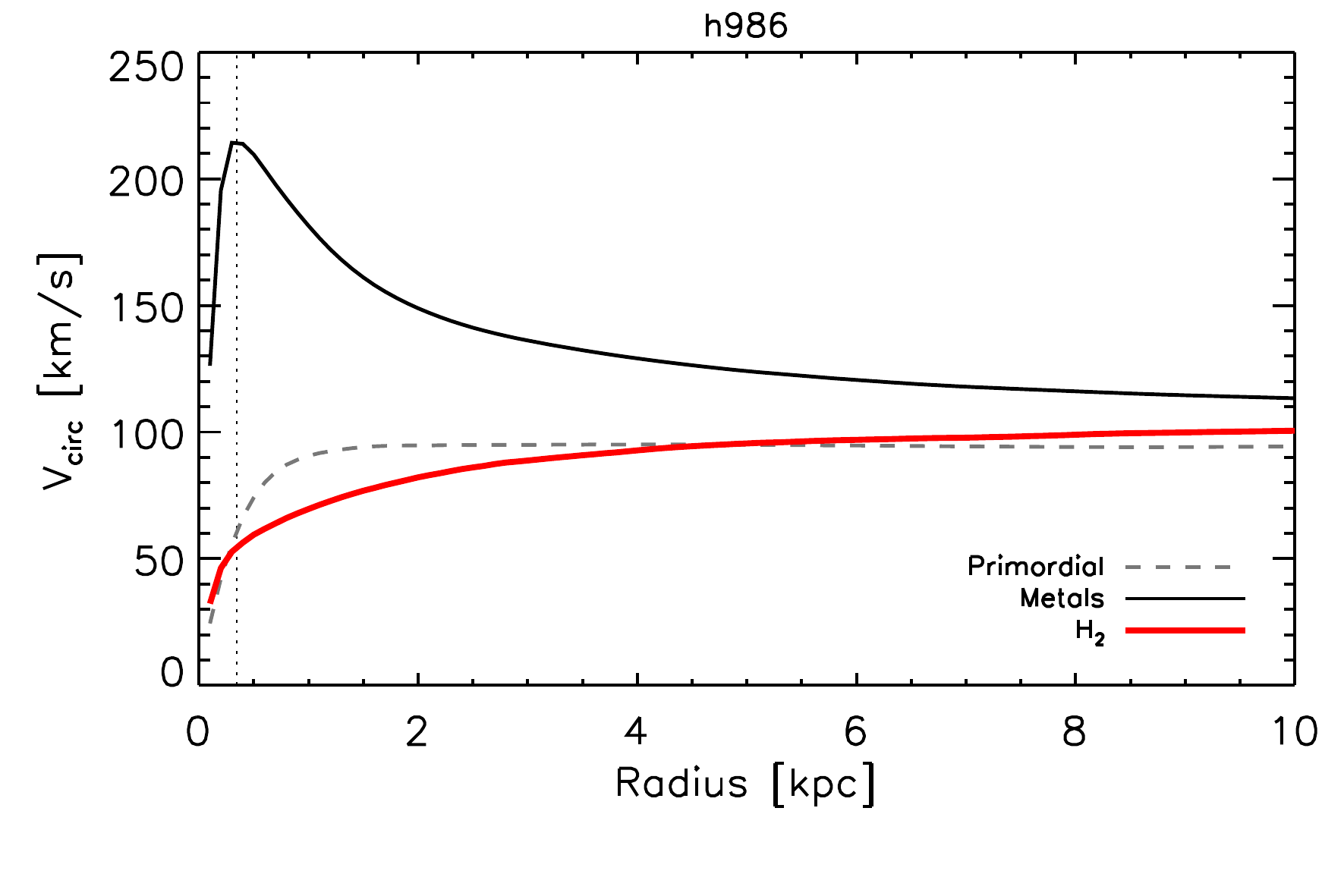}\includegraphics[width=0.45\textwidth]{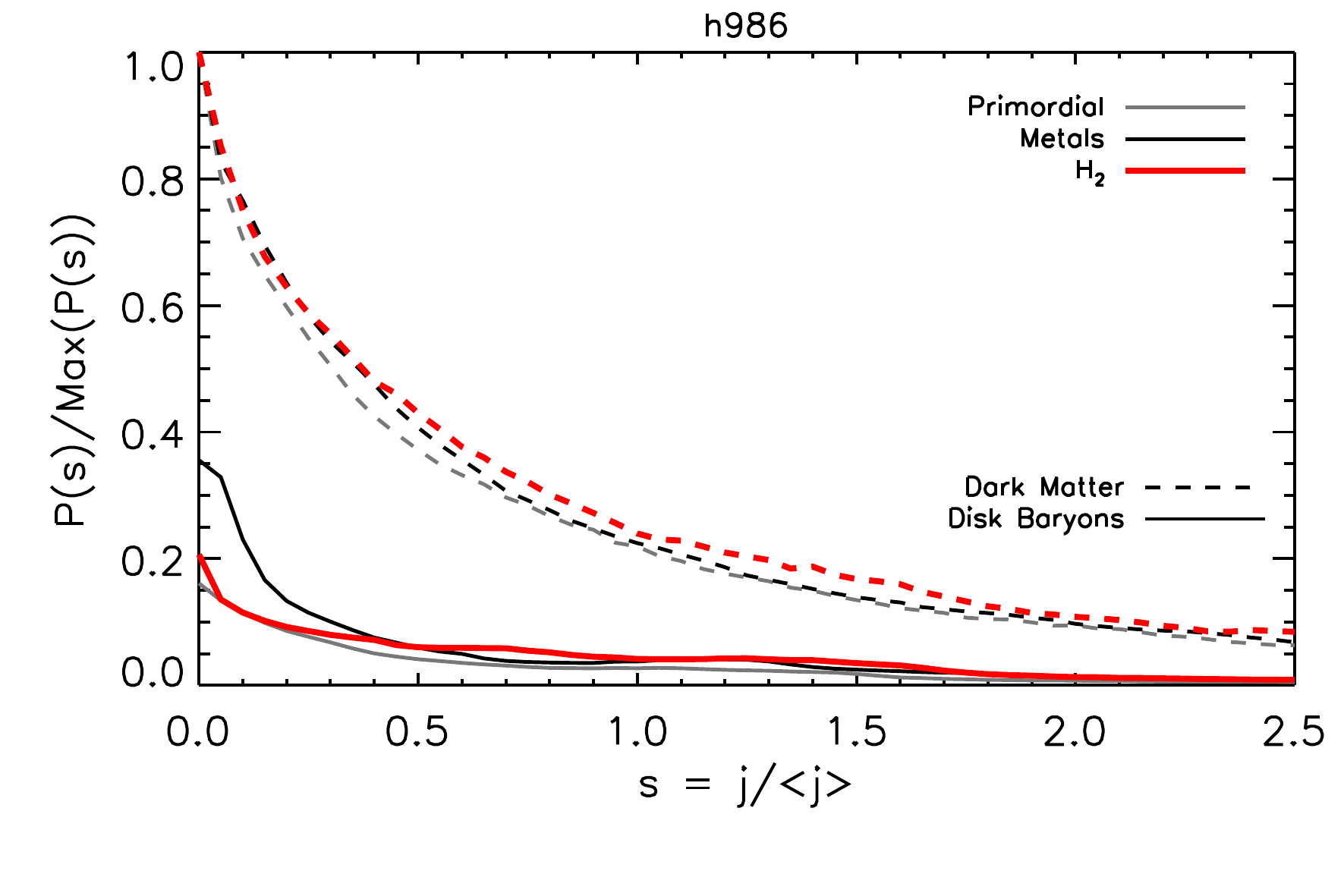}}
\centerline{\includegraphics[width=0.45\textwidth]{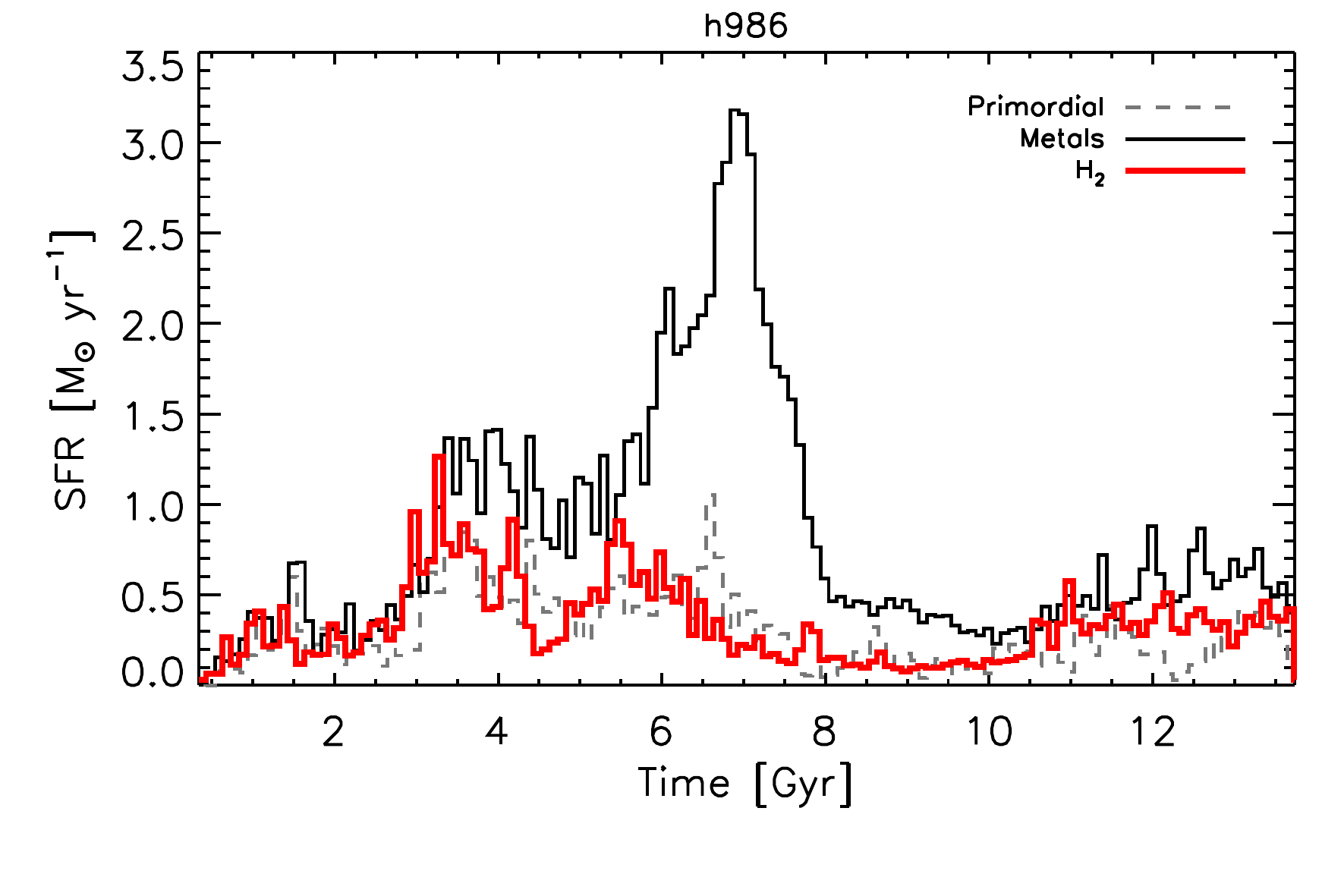}}
\caption{\footnotesize Top: Gas surface density and simulated
  observations of a spiral galaxy formed in a high-resolution zoom-in
  simulation. Top left: \HI\ gas surface density (gray) and H$_2$
  surface density (red). Top middle and right: optical images of
  stellar emission, showing the galaxy in face-on and edge-on
  orientations. Middle left: rotation curve for the same initial
  conditions, but with different sub-grid physics for treating cooling
  (primordial only vs. metal-line) and the ISM (\Htwo-based cooling
  and SF). Middle right: distribution of scaled specific angular
  momentum for the dark matter and baryons in these same
  simulations. Bottom: star formation histories for the same three
  models shown in the middle panels. Reproduced from
  \protect\citet{Christensen:2014}. }
\label{fig:nicedisk}
\end{figure}

The longstanding conventional paradigm to explain the origin of
galactic disks posits that gas accreting from the halo conserves its
specific angular momentum {\bf j}, thereby settling into a
disk~\citep{Fall:1980,Mo:1998}.  While modern cosmological simulations
support this basic paradigm, they suggest that the full story is much
more complicated.

The \emph{average} specific angular momentum of galactic disks is
indeed comparable to that expected from conserving the {\bf j} from
the halo~\citep{Dalcanton:1997,Dutton:2012}.  However, the {\it
distribution} of {\bf j} within disks predicted from simple infall
is strongly inconsistent with observations, in the sense that observed
galaxies have a strong deficit of low-{\bf j} gas, a mild deficit of
high-{\bf j} gas, and a large excess of intermediate-{\bf j} gas
\citep{vandenbosch:2001,Bullock:2001}.  This suggests that some
process removes low-{\bf j} gas and deposits it at intermediate-{\bf
  j}.

Early numerical hydrodynamic simulations of disk galaxy formation
suffered an even more severe ``angular momentum catastrophe'', as they
produced disks with much \emph{lower} average {\bf j} than the halo,
indicating that a large amount of {\bf j} was being lost during the
formation process. For many years, simulations were only able to
produce very compact disks with large spheroids, and were unable to
produce spirals even as late-type as the Milky Way
\citep{Steinmetz:1999,Sommer-Larsen:1999,Navarro:2000}.  Moreover,
these galaxies exhibited centrally-peaked rotation curves in
disagreement with observed flat rotation curves, did not lie on the
observed Tully-Fisher relation, and formed far too large a fraction of
their available baryons into stars. It was gradually realized that the
origin of this angular momentum catastrophe lay in too-efficient star
formation and gas consumption in small objects at high redshift.
These then assembled into low redshift galaxies via relatively
gas-poor mergers, which are very efficient at dissipating angular
momentum and building spheroids \citep{Maller:2002}.

Implementing more efficient star formation feedback has proven to be
the key to solving all of these problems
\citep[e.g.][]{Governato:2007,Guedes:2011}. Stellar-driven winds
preferentially remove low angular momentum gas from the centers of
galaxies, and deposit it in the disk outskirts after re-torquing in
the halo~\citep{Brook:2012,Ubler:2014}. Feedback also makes star
formation less efficient, keeping galaxies gas rich, which makes disks
more resiliant to mergers \citep{Robertson:2006a,Governato:2009}.
Finally, the baryonic mass of small infalling satellites, particularly
at early epochs, is greatly reduced, thereby mitigating early spheroid
growth via merging.

Fig.~\ref{fig:nicedisk} shows a state-of-the-art high resolution
zoom-in simulation of a disk galaxy using the GASOLINE code
\citep{Christensen:2012}. One can see that it is now possible to form
very late-type and even bulgeless galaxies
\citep{Christensen:2014}. The same simulations predict a $z=0$ $m_{\rm
  star}-M_{\rm halo}$ relation in agreement with observational
constraints~\citep{Munshi:2013}.  In dwarf galaxies, the same ``blast
wave'' feedback model can impulsively heat the dark matter, removing
the central cusp generically predicted in dark matter simulations, and
thereby producing rotation curves in better agreement with
observations~\citep{Governato:2010,Pontzen:2012,Oh:2011}. As a bonus,
\citet{Brooks:2014} have shown that destroying the central cusps in
dwarf galaxies leads to enhanced tidal stripping of satellites. The
combined effects of energetic stellar feedback and enhanced stripping
produces satellites with internal kinematics that agree with observed
dwarf spheroidal galaxies in the Local Group, plausibly resolving the
``too big to fail'' problem pointed out by
\citet{Boylan-Kolchin:2011}. While these successes may be specific to
a particular sub-grid model for feedback that may or may not be fully
accurate, nonetheless it suggests it is possible to form disk galaxies
with realistic properties within a $\Lambda$CDM universe provided that
the sub-grid treatment of stellar feedback and the ISM possess certain
key features. First, stellar feedback must be effective at keeping
galaxy-wide star formation efficiencies low, and stellar winds must
preferentially remove low-angular momentum material. Second, star
formation should occur only in very dense, highly clustered
environments like those that are expected to form GMCs, not smoothly
distributed over the whole disk, which helps to make stellar feedback
more efficient because the star formation is highly clustered as in
real galaxies.

Recent observations of disks during Cosmic Noon have presented new
challenges for models.  Disks at $z\sim 2$ are observed to be
substantially puffier, having rotation velocity $V_{\rm rot}$ divided
by gas dispersion $\sigma$ of $\sim 3$ as opposed to $\sim 10$ for
today's disks~\citep{Forster:2009}.  Many $z\sim 2$ disks also have
large, bright clumps that comprise a substantial portion of the disk
star formation~\citep{Guo:2012}, though significantly less of the
stellar mass \citep{Wuyts:2012}, and are generating
outflows~\citep{Genzel:2011}.  While most of these objects are
sufficiently massive by $z\sim 2$ to likely evolve into ellipticals
today, lower mass objects that will evolve into today's disks
generally have even higher $V_{\rm rot}/\sigma$.  Understanding the
origin of these properties and the evolution of the population to
$z=0$ has become a major cottage industry.

As discussed earlier, ISM pressurization was found to be necessary to
stabilize disks against fragmentation~\citep{Robertson:2004} and form
spiral galaxies like those observed today. With the discovery of
clumpy disks at $z\sim 2$ \citep{Elmegreen:2005,Forster:2009}, the
abundant clumps that formed in simulations without ISM pressurization
began to be touted as a ``feature''
\citep{Bournaud:2007,Dekel:2009b,Ceverino:2010}.  However, such models
produced overly high stellar fractions, typically $\ga 50\%$ whereas
abundance matching constraints suggest a value of $\sim
10-20\%$~\citep{Moster:2010,Behroozi:2010,Wake:2011}.  After
implementing more efficient stellar feedback due to radiation pressure
from young stars, \citet{Ceverino:2014} reduced the stellar fractions
by a factor of 2--3, but are they are still somewhat high
\citep{Moody:2014}.  Models with significant ISM pressurization,
either imposed~\citep{Genel:2012} or self-consistently
generated~\citep{Faucher:2013}, matched stellar fraction constraints
and still produced massive clumps, but these were less prominent and
quicker to disrupt.

The clump formation can be understood analytically in the context of
the \citet{Toomre:1964} $Q\equiv c_s\Omega/\pi G\Sigma$, where $c_s$
is the sound speed, $\Omega$ is the angular speed, and $\Sigma$ is the
local surface density.  If $Q<1$ gravitational collapse can overcome
shearing disruption, and the region is unstable.  Clumps are observed
to be regions with $Q\ll 1$~\citep{Genzel:2011}, and simulations
indicate that clumps are unstable regions self-regulated by
gravity~\citep{Ceverino:2010}.  The characteristic mass scale for
instability is $m_{\rm clump}\la 10^9\, M_\odot$, in good agreement
with observations suggesting clumps up to these masses~\citep{Genzel:2011}.
If this is the basic origin of clumps, then they are expected to
become less prominent in disks at later epochs, reducing in mass as
disks settle~\citep{Dekel:2009b}.  Simulations show that disks do
indeed settle towards $z=0$, in accord with
observations~\citep{Kassin:2014}.  If cosmological accretion drives
turbulence in the ISM, then the settling may be due to the decreasing
accretion rate at lower redshifts \citep{Genel:2012b}, though
\citet{Hopkins:2013b} argue that accretion does not drive the
turbulence in galactic disks. It remains to be demonstrated that a
single sub-grid ISM model can simultaneously reproduce the $\sim 10^9
\,M_\odot$ clumps in turbulent high-$z$ disks along with thin disks
with $\la 10^6\, M_\odot$ clumps today like the Milky Way. Nonetheless
there is at least a plausible description for the evolution of clumps
in disks across cosmic time.

\subsubsection{Formation of Spheroid-dominated Galaxies}

Since the seminal work of \citet{Toomre:1977}, it has been recognized
that nearly equal mass (``major'') mergers can efficiently remove
angular momentum from stellar disks, producing dispersion-dominated
spheroids
\citep{Toomre:1977,Barnes:1988,Barnes:1992,Hernquist:1992,Hernquist:1993b,Mihos:1996a}. Unequal
mass (``minor'') mergers down to mass ratios of $\sim$ 1:10 can thicken disks
and build up the spheroid component of galaxies
\citep{Walker:1996,Moster:2010b}. Mergers are expected to be ubiquitous
in the hierarchical CDM paradigm. Thus the most basic picture of the
origin of the two dominant classes of galaxy morphologies is that
smooth accretion of gas produces disks, and mergers destroy disks and
build spheroid-dominated galaxies. A merger-driven formation mechanism
for spheroid-dominated galaxies has been implemented in most SAMs
since the earliest such models
\citep{Kauffmann:1993,Kauffmann:1996b,Baugh:1996,SP:1999}, motivated
by the studies based on binary mergers simulated with numerical
hydrodynamics. These early works and others over the past decade
\citep[e.g.][]{Delucia:2006,Parry:2009} have shown that this picture
can qualitatively reproduce many of the observed correlations
pertaining to galaxy morphology, namely, spheroid dominated galaxies
are predicted to be more massive, more common in massive halos,
redder, and to have older stellar populations. In empirical support of
this picture, it has been shown that the observed rate of mergers
derived from pair counts and visually identified interacting galaxies
is in plausible statistical agreement with the build-up of the
quiescent, spheroid-dominated population
\citep{Robaina:2010,Hopkins:2008b}.

More recent work has led to a refinement of the merger
picture. Numerical hydrodynamic simulations showed that gas-rich
mergers do not drive efficient angular momentum loss, and so lead to
re-formation of disk-dominated galaxies
\citep{Robertson:2004,Robertson:2006a,Hopkins:2009a}. In addition,
following the formation of a spheroid via a merger, newly accreted gas
can re-form a disk. Thus, a picture has developed in which
\emph{morphological transformation and morphological demographics are
  intimately linked with feedback and quenching}. It is known that the
massive early-type galaxies in the local universe formed most of their
stars at least 8-10 Gyr ago, around $z\sim 2$--4
\citep{Trager:2000,Thomas:2005}. We also know that the average massive
star-forming galaxies at $z\sim 2$ are quite gas rich
\citep{Tacconi:2013,Genzel:2014}. Thus, in order to produce a
spheroid-dominated population at $z=0$, some process had to consume or
remove much of the gas from their progenitors \emph{before} they
merged, \emph{and} prevent significant amounts of new gas from
cooling. This appears to point qualitatively towards a combination of
``ejective'' and ``preventative'' feedback, perhaps linked with two
different modes of AGN feedback.

It has been suggested that spheroids may also form and grow \emph{in
  situ} due to internal gravitational instabilities.  There are two
different kinds of internal processes that may grow bulges, which are
frequently grouped together under the term ``disk instabilities'', but
which are physically quite distinct and are thought to produce
fundamentally different kinds of bulges. We have already discussed the
formation of giant clumps in Toomre-unstable disks (see
\S\ref{sec:structure:disks}), sometimes called ``violent disk
instabilities'' \citep{Dekel:2009b}. If these clumps survive and
migrate to the galaxy center, they may form a \emph{classical} bulge
\citep{Elmegreen:2008,Dekel:2009b,Bournaud:2011b}. However, there
remains some debate about the importance of clumps in feeding spheroid
growth. Simulations implementing kinetic feedback that were able to
match $m_{\rm star}-M_{\rm halo}$ constraints suggested that clumps
mostly disrupt before reaching the center
\citep{Genel:2012a}. \citet{Hopkins:2012b} also found that in
simulations of isolated disks (not cosmological) with a suite of
physically motivated stellar feedback physics, even large clumps
mostly blow themselves apart while in the disk, thereby only modestly
contributing to spheroid growth.  However, some recent simulations
suggest that clumps can survive substantially longer than a disk
dynamical time and grow a
spheroid~\citep{Mandelker:2014,Bournaud:2014}.  Comparing stellar and
SFR maps, \citet{Wuyts:2012} showed that clump lifetimes are $\sim
100-200$~Myr, which would suggest disruption unless inward migration
can occur on a single dynamical time or less, but radial age
gradients of clumps suggest somewhat longer
lifetimes~\citep{Genzel:2011}. Additionally, as the giant clumps orbit
within the disk, even if they disrupt before reaching the center, they
may drive inflows of gas into the galaxy nucleus, via the same sort of
physics as merger-induced nuclear inflows \citep{Bournaud:2011}.

The other process that is referred to as a ``disk instability'' is not
really a (global) instability at all. It involves the secular transfer
of mass into a compact, dynamically hot component via the formation of
a bar \citep{Toomre:1964,Hohl:1971,Ostriker:1973,Combes:1990}. The
topic of galactic bars is largely outside of the scope of this review,
but a few points are worth briefly noting. First, viewed side-on, bars
may be identified as ``boxy bulges'' (our Galaxy is a familiar
example), but if viewed face-on these structures would not be
identified as bulges \citep{Combes:1990}.  It is generally impossible
to robustly distinguish bars from bulges in distant galaxies. Second,
secular processes can redistribute angular momentum and mass within
the disk, building a \emph{pseudobulge}
\citep{Kormendy:2004,Kormendy:2013b}. In constrast to the violent disk
instabilities described above, \emph{the disk essentially remains in
  dynamical equilibrium during this secular evolution}. The
fundamental differences between \emph{classical bulges} and
\emph{pseudobulges} are briefly summarized in \S\ref{sec:structure},
and a much more complete discussion is given in
\citet{Kormendy:2004}. The stronger correlation between black hole
mass and \emph{classical} bulge mass recently emphasized by
\citet{Kormendy:2013} is presumably evidence that the processes that
build classical bulges (mergers and violent disk instabilities) are
most closely connected with black hole fueling.

\subsubsection{Demographics of Spheroid- and Disk-Dominated Galaxies}

Explaining the demographics of galaxies of different morphologies is
another challenge for theory. Detailed quantitative statistical
comparisons between the predictions of cosmological simulations and
observations of galaxy morphological demographics are difficult,
because up until now, most observational studies of galaxy morphology
have used classifiers that are not straightforward for models to
predict. Semi-analytic models predict the fraction of stellar mass or
light in a spheroid component ($B/T$), while most observational
studies use visual morphological classification or statistics such as
Sersic index or concentration.  A few observational studies have
carried out decompositions into spheroid and disk contributions
\citep{Simard:2011,Gadotti:2009,Bluck:2014}, but there are large
uncertainties in these decompositions as well \citep[see
  e.g.][]{Tasca:2011,Benson:2007}. There is a large dispersion in
observational estimates of galaxy morphological demographics derived
from different surveys and classification methods.

A number of studies have compared the predictions of SAMs with
observational estimates of luminosity or stellar mass functions
divided by galaxy morphology, or with the fraction of disk- or
spheroid-dominated galaxies as a function of stellar mass
\citep[e.g.][]{Benson:2007,Parry:2009,Guo:2011,Porter:2014}. These
studies all found fairly good agreement between the predictions of
these different SAMs and the observations, but interestingly the
dominant mechanism that drives spheroid growth is different in
different models, as we discuss further below. 

Although the details of the prescriptions differ, all semi-analytic
models that attempt to track galaxy morphology assume that mergers
destroy disks and build spheroids. However SAM-based studies have
found, to varying degrees, that non-merger related mechanisms for
spheroid growth may be needed. The most commonly invoked alternative
to mergers is a ``disk instability'' mode as described above. This is
assumed to occur when the mass in the disk exceeds a critical value
that depends on the angular momentum of the disk material. The
implementation of this process varies widely between models, leading
to significantly differing conclusions about its importance. The
GALFORM SAMs assume that when a disk becomes unstable, \emph{all} of
the stars and gas in the disk are moved to a spheroid component. They
find that these disk instabilities are the dominant channel for
spheroid growth except at the highest stellar masses
\citep{Parry:2009}. Other SAMs
\citep[][]{Delucia:2007,Guo:2011,Porter:2014} make a more moderate
assumption, that just enough stars or stars and gas are moved from the
disk to the spheroid to return the system to stability. These models
find that disk instabilities appear to be needed to reproduce the
observed numbers of spheroid dominated galaxies at intermediate masses
\citep{Porter:2014}, but are sub-dominant in driving spheroid growth
at all masses. An important and apparently robust prediction is that
models with a ``disk instability'' driven channel for spheroid growth
appear to form massive spheroids earlier than models in which
spheroids form only via mergers
\citep{Delucia:2011,Porter:2014}. These predictions can now be
confronted with observations from the new generation of medium-deep
surveys with HST (Brennan et al. in prep).

The flip side of producing enough spheroid dominated galaxies is the
challenge of producing galaxies that are close to pure disks, which
are perhaps surprisingly frequently observed in the real Universe
\citep{Kormendy:2010,Fisher:2011}. If mergers destroy disks and build
spheroids, and nearly all halos of all masses have experienced mergers
during the course of their formation history, as predicted by \LCDM,
is it possible to reconcile the existence of these objects with the
\LCDM\ picture? \citet{Hopkins:2009b} showed that including the
suppression of disk destruction in mergers with high gas fraction
progenitors alleviates this problem, bringing predictions into
agreement with observations in a semi-analytic model --- the majority
of mergers occur at high redshift when galaxy gas fractions are
expected to have been fairly high. Furthermore, Moster et al.
(\citeyear{Moster:2010b,Moster:2012}) showed that accounting for the
presence of both cold gas in the disk and hot gas in the halo
decreases disk heating due to minor mergers by a factor of 2--3
relative to previous calculations that included dissipationless
components (stars and dark matter) only. However, \citet{Porter:2014}
showed that adding a disk instability driven channel for spheroid
formation, tuned to reproduce the abundances of spheroid-dominated
galaxies, may leave behind too few objects with extremely low $B/T\la
0.2$. Detailed studies with larger samples of galaxies simulated at
high resolution in a full cosmological context are required to
determine whether this is truly a fundamental problem for \LCDM, but
it remains a serious concern.

Extensive detailed predictions on morphological demographics from
numerical cosmological simulations have not yet appeared in the
literature. Such studies should be possible with the new generation of
simulations, and detailed analysis of these simulations should help
shed light on the physical mechanisms that are responsible for shaping
galaxy morphology.

\subsubsection{Structural Scaling Relations}
\label{sec:results:structure}

The existence of structural scaling relations for galaxies, the
relationship between the structure of disks and spheroids at a given
mass scale, and the evolution of these relations over cosmic time,
encode crucial information about galaxy formation and provide
stringent constraints for models.

What physics determines the internal structure of galaxies? The most
basic picture is that dark matter and diffuse gas acquire angular
momentum through tidal torques and mergers
\citep{Peebles:1969,Vitvitska:2002}, leading to dark matter and
gaseous halos with a broad log-normal distribution of \emph{spin
parameters}. The dimensionless spin parameter is usually defined as
\[ \lambda \equiv \frac{J|E|^{1/2}}{GM^{5/2}} \]
where $M$, $J$, and $E$ are the mass, angular momentum, and total
energy of the system, respectively (MvdBW, p. 502). If we assume,
perhaps na\"{i}vely, that the halo gas conserves its angular momentum
as it cools and collapses to form a disk, and that the post-collapse
disk surface density profile has an exponential form, then the disk
scale radius will be given by
\[ r_s = \frac{1}{\sqrt{2}} \lambda r_{\rm vir} F^{-1}_R F^{-1/2}_E \]
where $r_{\rm vir}$ is the virial radius of the halo, and $F_R$ and
$F_E$ are functions that account for the initial density profile of
the dark matter halo and the contraction of the inner halo due to the
increased gravitational force after the gas falls in
\citep{Mo:1998}. The rotation velocity can then be calculated by
adding the contribution of the exponential disk and the contracted
halo in quadrature.

In spite of its simplicity, this model does remarkably well at
reproducing the size-mass relation for disk galaxies and its evolution
since $z\sim 2$
\citep{Somerville:2008a,Firmani:2009,Dutton:2011}. Recently,
high-resolution numerical hydro simulations have also been shown to be
quite successful at reproducing the size-mass relation for galactic
disks and its evolution \citep{Brooks:2011,Aumer:2013b}. As discussed
above, hydro simulations have only recently been able to successfully
reproduce disk sizes, and including strong star formation driven
outflows that preferentially remove the low angular momentum material
appears to be a crucial component of this success.  Recent simulations
suggest that accretion by ``cold streams'' may bring most of the gas
into galaxies, with an average specific angular momentum that is a
factor of $\sim 2$--3 \emph{higher} than that of the dark matter halo
\citep{Stewart:2013}. About a factor of 2--3 of this angular momentum
is then lost via torques within the mis-aligned disk and via outflows
\citep{Danovich:2014}. The success of the simple model for predicting
disk sizes may therefore be simply a happy accident.

Much of the convincing evidence for the importance of mergers in
producing spheroid-dominated galaxies comes from the success of merger
simulations in reproducing structural properties of classical bulges
and elliptical galaxies. For example, early work
\citep{Hernquist:1992,Barnes:1992} showed that mergers transform
rotationally supported disks with exponential light profiles into
slowly rotating remnants with luminosity profiles that are
well-described by an $r^{1/4}$ form over a large radial range. More
recently, it has been shown that remnants of binary disk mergers lie
on the observed fundamental plane
\citep{Robertson:2006b,Hopkins:2009c}.

A striking recent observation is that, at fixed stellar mass,
spheroid-dominated galaxies at $z\sim 2$ have much smaller sizes and
central densities higher by orders of magnitude compared to today's
\citep[e.g.][]{Trujillo:2006,vanDokkum:2008,vanDokkum:2014,vanderwel:2014,Barro:2013}.
For dissipationless (dry) mergers on parabolic orbits (hence with
small orbital energy), energy conservation and the virial theorem can
be used to show that, given a progenitor mass ratio $\eta\leq 1$ and
ratio of squares of their velocity dispersions of $\epsilon\leq 1$,
the ratio of final to initial radius is given by
\begin{equation}
\frac{r_f}{r_i} = \frac{(1+\eta)^2}{1+\epsilon\eta}
\end{equation}
\citep{Naab:2009}.  For a 1:1 merger, $\eta=\epsilon=1$, hence
$r_f/r_i=2$, which leads to a modest surface density reduction of a
factor of four.  One can show that, for a given total mass increase,
the size is increased much more by a series of minor mergers than by a
single major one.  Numerical simulations confirm such a size increase
in dissipationless mergers, which generally move galaxies along the
mass-size relation \citep{Boylan-Kolchin:2005}.  This can reproduce
the observed size increase and central density
reduction~\citep{Naab:2009,Oser:2012} since $z\sim 2$ for
cosmologically-plausible merger histories~\citep{Gabor:2012}, via
minor mergers depositing material predominantly in the outskirts.

If disks (star forming galaxies) are continuously being transformed
into spheroids (quiescent galaxies), as the demographic observations
indicate (see \S\ref{sec:intro:obs}), how then can we understand the
very different slopes and evolution of the size-mass relationship for
disks and spheroids? Several recent works have pointed out that
accounting for the effects of \emph{dissipation} in gas-rich mergers,
can lead to important changes in the scaling relations
\citep{Covington:2008,Hopkins:2009c,Hopkins:2010b,Shankar:2010,Shankar:2013,Covington:2011,Porter:2014}.
In the presence of gas,
energy is dissipated, which can lead to merger remnants that are
\emph{smaller and denser} than their progenitors.  \citet{Porter:2014}
implemented a recipe for computing spheroid sizes and velocity
dispersions based on a simple analytic model including the effects of
gas dissipation, tuned to binary merger simulations, self-consistently
within the Santa Cruz SAM. They showed that without any tuning the
model predicts rapid size evolution of spheroid-dominated galaxies
since $z\sim 2$, along with the weaker evolution in the Faber-Jackson
relation, in very good quantitative agreement with the observed
structural relations.

In this picture, dissipation plays a major role in explaining the
different slope, scatter, and evolution of the size-mass relation for
spheroid dominated (quiescent) galaxies relative to disks. Lower mass
spheroids have lower mass progenitors, which have higher gas fractions
at all redshifts. More gas means more dissipation and smaller
remnants, thus a steeper size-mass relation. Progenitors at higher
redshifts have higher gas fractions than those at lower redshift, so
the size-mass relation for spheroids ``tilts away'' from that for
disks more, contributing to more rapid size evolution especially for
the lower-mass spheroids. The decrease in scatter occurs because disks
with higher angular momentum have larger radii and lower gas
densities, resulting in less efficient star formation. These large
radius disks therefore end up with higher gas fractions, and
experience more dissipation when they merge, producing smaller
remnants.
Similarly, the observed tilt of the Fundamental Plane can be explained
by the expected trends in galaxy gas content with mass and redshift,
and the physics of gas dissipation in mergers
\citep{Hopkins:2009c,Covington:2011,Porter:2014}.

\section{Summary and Outlook}
\label{sec:summary}

Galaxy formation models set within the hierarchical CDM paradigm have
made remarkable progress over the past decade.  In this review, we
have focused on the methods and phenomenology of models that attempt
to track astrophysical processes and predict galaxy properties within
a cosmological framework. We identified a set of key observations that
current models strive to reproduce, and which describe the assembly of
galaxies from Cosmic Noon ($z\sim 2$--3) to the present. These
observations include distribution functions of global properties such
as stellar mass functions and global scaling relations such as those
between stellar mass and SFR, gas fraction, and ISM metallicity. In
addition, observations are now starting to provide measurements of
galaxy demographics, how the break-down of the galaxy population in
terms of star-forming and quiescent, and disk and spheroid dominated
objects, has evolved over this time period. The observed relationships
between \emph{global} and \emph{structural} properties (such as light
profile shape, size or internal density, and kinematics) and their
evolution provide even stronger constraints on models. We described
how well current state-of-the-art galaxy formation models are able to
reproduce these observations, and what we have learned from their
successes and failures about the physics of galaxy formation.

Although many discrepancies with observations remain, overall we would
give today's suite of galaxy formation models a passing
grade. Summarizing the scorecard we have discussed in detail in this
article:
\begin{itemize}

\item Qualitatively, hierarchical models correctly predict the
  build-up of stellar mass over cosmic time, with massive galaxies
  forming earlier and more rapidly than low-mass
  galaxies. Quantitatively, most models agree with observed galaxy
  number densities from $z\sim 4$--0 at least at the factor of 2--3
  level. However, models tend to predict that galaxies have nearly
  self-similar star formation histories, while observations imply a
  stronger mass dependence for these histories (sometimes known as
  ``downsizing''). This is part of a set of linked discrepancies
  connected with low-mass galaxies that we termed the ``dwarf galaxy
  conundrum'', which remains an open puzzle for models.

\item Models predict qualitatively the right slope and evolution of
  mean global scaling relations between stellar mass and SFR, gas
  fraction, or gas phase metallcity. These linked correlations can be
  understood at the most basic level via a very simple flow model
  describing an approximate equilibrium between galactic inflows and
  outflows. Quantitatively, compared with our current observational
  estimates, models tend to predict a SFMS and MZR that are too steep,
  and possibly gas fractions that are too low at intermediate
  redshift, in galaxies with low stellar masses. These are additional
  symptoms of the dwarf galaxy conundrum mentioned above. Also,
  models have difficulty reproducing the observed redshift dependence
  of the sSFR at any mass, indicating that real galaxies deviate from
  the simple equilibrium model.

\item Models can qualitatively explain the existence of two basic
  morphological types, disks and spheroids, via two different assembly
  modes. Disks are formed via smooth accretion of diffuse gas, which
  largely conserves its angular momentum, while spheroids are formed
  via gas-poor mergers that efficiently transfer angular
  momentum. Recently, numerical simulations demonstrated the ability
  to form pure disks in at least some cases -- a major achievement as
  previous generations of simulations were only able to form
  spheroid-dominated galaxies.  The strong feedback in such models
  also results in rotation curves and Local Group satellite
  demographics in better agreement with observations, which had
  previously been identified as a fundamental challenge to cold and
  collisionless dark matter.  However, it is still unclear how well
  models match observed morphological demographics and their evolution
  in detail. There is still much debate about how efficiently mergers
  can build spheroids, how this depends on the parameters of the
  merger and the gas fraction of the progenitors, and the role of
  other processes such as secular evolution and violent disk
  instabilities.

\item Models predict the correct qualitative trends between stellar
  population demographics (the fraction of SF and quiescent galaxies)
  and internal properties such as stellar mass: galaxies with higher
  stellar masses, higher spheroid fractions, and higher central
  densities have a higher probability of being quiescent. Some models
  correctly reproduce the dependence of quiescent fraction on
  environmental parameters such as large scale density as well, but
  the physics specific to the quenching of satellite galaxies remains
  imperfectly understood. Quantitatively, models still have difficulty
  reproducing observed color or sSFR distributions in detail.  Models
  have not yet extensively confronted the emerging measurements of
  stellar population demographics at high redshift.

\item Many models are able to at least qualitatively reproduce the
  observed sizes and internal velocities of observed galaxies, and
  scaling relations such as the Kormendy (size-mass), Tully-Fisher
  (mass-velocity), and Fundamental Plane relations. Reproducing
  observed disk sizes in numerical simulations has been a multi-decade
  struggle, and the solution has emerged through a combination of
  greatly increased resolution, more physical treatment of the ISM,
  and the effective implementation of stellar winds that
  preferentially remove low-angular momentum gas. Correctly
  reproducing the structural scaling relations and their evolution for
  \emph{both disks and spheroids}, as well as the correct overall
  evolution of the number densities of these two populations, remains
  an open challenge for models. 

\end{itemize}

Although there remain a wide range of models, and a healthy
diversity of computational methods, virtually all models implement a
qualitatively similar set of core physical processes.  While it is
possible that all models are being led down the garden path due to
their reliance on phenomenology, the concordance among models using
different methods is encouraging, and strongly suggests that we are
making fundamental progress in at least identifying the main physical
players involved. Some of the core processes identified include the
prevalance of cold smooth accretion in building disks and fueling
star-forming galaxies, the ubiquity and efficiency of star
formation-driven outflows, the importance of black hole-related
feedback in quenching star formation in massive galaxies,
merger-driven morphological evolution that depends on the gas content
of progenitors, and various physical processes that uniquely impact
satellite galaxies once they fall into a larger halo containing hot
gas. 
In addition, the convergence towards a similar qualitative view of the
\emph{types} of processes that are needed in different circumstances,
based on more empirical considerations (e.g., preventative
vs. ejective feedback, internal vs. environmental quenching, etc.) is
also encouraging.

Many of these processes connect stellar scales to cosmological scales,
making {\it ab initio} modeling nearly impossible, and forcing models
to rely on phenomenological prescriptions to describe sub-grid
physics, which must be calibrated in some way by observations. It is
clear that many model results are sensitive to the details of these
sub-grid recipes and their implementation, leading to a valid concern
that these models may have little genuine predictive power
\citep{Haas:2013a,Haas:2013b}. There are perhaps two ways to combat
this concern. First, although the sub-grid recipes and their
parameters are tuned to match a subset of observations, the current
suite of available observations is diverse and rich enough that by
confronting models with as wide as possible a set of complementary
constraints, and by exploring different sub-grid recipes and
implementations, one can isolate the approach that satisfies the
broadest set of constraints.  Second, by studying ``small scale''
simulations (for example, of the ISM and the formation of individual
stars, or regions near SMBH), one may hope to place the sub-grid
recipes used in our cosmological simulations on a physically grounded
foundation. Zoom techniques are now enabling simulations that are
starting to bridge the gap between the scales of individual stars and
SMBH and galactic scales. Although it will not be feasible to simulate
cosmological volumes with these techniques in the near future, they
will allow us to learn much about the interface between the
``micro''-scales of stars and BH and the ``macro'' scales of galaxies.

In addition, there are physical processes that may be important in
regulating galaxy formation, but which are not commonly included in
current ``mainstream'' models. These include turbulence, magnetic
fields, cosmic rays, and self-consistent radiative transfer. It is
important to carry out experiments to determine the importance of
these processes in shaping the observable properties of galaxies, and
there has been significant recent progress on this front as well
\citep[e.g.][]{Scannapieco:2010,Kotarba:2011,Wise:2011,Mendygral:2012,Hanasz:2013,Pfrommer:2013}.

Ideally, we would obtain direct observational confirmation (or
refutation) of the set of core processes that models currently
invoke. However, in many cases this is challenging.  Smooth gas
accretion (i.e. in small enough lumps that adiabatically add to the
fuel supply without disrupting galactic structure) is expected to be
very diffuse and in a phase that is difficult ($T\sim 10^4$ K) to
nearly impossible ($T\sim 10^5$ K) to detect.  The key parameter
characterizing outflow efficiency in models is the mass loss rate, but
since outflows are highly multi-phase it is difficult to account for
all the mass~\citep{Veilleux:2005}.  We observe the signatures of
black hole activity in the form of AGN and jets associated with
massive galaxies, but it is difficult to observationally constrain how
efficiently this energy couples to surrounding gas to enact
quenching. We can observe signposts and signatures of mergers in the
form of close pairs and morphologically disturbed galaxies, but their
rate is difficult to quantify precisely and their effect is difficult
to directly constrain observationally. We can measure the statistics
of galaxies in different environments, but it has been difficult with
existing samples to disentangle the correlations between environment
and internal properties, and to locate the environments at high
redshift that are the progenitors of typical groups and clusters in
the local Universe.

However, there are several important observational developments taking
place now, or on the horizon, that will challenge and help to refine our
models of galaxy formation. First, a new generation of sub-mm and
radio interferometers (including ALMA, NOEMA, JVLA, Apertif, ASKAP,
MeerKAT, and the SKA) will literally revolutionize our ability to
characterize the cold gas in the ISM of galaxies out to high redshifts
\citep{Carilli:2013}. Second, high-resolution spectroscopy in the
rest-frame UV is now able to probe the diffuse gas and metals in the
circumgalactic medium of galaxies for galaxy-targetted sightlines
spanning a diverse range of galaxy types, from nearby galaxies to
$z\sim 2$--3
\citep[e.g.][]{Rudie:2012,Prochaska:2013,Tumlinson:2013,Peeples:2014}.
This provides constraints on the gas and metals that have been ejected
by the winds invoked by our models, which probably comprise a much
larger fraction of the halo baryon budget than the stars and cold ISM
within galaxies. Third, Integral Field Unit spectrographs on ground
based telescopes and on JWST will allow us to better characterize
stellar and AGN driven winds and to study spatially resolved stellar
population parameters and kinematics for large samples of nearby and
high-redshift galaxies. Finally, high-resolution, wide-field
multi-wavelength imaging such as will be possible with WFIRST will
enable us to study galaxy internal properties and demographics over a
much larger range of environments, allowing us to better disentangle
internal and environmental forces and accumulating better statistics
for rare events such as mergers and luminous AGN.

We thus live in interesting times where modelers are now offering some
specific and non-trivial challenges to observers to go out and
confirm, or rule out, key physical processes.  Just because a given
mechanism is not observed does not mean it is not occuring; one must
carefully assess whether that mechanism is expected to be observable.
A general trend is that models make the most direct predictions about
gas-related processes, particularly inflows and outflows in the baryon
cycle, with the growth of stellar and black hole components being
almost a side-effect.  Hence, in principle, observations that trace gas
processes directly offer the greatest potential for new advances and
constraints.  Modelers and observers must work together to identify
key tests that can be conducted with present and upcoming facilities
in order to constrain the core physical processes.  The emerging
interplay between galaxy formation models and state-of-the-art
telescopes is the hallmark of a healthy and vibrant area of research.

The way forward for galaxy formation models is fairly clear, but
immensely challenging.  As a blueprint, consider the Lyman-$\alpha$
forest: several decades ago, studying the interplay of gas dynamics
with cosmological structure formation led to a revolution in our
understanding that eventually resulted in the Ly$\alpha$ forest
becoming a pillar of precision cosmology.  Our goal should be to
equivalently turn galaxy formation into a precision field, where
parameterized recipes are tied to the physics of small scale processes
in such a way that the parameters no longer need to be empirically
tuned, but are constrained by our physical understanding of those
processes (e.g. stellar evolution models, or BH accretion disk
models).  Numerical simulations on different scales (zooms and
cosmological volumes) and semi-analytic models have crucial and
complementary roles to play in this process, helping to better
understand the physics in detail as well as to synthesize and
parameterize it within a \LCDM\ context.  It is almost surely the case
that the physical processes included in models so far will not be
sufficient to fully describe galaxy evolution, and there will be many
twists and surprises forthcoming.  Hence there is much work to be
done, but it appears that cosmological models of galaxy formation are
on a secure foundation for the exciting journey ahead.

\section*{Acknowledgements}
It would take many more pages to thank all the colleagues who have
provided valuable insights and participated in discussions that have
shaped this work, and we apologize for the inevitable choices we had
to make to review this vast topic while conforming to page limits.
But we would particularly like to thank Andrew Benson, Richard Bower,
Rob Crain, Darren Croton, Michelle Furlong, Violeta Gonzalez-Perez,
Bruno Henriques, Yu Lu, Joop Schaye, Paul Torrey, Mark Vogelsberger,
and their collaborators for providing the data from their models and
simulations and for constructive comments on this article.  We also
thank Avishai Dekel, Thorsten Naab, and Gerg\"{o} Popping for
comments. We especially thank our Scientific Editor, John Kormendy,
for his thorough reading of the paper, and for comments and
suggestions that improved the article. rss gratefully acknowledges the
generous support of the Downsbrough family.  RD acknowledges support
from the South African Research Chairs Initiative and the South
African National Research Foundation.  This work was supported in part
by NASA grant NNX12AH86G.

\section*{Glossary of Acronyms}

\noindent AGB: asymptotic giant branch\\
\noindent AGN: active galactic nucleus\\
\noindent ALMA: Atacama Large Millimeter/submillimeter Array\\
\noindent AMR: adaptive mesh refinement\\
\noindent ASKAP: Australian Square Kilometre Array Pathfinder\\
\noindent BH: black hole\\
\noindent BLR: broad line region \\
\noindent $B/T$: bulge to total ratio\\
\noindent CDM: cold dark Matter\\
\noindent ckpc: comoving kilaparsec\\
\noindent cMpc: comoving megaparsec\\
\noindent EC-SPH: entropy-conserving SPH\\
\noindent EoR: epoch of reionization\\
\noindent DI-SPH: density-independent SPH\\
\noindent FOF: friends of friends\\
\noindent GMC: giant molecular cloud\\
\noindent GR: General Relativity\\
\noindent \HI: neutral hydrogen\\
\noindent HOD: halo occupation distribution\\
\noindent HST: Hubble Space Telescope\\
\noindent IGM: intergalactic medium\\
\noindent IMF: initial mass function\\
\noindent IR: infrared\\
\noindent ISM: interstellar medium\\
\noindent JVLA: Jansky Very Large Array\\
\noindent JWST: James Webb Space Telescope\\
\noindent LF: luminosity functions\\
\noindent MeerKAT: http://www.ska.ac.za/meerkat/index.php\\
\noindent MZR: mass-metallicity relation\\
\noindent NFW: Navarro-Frenk-White\\
\noindent NOEMA: NOrthern Extended Millimeter Array; http://iram-institute.org/EN/noema-project.php\\
\noindent PE-SPH: pressure-entropy SPH\\
\noindent PM: particle-mesh\\
\noindent PPM: Piecewise Parabolic Method\\
\noindent SAM: semi-analytic model\\
\noindent SDSS: Sloan Digital Sky Survey\\
\noindent SED: spectral energy eistribution\\
\noindent SF: star formation\\
\noindent SFE: star formation efficiency\\
\noindent SFMS: star forming main sequence\\
\noindent SFR: star formation rate\\
\noindent SKA: Square Kilometer Array\\
\noindent SMBH: supermassive black hole\\
\noindent SMF: stellar mass function\\
\noindent SN: supernova\\
\noindent sSFR: specific star formation rate\\
\noindent SHAM: sub-halo abundance matching\\
\noindent SO: spherical overdensity\\
\noindent SPH: smoothed particle hydrodynamics\\
\noindent ULIRG: ultra-luminous infrared galaxies\\
\noindent UV: ultraviolet\\
\noindent WFIRST: Wide-Field Infrared Survey Telescope\\
\noindent \LCDM: cold dark matter with a cosmological constant ($\Lambda$)\\

\bibliography{araa}
\end{document}